\documentclass[%
reprint,
amsmath,amssymb,
aps,
rmp,
]{revtex4-2}
\usepackage{graphicx}
\usepackage{dcolumn}
\usepackage{bm}
\usepackage[mathlines]{lineno}

\usepackage[english]{babel} 
\usepackage{cuted}
\usepackage[colorlinks=true,citecolor=blue,urlcolor=blue]{hyperref}
\hypersetup{colorlinks,linkcolor=blue,filecolor=blue,urlcolor=blue,citecolor=blue,}
\usepackage{multirow}

\usepackage{caption}
\usepackage{ragged2e}

\usepackage{natbib}

\begin{document}


\captionsetup[figure]{justification=justified,labelfont={bf},name={Fig.}}
\preprint{APS/123-QED}

\title{Quantum Correlation Sharing: A Review On Recent Progress From Nonlocality To Other Non-Classical Correlations}

\author{Zinuo Cai}
\affiliation{Key Laboratory of Low-Dimension Quantum Structures and Quantum Control of Ministry of Education, Synergetic Innovation Center for Quantum Effects and Applications, Xiangjiang-Laboratory and Department of Physics, Hunan Normal University, Changsha 410081, China}
\author{Tianfeng Feng}
\affiliation{QICI Quantum Information and Computation Initiative, Department of Computer Science, The University of Hong Kong, Pokfulam Road, Hong Kong}
\author{Changliang Ren}\thanks{Corresponding author: renchangliang@hunnu.edu.cn}
\affiliation{Key Laboratory of Low-Dimension Quantum Structures and Quantum Control of Ministry of Education, Synergetic Innovation Center for Quantum Effects and Applications, Xiangjiang-Laboratory and Department of Physics, Hunan Normal University, Changsha 410081, China}

\author{Xiaoqi Zhou}
\affiliation{State Key Laboratory of Optoelectronic Materials and Technologies and School of Physics, Sun Yat-sen University, 510006, Guangzhou, China}

\author{Jingling Chen}
\affiliation{Theoretical Physics Division of Chern Institute of Mathematics, Nankai University, Tianjin 300071, China}

\date{\today}

\begin{abstract}
This review offers a comprehensive exploration and synthesis of recent advancements in the domain of quantum correlation sharing facilitated through sequential measurements. We initiate our inquiry by delving into the interpretation of the joint probability, laying the foundation for an examination of quantum correlations within the context of specific measurement methods. The subsequent section meticulously explores nonlocal sharing under diverse measurement strategies and scenarios, with a specific focus on investigating the impact of these strategies on the dissemination of quantum nonlocality. Key perspectives such as ``asymmetry'' and ``weak value'' are scrutinized through detailed analyses across various scenarios, allowing us to evaluate the potential of nonlocality sharing. We also provide a retrospective overview of experimental endeavors associated with this phenomenon. The third part of our exploration presents research findings on steering sharing, offering clarity on the feasibility of steering sharing and summarizing the distinctive properties of quantum steering sharing in different scenarios. Continuing our journey, the fourth section delves into discussions on the sharing of diverse quantum correlations, encompassing network nonlocality, quantum entanglement, and quantum contextuality. Moving forward, the fifth section conducts a comprehensive review of the progress in the application of quantum correlation sharing, specifically based on sequential measurement strategies. Applications such as quantum random access coding, random number generation, and self-testing tasks are highlighted. Finally, we discuss and list some of the key unresolved issues in this research field, and conclude the entire article.
\end{abstract}
\maketitle
\setcounter{page}{0}
\thispagestyle{empty}

\tableofcontents

\section{Introduction}

{\color{black} Quantum mechanics (QM) stands as one of the most groundbreaking developments in twentieth-century physics, providing access to the underlying laws governing the microscopic realm. Despite its significance, the interpretation of QM has been a persistent source of controversy, epitomized by the Einstein-Podolsky-Rosen (EPR) Paradox presented in 1935 \cite{EPR1935PhysRev.47.777}. This paradox challenges the completeness of QM, positing that quantum entanglement results in instantaneous effects at a distance. The characterization of nonclassical correlations inherent in quantum entanglement has been a long-standing challenge, with questions about the existence of physical systems possessing such properties and the lack of effective differentiation methods.

Over a decade of investigation has revealed that aligning quantum correlation predictions with a specific local hidden variables theory can resolve the controversy. John Bell's introduction of the ``Bell inequality'' \cite{J.S.Bell.PhysicsPhysiqueFizika.1.195} in 1964 established a constant boundary that local hidden variables theories must adhere to, but QM predictions may exceed this boundary in certain cases. \cite{chshPhysRevLett.23.880} proposed a more experimentally accessible inequality, initiating a series of experiments to test the existence of quantum correlation phenomena. In 1982, Aspect's experiments provided the first detection of Bell inequality violation, signifying the presence of Bell nonlocality \cite{Aspect.PhysRevLett.49.1804}.

While Bell nonlocality and quantum entanglement were initially considered synonymous, Werner's work in 1989 demonstrated that quantum entanglement in mixed states is a weaker quantum correlation than Bell nonlocality \cite{Werner.PhysRevA.40.4277}. In 2007, Wiseman et al. introduced another type of nonclassical quantum correlation derived from the EPR paradox, known as quantum steering \cite{Wiseman2007PhysRevLett.98.140402}, which relates quantum entanglement to Bell nonlocality. The understanding of quantum correlation has steadily evolved over more than 70 years.

Experimental verification of various quantum correlations has flourished over the past four decades (\citealp{Weihs.PhysRevLett.81.5039};
\citealp{Rowe.MA.Nature409.791};
\citealp{Hofmann.Science.337.72_2012};
\citealp{Giustin.Nature497.227_2013};
\citealp{Christensen.PhysRevLett.111.130406}),
with milestone experiments such as the vulnerability-free verification of Bell nonlocality in 2015 \cite{hensen2015loophole,Giustina_PhysRevLett.115.250401,Shalm.PhysRevLett.115.250402}. Several systematic review articles cover quantum entanglement, quantum steering, and Bell nonlocality 
(\citealp{horodecki2009quantum};
\citealp{Brunner.RevModPhys.86.419};
\citealp{Cavalcanti.Rep.Prog.Phys.80.024001_2017};
\citealp{Uola.RevModPhys.92.015001}). Ongoing research efforts focus on characterizing quantum correlations in more intricate networks \cite{tavakoli2022bell,Pozas_2022_PhysRevLett.128.010403_full} and temporal correlations 
(\citealp{kuhuanyu_2018_PhysRevA.98.022104};
\citealp{Fabio_2018_PhysRevA.98.012328};
\citealp{zhaozhikuan_2018_PhysRevA.98.052312};
\citealp{ZhangTian_2020_NewJ.Phys}).

Initially a fundamental topic in QM, the exploration of quantum correlation has paved the way for the emergence of Quantum Information Science \cite{Bennett_1984Quantum,bennett2000quantum,bouwmeester2000physics}. This interdisciplinary field aims to develop practical applications based on advancements in quantum correlation, encompassing quantum communication 
(\citealp{bouwmeester1997experimental};
\citealp{gisin2007quantum};
\citealp{prl2016.117.210501};
\citealp{ren2017ground};
\citealp{prl2020.125.070502};
\citealp{prl2021.126.010503}),
quantum computing (\citealp{wang2017high};
\citealp{XIN201817};
\citealp{PhysRevLett.120.040501};
\citealp{arute2019quantum};
\citealp{PhysRevLett.125.260404};
\citealp{PhysRevLett.130.090602}),
quantum precision measurement
(\citealp{Yamamoto_1990_87precision};
\citealp{Braunstein_1992_PhysRevLett.69.3598};
\citealp{Vittorio_2006_PhysRevLett.96.010401};
\citealp{Nicholas_2011_PhysRevLett.107.113603}), and other related topics.

Although there have been many reviews about different quantum correlations
(\citealp{horodecki2009quantum};
\citealp{Brunner.RevModPhys.86.419};
\citealp{Cavalcanti.Rep.Prog.Phys.80.024001_2017};
\citealp{Uola.RevModPhys.92.015001}), we will introduce and review the recent progress in studying quantum correlations based on sequential measurements from a different research perspective, which has gained increasing attention in recent years. The investigation into this research theme is closely tied to the repetitive utilization of quantum resources. Indeed, there have been some studies on quantum correlations based on temporal measurements before (\citealp{nogues_1999nature_seeing};
\citealp{gudder_2001_sequential};
\citealp{Pryde_2004_PhysRevLett.92.190402};
\citealp{Sciarrino_2006_PhysRevLett.96.020408};
\citealp{Radim_2011_PhysRevA.83.032311};
\citealp{nagali_2012_SciRep_testing};
\citealp{Burgarth_2015_NewJPhys113055};
\citealp{westerbaan_2016_universal}). However, \citeauthor{Silva.Ralph_PhysRevLett.114.250401_2015} approached this issue from a fresh perspective in 2015. In a sense, this serves as a seed work for this review.
They reported a counterintuitive result, namely that weak measurements can be used to share nonlocality between two or more observers. The scenario involves allocating a pair of a maximally entangled states to three observers:  Alice, Bob$_1$, and Bob$_2$. Alice accesses one qubit, while the two Bobs access the other. Alice performs a projective measurement on her qubit, and Bob$_1$ performs a weak measurement on his qubit, then transfers it to Bob$_2$, who eventually performs a projective measurement. The measurement results indicate that both Bob$_1$ and Bob$_2$ may simultaneously violate the CHSH inequality with Alice. So far, along this path, a series of effective theoretical studies about nonlocality sharing have been investigated
(\citealp{Rodrigo.Gallego_New.J.Phys.16.033037_2014};
\citealp{Mal.Shiladitya_math4030048_2016};
\citealp{Matteo.Schiavon_Quantum.Sci.Technol.2.015010_2017};
\citealp{Florian.J.Curchod_LIPIcs.2018.8580_2018};
\citealp{Tavakoli_2018_PhysRevA.97.032131};
\citealp{Li.Xiao.Gang_Quantum.Inf.Process.s11128.018.1888.8_2018};
\citealp{Datta.Shounak_PhysRevA.98.042311_2018};
\citealp{Das.Debarshi_PhysRevA.99.022305_2019};
\citealp{Saha.Sutapa_Quantum.Inf.Process.s11128.018.2161.x_2019};
\citealp{Ren.Changliang_PhysRevA.100.052121_2019};
\citealp{Brown.Peter.J_PhysRevLett.125.090401_2020};
\citealp{Cheng.Shuming_PhysRevA.104.L060201_2021};
\citealp{Zhang.Tinggui_PhysRevA.103.032216_2021};
\citealp{cabello_2021_bell};
\citealp{Ren.Changliang_PhysRevA.105.052221_2022};
\citealp{Anna._PhysRevLett.129.230402_2022};
\citealp{Zhu.Jie_PhysRevA.105.032211_2022};
\citealp{Ming.Liang.Hu_SCI.CHINA.PHYS.MECH.s11433.022.1892.0_2022LimitsOS};
\citealp{Zhang.Tinggui_Quantum.Inf.Process.s11128.022.03699.z_2022};
\citealp{Cheng.Shuming_PhysRevA.105.022411_2022};
\citealp{Ya.Xi_arXiv.2207.00296v1_2022};
\citealp{Xi_Ya_PhysRevA.107.062419};
\citealp{Sasmal_2023_arXiv_unbounded}),
and various experimental demonstrations have also been carried out (\citealp{Schiavon._Quantum.Sci.Technol.2_015010_2017};
\citealp{Hu.Meng.Jun_Quantuminf.s41534-018-0115-x_2018};
\citealp{Feng.Tianfeng_PhysRevA.102.032220_2020};
\citealp{Foletto.Giulio_PHYS.REV.APPL.13.044008_2020};
\citealp{Foletto_PhysRevApplied.13.069902};
\citealp{Xiao.Ya_arXiv.2212.03815_2022}).

These findings not only offer fresh insights into the interplay between nonlocality and quantum measurement, particularly the manifestation of nonlocality sharing through weak measurements, but also stimulate further investigation into analogous concepts such as quantum entanglement, quantum coherence, quantum discord, and quantum steering. Moreover, researchers have delved into the broader concept of sequential measurements as an alternative to weak measurements, prompting explorations into the properties of quantum correlation within a comprehensive framework of sequential measurement. This has led to the discovery of various applications of sequential measurements in understanding quantum correlation, such as, the steering sharing in theoretical (\citealp{Sasmal.Souradeep_PhysRevA.98.012305_2018};
\citealp{YaoDan.PhysRevA.103.052207_2021};
\citealp{Shashank._PhysRevA.103.022421_2021};
\citealp{Zhu.Jie_PhysRevA.105.032211_2022};
\citealp{Han.XinHong_arXiv.2303.05954_2023})
and experimental investigation 
(\citealp{Xiang.Yu_PhysRevA.99.010104_2019};
\citealp{Shenoy.H._PhysRevA.99.022317_2019};
\citealp{Yeon.Ho.Choi._optica.394667_2020};
\citealp{Paul_PhysRevA.102.052209_2020};
\citealp{Han.Xinhong_Quantum.Inf.Process.s11128-021-03211-z_2021};
\citealp{Kun.Liu_arXiv.2102.12166_2021};
\citealp{Tong.Jun.Liu_OPT.OE.470229_2022};
\citealp{Han.Xin-Hong_PhysRevA.106.042416_2022};
\citealp{Lijian_2022EPR_steering};
\citealp{Qiao-Qiao.Lv_Phys.AMath.Theor.2307.09928_2023};
\citealp{Chen.Yuyu.Chin.Phys.B.32.040309_2023}),
network nonlocality sharing (\citealp{Hou.Wenlin_PhysRevA.105.042436_2022};
\citealp{Wang.Jian.Hui_PhysRevA.106.052412_2022};
\citealp{Pritam_2022_PhysRevA.106.052413};
\citealp{Mahato.Shyam.Sundar._PhysRevA.106.042218_2022};
\citealp{Zhang.Tinggui_FrontPhys.s11467.022.1242.6_2023};
\citealp{Kumar.Rahul_Quantum.Studies.s40509-023-00300-9_2023};
\citealp{MaoYaLi._PhysRevResearch.5.013104_2023}), 
entanglement sharing 
(\citealp{Bera.Anindita_PhysRevA.98.062304_2018};
\citealp{Maity.Ananda.G_PhysRevA.101.042340_2020};
\citealp{Foletto.Giulio_PHYS.REV.APPL.13.044008_2020};
\citealp{Srivastava.Chirag_PhysRevA.103.032408_2021};
\citealp{Pandit.Mahasweta_PhysRevA.106.032419_2022};
\citealp{Srivastava.Chirag_PhysRevA.105.062413_2022};
\citealp{Arun.Kumar.Das_Quantum.Inf.Process.s11128-022-03728-x_2022};
\citealp{Srivastava_2022sequential};
\citealp{HuMingLiang_PhysRevA.108.012423};
\citealp{limaosheng_2023arXiv_sequentially}), 
contextuality sharing
(\citealp{Kumari.Asmita_PhysRevA.100.062130_2019};
\citealp{Anwer.Hammad_Quantum.q-2021-09-28-551_2021};
\citealp{Chaturvedi.Anubhav_Quantum.q-2021-06-29-484_2021};
\citealp{Kumari.A_PhysRevA.107.012615_2023}).

Furthermore, it has been observed that these research outcomes are closely connected to numerous quantum information tasks, leading to a gradual exploration of issues related to sequential measurements, including, self-test (\citealp{Karthik.Mohan_NEW.J.PHYS.10.1088.1367-2630.ab3773_2019};
\citealp{Chirag_PhysRevA.103.032408_2021};
\citealp{Miklin.Nikolai_PhysRevResearch.2.033014_2020};
\citealp{Armin.Tavakoli_sciadv.aaw6664_2020};
\citealp{Mukherjee.Sumit_PhysRevA.104.062214_2021};
\citealp{Pan.A.K_PhysRevA.104.022212_2021};
\citealp{Prabuddha.Roy_NEW.J.PHYS.10.1088.1367-2630.acb4b5_2023};),
random number generation 
(\citealp{Curchod.F.J_PhysRevA.95.020102_2017};
\citealp{BrianCoyle_EPTCS.273.2_2018};
\citealp{Xue-Bi.An_Opt.Lett.ol-43-14-3437_2018};
\citealp{Bowles_Quantum.q-2020-10-19-344_2020};
\citealp{Pan.A.K_PhysRevA.104.022212_2021};
\citealp{Foletto.Giulio_PhysRevA.103.062206_2021}), 
and quantum random access codes 
(\citealp{Li.Hong-Wei_Commun.Phys.s42005-018-0011-x_2018};
\citealp{Karthik.Mohan_NEW.J.PHYS.10.1088.1367-2630.ab3773_2019};
\citealp{Anwer.Hammad_PhysRevLett.125.080403_2020};
\citealp{Foletto.Giulio_PhysRevResearch.2.033205_2020};
\citealp{Shihui.Wei_NEW.J.PHYS.1367-2630/abf614_2021};
\citealp{Das.Debarshi_PhysRevA.104.L060602_2021};
\citealp{Xiao.Ya_PhysRevResearch.3.023081_2021};
\citealp{Xiao.Yao_Quantum.Inf.Process.s11128-023-03924-3_2023}) etc.

The structure of this article is outlined as follows: in the subsequent part of the introduction, we delve into the nature of EPR-quantum correlations, elucidating key definitions and elementary criteria from the perspective of joint probabilities. Fundamental concepts are introduced, including a brief comparison of von Neumann measurements, weak measurements, and POVM, as these form the core of subsequent discussions. Section \ref{nonlocality_sharing_section} is dedicated to a comprehensive exploration of nonlocality sharing under various measurement strategies and scenarios. Specifically, we investigate the impact of measurement strategies on quantum nonlocality sharing, examining perspectives such as ``asymmetry'' and ``weak value''. Through detailed analyses in diverse scenarios, we assess the capability of nonlocal sharing, accompanied by a retrospective overview of experimental efforts related to nonlocal sharing.
Section \ref{steering_sharing_section} introduces steering sharing, elucidating the shareability of steering, with a focus on directional aspects in sequential scenarios. Various measurement strategies for steering sharing are demonstrated.
The sharing of other types of quantum correlations, including network nonlocality, quantum entanglement, and quantum contextuality, are discussed in Section \ref{sharing_othercorrelation_section}. It is noteworthy that these extended considerations offer new perspectives for sharing quantum physical resources in the realm of quantum physics.
In Section \ref{application_section}, we review the sharing of quantum correlations under quantum sequential measurement strategies in applications such as quantum random access coding, random number generation, and self-testing tasks. Furthermore, it provides updates on the latest developments.
Finally, Section \ref{conclusion_section} presents our conclusions, along with prospects and outlooks for the field of quantum resource sharing derived from sequential measurements.

{\color{black}
\subsection{ Quantum Correlations Vs Joint Probability From Sequential Measurements}
The research on different quantum correlations has developed for a long time, and a multitude of different correlation criteria have been proposed, such as, for Bell nonlocality (\citealp{J.S.Bell.PhysicsPhysiqueFizika.1.195};
\citealp{chshPhysRevLett.23.880};
\citealp{Mermin.PhysRevLett.65.1838};
\citealp{Ardehali.PhysRevA.46.5375};
\citealp{A.V.BelinskiPhys.Usp.36653_1993};
\citealp{Collins.PhysRevLett.88.040404};
\citealp{M.Zukowski.PhysRevLett.88.210401};
\citealp{Brukner.PhysRevLett.89.197901};
\citealp{Brunner.RevModPhys.86.419}), 
quantum entanglement (\citealp{Micha.Horodecki.Phys.Lett.A.223.1_1996};
\citealp{Barbara.M.Terhal.Phys.Lett.A.271.319_2000};
\citealp{bruss.J.Mod.Opt.49.1399_2002};
\citealp{guhne.J.Mod.Opt.50.1079_2003};
\citealp{Uffink_Phys.Lett.A.372.1205_2008};
\citealp{Lougovski.PhysRevA.80.034302_2009};
\citealp{Ryszard_2009_RevModPhys.81.865};
\citealp{Otfried_2009_Phys.Rep.474.1_2009};
\citealp{chruscinski.J.Phys.A.Math.Theor.47.483001_2014}), 
and quantum steering
(\citealp{Cavalcanti.PhysRevA.80.032112_2009};
\citealp{Reid.PhysRevA.88.062108};
\citealp{Cavalcanti.Opt.Soc.Am.B32.A7415_2015};
\citealp{Costa.PhysRevA.93.020103_2016}). 
All of these criteria can be expressed in terms of joint probability distribution. Hence, the EPR quantum correlation problem can be summarized as ``a thought experiment based on real joint probability''. This perspective allows us to explain almost all of the criteria and experimental demonstrations of quantum correlations to date. This review aims to comprehensively summarize the recent research on the nature of quantum correlations in the context of sequential measurements. We start by elucidating the inherent reasons for quantum correlation sharing through sequential measurements, focusing on the change of conditions of joint probability.

For a quantum state system comprising N particles, there are N observers, each assigned one particle, and each observer has the flexibility to employ multiple measurement methods. However, during each measurement cycle, each observer will only perform a single measurement on the particle they hold. This cyclic process generates a joint probability distribution based on N measurements. Subsequently, combinations of these joint probabilities, or derived physical quantities (such as averages), are utilized to showcase conflicts with classical models. For instance, in a 2-qubit system, the joint probability involves two measurements denoted as $P(a,b|A, B,\rho)$. It's important to note that the quantum state measurements exhibit the characteristic that each observer measures their particle only once. Nevertheless, this characteristic should not be regarded as an indispensable criterion. Instead, this joint probability may originate from a higher dimensional joint probability based on a more extensive set of measurements \cite{Masanes.PhysRevA.73.012112,Toner_2009_monogamy}.

\begin{widetext}
Considering a straightforward example involving a 2-qubit system, if each qubit is sequentially measured by two observers in their possession during each round—using measurement operators A$_1$ and A$_2$ for one qubit, and B$_1$ and B$_2$ for the other—one arrives at the joint probability $P(a_1, a_2, b_1, b_2 | A_1, A_2, B_1, B_2, \rho)$. Describing the correlation information of a 2-qubit quantum state through joint probability provides a more comprehensive perspective. This is due to its ability to consistently revert to the previous joint probability, as illustrated by the equation,
\begin{equation}
P(a_1, b_1 | A_1, B_1, \rho) = \sum{a_2, b_2} P(a_1, a_2, b_1, b_2 | A_1, A_2, B_1, B_2, \rho),
\end{equation}
where $\rho$ is the density matrix of the quantum state. The previous research overlooked the above-discussed approach mainly due to the measurements. Currently, the most frequently used quantum correlation criteria approach is where Alice and Bob use strong measurements for their first measurements. And if the observer makes a subsequent second measurement, the joint probability can always be expressed as,
\begin{equation}
P(a_1,a_2,b_1,b_2|A_1,A_2,B_1,B_2,\rho)=P(a_1,b_1|A_1,B_1,\rho)P(a_2|A_2,\Pi^{a_1}_{A_1})P(b_2|B_2,\Pi^{b_1}_{B_1}), 
\end{equation}
where $\Pi^{a_1}_{A_1}$ and $\Pi^{b_1}_{B_1}$ represent the eigenstates corresponding to the first measurement operator. In this case, any subsequent measurements hold no significance as they do not provide any correlation information about the initial state, due to $P(a_2, b_2 | A_2, B_2)=P(a_2|A_2,\Pi^{a_1}_{A_1})P(b_2|B_2,\Pi^{b_1}_{B_1})$. However, if the decomposition of the joint probability mentioned above is not satisfied, the joint probability $P(a_2, b_2 | A_2, B_2)$ cannot be separated into the product of two probability densities. At this stage, subsequent measurements cease to be ordinary. The non-classical correlations originally present in the density matrix $\rho$ of a composite quantum system may still be concealed within the outcomes of this joint probability.
\end{widetext}

From a physical perspective, meeting the above requirements involves minimizing the disruption of the initial state's correlation information during the forward measurements. With the development of quantum non-destructive measurement theory and technology, the mentioned process is achievable, such as weak measurement processes 
(\citealp{Aharonov_1964_PhysRev.134.B1410};
\citealp{Aharonov_1988_PhysRevLett.60.1351};
\citealp{Aharonov_2005_Quantum_Paradoxes};
\citealp{Gisin.arXiv.1407.8122v1._2016}). 
Weak measurements do not fully disturb or only partially disturb the state of the system, where the trade-off between information and disturbance has been studied (\citealp{Fuchs_1996_PhysRevA.53.2038};
\citealp{Sciarrino_2006_PhysRevLett.96.020408};
\citealp{Banaszek_2006_OpenSyst.Inf.Dyn.13.1};
\citealp{Francesco_Open.Syst.Inf.Dyn.16.29_2009};
\citealp{nagali_2012_SciRep_testing};
\citealp{Sparaciari.2014_ProbingQB}). 
Besides, the generalized Positive Operator-Valued Measurement (POVM)
(\citealp{Busch.PhysRevD.33.2253_1986};
\citealp{Paul.Springer.Berlin_1997};
\citealp{yuan_2022_brief}) may also satisfy the above conditions. Quantum correlation sharing is related to the strategy of measurements, which we will unfold this review around.

}

\subsection{The Definition And Simple Criterion Of EPR Quantum Correlation}

{\color{black}{Here, we will revisit the definitions of quantum correlations, particularly quantum entanglement, quantum steering, and Bell non-locality, derived from the EPR paradox. Additionally, we will introduce the definition of network quantum correlations which is a recent and emerging topic. The distinctions among these definitions will be elucidated within a straightforward scenario.}}


Considered a scenario involving two observers, Alice and Bob, who jointly possess the state $\rho_{AB}$ of a 2-qubit quantum system. Alice performs a measurement $\hat{A}$, yielding the result `a'. Similarly, Bob measures $\hat{B}$ with the result `b'. Without loss of generality, we obtain the joint probability distribution $P(a, b|\hat{A}, \hat{B}, \rho_{AB})$ in the real-world context.

For this joint probability to accurately align with the predictions of quantum mechanics, it must adhere to the following criterion,
\begin{eqnarray}
		P(a,b| \hat{A},\hat{B},\rho_{AB})=\mathrm{Tr}\bigl[\bigl(\hat{E}^{a}_{A}\otimes\hat{E}^{b}_{B}\bigr)\rho_{AB}\bigr],
\end{eqnarray}
where ${\hat{E}^{a}_{A}}$ represents the density matrix associated with the eigenstate of the operator $\hat{A}$ and eigenvalue `$a$', and $\{\hat{E}^{b}_{B}\}$ represents the density matrix associated with the eigenstate of the operator $\hat{B}$ and eigenvalue `$b$'.

For any arbitrary measurements, if the above joint probability satisfies the following decomposition,
\begin{eqnarray}\label{lhv}
P(a,b| \hat{A},\hat{B},\rho_{AB})=\int P(a| \hat{A},\xi)P(b|\hat{B},\xi)P_{\xi}d\xi,
\end{eqnarray}
which indicates the existence of a local hidden variables (LHV) model \cite{Brunner.RevModPhys.86.419} capable of explaining the given joint probability.
Here, $\xi$ represents the hidden variable, $P_\xi$ is the distribution of the hidden variable, and $P(a| A,\xi)$ denotes the probability of measuring $\hat{A}$ and obtaining result `$a$' under the specific hidden variable $\xi$, and a similar definition for  $P(b|\hat{B},\xi)$ as well. If Eq. (\ref{lhv}) lacks a solution, it signifies that no LHV model can elucidate the joint probability distribution in the described scenario, indicating the presence of Bell nonlocality in the system.


For arbitrary measurements, if the joint probability and marginal probability satisfy\cite{Wiseman2007PhysRevLett.98.140402},
\begin{eqnarray}\label{lhs1}
	&P(a,b|\hat{A},\hat{B},\rho_{AB})=\int P(a|\hat{A},\xi)P_{Q}(b| \hat{B},\xi)P_{\xi}\mathrm{d}\xi,\\&
	P_{Q}(b| \hat{B},\xi)=\mathrm{Tr}[\hat{E}^{b}_{B}\rho_{\xi}^{B}],\label{lhs2}
\end{eqnarray}
which indicates the presence of a local hidden state (LHS) model interpretation for the joint probability. Here, $\rho_{\xi}^{B}$ represents the density matrix of the subspace controlled by Bob, satisfying 
$\int P_{\xi}\rho_{\xi}^{B}d\xi=\mathrm{Tr}_{A}[\rho_{AB}]$. $P_{Q}(b| \hat{B},\xi)$ denotes the probability of obtaining the result `$b$' when a true quantum state $\rho_{\xi}^{B}$ measured by $\hat{B}$.
Eq. (\ref{lhs1}) implies that Alice does not influence Bob's state. If no solution exists, it implies that no LHS model can elucidate the joint probability distribution in the scenario, indicating the presence of steering properties: Alice can steer Bob.


For any measurement, if the joint probability and the marginal probability adhere to the conditions,
\begin{eqnarray}\label{sm}
	&P(a,b|\hat{A},\hat{B},\rho_{AB})=\int P_Q(a|\hat{A},\xi)P_Q(b|\hat{B},\xi)P_\xi d\xi,\\&
	P_Q(a| \hat{A},\xi)=\mathrm{Tr}[\hat{E}^{a}_{A}\rho_\xi^A],\\&
	P_Q(b| \hat{B},\xi)=\mathrm{Tr}[\hat{E}^{b}_{B}\rho_\xi^B],
\end{eqnarray}
it signifies the presence of a Separable Model (SM) interpretation for the joint probability \cite{horodecki2009quantum}. Here, $\rho_{\xi}^{A}$ represents the density matrix of the subspace controlled by Alice, satisfying $\int P_{\xi}\rho_{\xi}^{A}d\xi=\mathrm{Tr}_{B}[\rho_{AB}]$. 
Eq. (\ref{sm}) is a formulation that satisfies the relevant probabilities of separable states in quantum mechanics. In the absence of a solution, it implies that the Separable Model (SM) fails to explicate the joint probabilities of the scenario, signifying the presence of entanglement in the given scenario.



According to the provided definition, excluding Local Hidden Variables (LHV) models, Local Hidden State (LHS) models, or Separable Model (SM) interpretations represents a significant challenge. In response, Bell's innovative approach is to outline the constraints of classical models rather than delving into their specific forms. When the predictions of quantum mechanics surpass the constraints imposed by classical models, the existence of non-classical correlations is affirmed. Bell intuitively proposed a method, known as Bell's inequality \cite{J.S.Bell.PhysicsPhysiqueFizika.1.195}, which will be always obeyed by any classical correlations. Since then, researchers have extensively explored various methods to elucidate quantum correlations. Actually, all of these criteria can be expressed in
terms of the joint probability distribution. Without loss of generality, we review and introduce the simplest and arguably the most renowned inequality criterion for assessing Einstein-Podolsky-Rosen (EPR) quantum correlations.

In a 2-qubit system, Alice and Bob can randomly select one from $m$ possible operators to measure their received qubits, respectively. Alice's measurements are labeled as $\{\hat{A}_x\}=\hat{n}_{x}\cdot\hat{\sigma}$, for Bob as $\{\hat{B}_y\}=\hat{n}_{y}\cdot\hat{\sigma}$. where $x\in\{0,1..m-1\}$, $y\in\{0,1..m-1\}$, $\hat{\sigma}=(\sigma_x,\sigma_y,\sigma_z)$ are three Pauli matrices. Through multiple repeated experiments, one can obtain the joint probability, denoted as $P(a,b|x,y)$. Further, the average value of the combination of any arbitrary operators can be given from the joint probability, $\langle \hat{A}_x\hat{B}_y\rangle=\sum_{a,b}(-1)^{a+b}P(a,b|x,y)$.
All Bell inequalities can be constructed by the average values, such as Clauser-Horne-Shimony-Holt (CHSH) inequality \cite{chshPhysRevLett.23.880},
\begin{align}
	S=\langle \hat{A}_0\hat{B}_0\rangle+\langle \hat{A}_0\hat{B}_1\rangle+\langle \hat{A}_1\hat{B}_0\rangle-\langle \hat{A}_1\hat{B}_1\rangle\le2.
\end{align}
This inequality can be violated under specific conditions, thereby illustrating the non-locality inherent in quantum systems.
For instance, when the initial state is the maximally entangled state, $|\psi\rangle=\frac{1}{\sqrt2}(|00\rangle+|11\rangle)$, Alice chooses $\hat{A}_0=\sigma_z$ and $\hat{A}_1=\sigma_x$, Bob chooses $\hat{B}_0=\frac{\sigma_z+\sigma_x}{2}$ and $\hat{B}_1=\frac{\sigma_z-\sigma_x}{2}$, the resulting CHSH value is given by $S=2\sqrt2\ge2$, representing the quantum-theoretical upper bound of the CHSH inequality.



Similar to the determination of Bell nonlocality,  if the joint probability distribution in the given scenario defies characterization by a Local Hidden State (LHS) model, Bob must concede the possibility that Alice might steer her state through certain ``remote actions'' within her subspace.  In such circumstances, Bob's state is considered steerable. It is important to note that ``steerable'' is asymmetric. Such as Alice can steer Bob in certain states, but the reverse may not hold. 
Similarly, we can utilize the joint probability distribution of the scenario to derive inequalities for assessing steering. A classical exemplary illustration is the linear steering inequality
\cite{Cavalcanti.PhysRevA.80.032112_2009},
\begin{align}\label{linear_inequality}
	S^m=&\sum_{i=1}^{m}|\left \langle \hat{A}_i\otimes \hat{B}_i \right \rangle  |\le \mathcal{B}_m,\\
       \mathcal{B}_{m} =&\max_{\{a_i=\pm 1\}}\left\{\frac{1}{\sqrt{m}}\lambda_{\max}\left(\sum_{i=1}^ma_i\hat{B}_i\right)\right\},\nonumber
\end{align}
where $\mathcal{B}_{m}$ represents the classical bound, $\lambda_{\max}(\hat{\mathrm{X}})$ denotes the maximum eigenvalue of $\hat{\mathrm{X}}$. The parameter $m$ signifies the number of measurement choices. Quantum steering can be determined by $S^m > \mathcal{B}_m$. Notably, $\mathcal{B}_m$ varies with alterations in the measurement settings, for instance, $\mathcal{B}_2=\frac{1}{\sqrt{2}}\approx0.7071$, $\mathcal{B}_3=\mathcal{B}_4=\frac{1}{\sqrt{3}}\approx0.5773$, $\mathcal{B}_6\approx0.5393$. Specifically, for $m=2,3$, the linear steering inequality can be expressed as $S^m=\frac{1}{\sqrt{m}}\sum_{i=1}^{m}|\left \langle A_i\otimes B_i \right \rangle  |\le 1$.



In the determination of entanglement, a commonly employed method is the entanglement witness \cite{horodecki2009quantum}. This approach relies on the expected values of specific Hermitian operators to discern entangled states from separable ones. This method of entanglement detection leverages the Hahn-Banach theorem, asserting that for any element outside a closed convex set in a normed linear space, there always exists a functional on that space that ``separates'' the element from the closed convex set.
The entanglement witness is represented by the operator W \cite{Barbara.M.Terhal.Phys.Lett.A.271.319_2000}, where $\left\langle \mathrm{W} \right\rangle_{\rho}$ denotes the expected value of the Hermitian operator W concerning the state $\rho$.
For instance, for all separable states $\rho_{o}$, it holds that $\left\langle \mathrm{W} \right\rangle_{\rho_{o}} \ge0$. Conversely, there exists at least one entangled state $\rho_{e}$ for which $\left\langle \mathrm{W} \right\rangle_{\rho_{e}} < 0$.



With the deepening exploration of quantum correlations, quantum phenomena within networks have gradually become a focal point of attention \cite{tavakoli2022bell}. The network correlation manifests in scenarios where there are multiple independent sources, representing a form of non-locality that transcends Bell's theorem 
(\citealp{M._1993_PhysRevLett.71.4287};
\citealp{Zukowski.Marek_New.York.Acad.Sci.755.91_1995};
\citealp{Panjianwei_PhysRevLett.80.3891_1998};
\citealp{Bose_PhysRevA.60.194_1999};
\citealp{Thomas_PhysRevLett.88.017903_2001};
\citealp{Branciard_2010_PhysRevLett.104.170401};
\citealp{Fritz_2012_beyond};
\citealp{Branciard_2012_PhysRevA.85.032119};
\citealp{Gisin_2017_PhysRevA.96.020304};
\citealp{Francesco_2017_NewJ.Phys.19.113020_2017};
\citealp{Tavakoli_2021_PhysRevLett.126.220401}). 
Numerous studies have delved into the investigation of these phenomena, expanding their scope to more intricate network structures
(\citealp{Tavakoli_2014_PhysRevA.90.062109};
\citealp{Mukherjee.QuantumInf.Proc.14.2025_2015};
\citealp{fritz_2016_beyond};
\citealp{Mukherjee_2017_PhysRevA.96.022103};
\citealp{Fraser.PhysRevA.98.022113_2018};
\citealp{wolfe_2019_inflation};
\citealp{Renou.PhysRevLett.123.140401_2019};
\citealp{Mukherjee.PhysRevA.101.032328_2020};
\citealp{Renou.Nature600.625_2021};
\citealp{Yang.PhysRevA.104.042405_2021};
\citealp{Pozas_2022_PhysRevLett.128.010403_full};
\citealp{Renou.PhysRevLett.128.060401_2022};
\citealp{Munshi.PhysRevA.105.032216_2022};
\citealp{Tavakoli.Rep.Prog.Phys.85.056001_2022};
\citealp{Brunner.PhysRevA.105.022206_2022}).

Considered a simple network scenario with two sources, each associated with a hidden variable $\xi_i$, $i\in\{1,2\}$. The two sources are shared between Alice and Bob, as well as between Bob and Charlie respectively. 
The initial state can be given by $\rho_{ABC}=\rho_{AB}\otimes\rho_{BC}$.
Alice performs a measurement $\hat{A}$ in her subspace, corresponding to the outputs $a$, and Charlie performs a measurement $\hat{C}$ in his subspace, corresponding to the outputs $c$. Bob carries put a Bell state measurement $\hat{B}$ on the received two particles with the outcome $b=b_0b_1$. The joint probability obtained in this scenario can be expressed as $P(a,b,c|\hat{A},\hat{B},\hat{C},\rho_{ABC})$. For any measurement, if the joint probability can be factorized \cite{Branciard_2010_PhysRevLett.104.170401},
\begin{align}\label{blhv}
	P(a,b,c|\hat{A},\hat{B},\hat{C},\rho_{ABC})
	=\int_{\xi_1} d\xi_1 \int_{\xi_2}d\xi_2 P_1(\xi_1)P_2(\xi_2)\nonumber \\
	\times P(a|\hat{A},\xi_{1})P(b|\hat{B},\xi_{1},\xi_{2})P(c|\hat{C},\xi_{2}),
\end{align}
which signifies that the joint probability is explicable through a bilocal hidden variable (BLHV) model. Here, $P_1(\xi_1)$ represents the hidden variable distribution of the first source, and similarly, $P_2(\xi_2)$ pertains to the second source. The term $P(b|\hat{B},\xi_{1},\xi_{2})$  denotes the probability distribution linked to a measurement $\hat{B}$ yielding the outcome $b$, conditioned on the determined hidden variables $\xi_1$ and $\xi_2$.
Furthermore, $P(a|\hat{A},\xi_{1})$ characterizes the probability distribution associated with a measurement $\hat{A}$ resulting in the outcome `$a$' under the influence of a determined hidden variable $\xi_1$. Analogously, the term $P(c|\hat{C},\xi_{2})$ represents the probability distribution for a measurement $\hat{C}$ resulting in the outcome $c$ under a specified hidden variable $\xi_2$.
Eq. (\ref{blhv}) delineates the local properties within the network scenario. 
If no solution exists, it implies that the joint probability for such a network scenario cannot be described by the BLHV model. This substantiates the nonlocal correlation within the network scenario.


Assuming that Alice and Charlie randomly select one of the m possible measurements to perform on the received particle, with Alice's measurement denoted as $\hat{A}_x$ and Charlie's measurement as $\hat{C}_z$, where $x,z\in\{0,1..m\}$. The joint probability $P(a,b,c|x,z)$ can be obtained. Consequently, its average value can be expressed as $\langle A_{x}{B}^{b_k}{C}_{z}\rangle=\sum_{a,b_0b_1,c}(-1)^{a+b_k+c}P(a,b_0b_1,c|x,z)$.
For the case of $m=2$, the joint probability can construct an inequality serving as a criterion for network nonlocality, such as the Branciard-Rosset-Gisin-Peronio (BRGP) inequality \cite{Branciard_2010_PhysRevLett.104.170401},
\begin{eqnarray}
	\sqrt{\mid\mathrm{I}\mid}+\sqrt{\mid \mathrm{J}\mid}\leqslant1,
\end{eqnarray}
where 
\begin{align}
	\mathrm{I}=\frac{1}{4}\sum_{x,z=0,1}\langle A_{x}B^{b_0}C_{z}\rangle,\nonumber \\
	\mathrm{J}=\frac14\sum_{x,z=0,1}(-1)^{x+z}\langle{A}_{x}{B}^{b_1}{C}_{z}\rangle.\nonumber
\end{align}
The violation of this equality transcends the constraints of locality in network scenarios, showcasing a nonlocal property inherent in the network.



The above description pertains to the most typical criteria concerning Bell nonlocality, steering, and network nonlocality. Discussions regarding these criteria take place within the framework of the simplest Bell scenarios or bilocal scenarios. For more complex scenarios, corresponding criteria can be constructed by extending classical standards, utilizing joint probabilities obtained from the measurement process. Further elaboration on this topic is presented in Sections \ref{nonlocality_sharing_section}, \ref{steering_sharing_section}, and \ref{sharing_othercorrelation_section}.

\subsection{Von Neumann Measurement, Weak Measurement, POVM}



The exploration of quantum correlations intricately intertwines with quantum measurements \cite{Aharonov_1988_PhysRevLett.60.1351,
Aharonov_2005_Quantum_Paradoxes}. Among the pivotal quantum measurements, notable ones include the von Neumann measure \cite{Busch_1996_quantum,Nielsen_2010_quantum,wheeler_2014_quantum} and the positive-operator valued measure (POVM)
(\citealp{Davies_1970_operational};
\citealp{Busch.PhysRevD.33.2253_1986};
\citealp{yuan_2022_brief}), among others.
The most basic form of quantum measurement is the von Neumann measure, also referred to as the projection-valued measure (PVM) \cite{Nielsen_2010_quantum}. It is defined through a Hermitian operator that can be decomposed as $\hat{M}=\sum_i\lambda_i|i\rangle\langle i|=\sum_i\lambda_i \hat{P}_i$. Here, $\lambda_i$ denotes the eigenvalue of the operator, $|i\rangle$ represents an eigenvector with an eigenvalue of $\lambda_i$, and $\hat{P}_i=|i\rangle\langle i|$ is designated as a von Neumann measurement operator within the Hilbert space $\mathcal{H}$. The constraints associated with this measurement are as follows,
\begin{equation}
	\begin{matrix}(i).\hat{P}_i=\hat{P}^{\dagger}_i;&& 	(ii).\hat{P}_i\ge0; && (iii).\hat{P}_i^2=\hat{P}_i;\nonumber 
 \end{matrix}
\end{equation}
\begin{equation} \label{pvm}
\begin{matrix}
(iv).\hat{P}_i\hat{P}_j=\delta_{ij},i\ne j; &&(v).\sum_{i\in M}\hat{P}_i=\mathbb{I}_{\mathrm{H}}.
\end{matrix}
\end{equation}


In a von Neumann measurement, when the measurement yields the result $\lambda_i$, the quantum state of the system undergoes collapse to the eigenstate corresponding to $\hat{P}_i$. Consequently, the post-measurement state becomes:  $|{{\psi}^{(i)}_{\mathrm{post}}}\rangle=\frac{\hat{P}_i|\psi\rangle}{\sqrt{\langle\psi|\hat{P}_i|\psi\rangle}}$, which requires normalization to ensure a total probability of 1. The Von Neumann measurements constitute a probabilistic process, and the outcomes are not deterministic but are contingent upon the probability distribution inherent in quantum states. This distribution is dictated by the inner product of quantum states and measurement operators: $p(i|\psi\rangle)=\langle\psi|\hat{P}_i|\psi\rangle$. 
The Von Neumann measurements form the foundational basis for the quantum mechanics measurement theory, providing a mathematical framework elucidating the measurement of quantum states and their evolution post-measurement.


POVM (Positive Operator-Valued Measure) is a generalized form of quantum measurements, which extends the principles of von Neumann measurements. It allows for a broader range of measurement methods, encompassing not only standard projective measurements but also non-standard ones. Within the von Neumann measurement framework, POVM measurements relax the constraints on the measurement operator regarding the idempotence condition (Eq. (\ref{pvm}. iii)) and orthogonality constraint (Eq. (\ref{pvm}. iv)), denoted as $\{\hat{Q}_i, i \in M\}$. The associated constraints are,
\begin{equation}
	\begin{matrix}
		(i).\hat{Q}_i=\hat{Q}^{\dagger}_i;& 	(ii).\hat{Q}_i\ge0; &(iii)\sum_{i\in M}\hat{Q}_i=\mathbb{I}_{\mathrm{H}}.
	\end{matrix}
\end{equation}

POVM measurements employ a set of positive operator-valued operators to delineate the measurement process, rather than solely relying on standard projective operators. These operators are Hermitian but not necessarily orthogonal. Diverging from von Neumann measurements, POVM measurements enable a broader and more generalized probability distribution. In POVM measurements, the probability of obtaining a measurement result $\lambda_i$ is determined by the inner product between the POVM elements and the quantum state $\rho$ before measurement, expressed as $p(i|\rho\rangle)=\mathrm{Tr}(\hat{Q}_i\rho)$. The significance of POVM measurements lies in their provision of a more flexible approach to measurements, enabling researchers to design various types of measurements, including those accommodating a certain degree of uncertainty or noise. This flexibility proves highly valuable in applications such as quantum information processing, quantum communication, and quantum measurements. Furthermore, POVM measurements play a pivotal role in the study of quantum measurement theory, furnishing a robust tool for describing and analyzing intricate measurement processes.

It is worth noting that, due to the absence of orthogonal constraints, directly determining the post-measurement state in POVM measurements is not feasible. In this context, the Kraus operators associated with the $\hat{Q}_i$, denoted as $\{\hat{A}_i,i\in M\}$, play a crucial role, enabling the implementation of a general measurement approach\cite{Nielsen_2010_quantum}.
The Kraus decomposition for each $i\in M$ can be expressed as $\hat{Q}_i=\hat{A}_i^{\dagger}\hat{A}_i$. 
However, the Kraus decomposition of the measurement operator $\hat{Q}_i$ lacks uniqueness, as exists any unitary operator $\hat{U}$, $\hat{U}\hat{A}_i$ still belongs to the Kraus operator. Employing the Kraus operator $\hat{A}_i=\sqrt{\hat{Q}_i}$ offers a method to describe the post-measurement state, expressed as $\rho^{(i)}_{\mathrm{post}}=\frac{\hat{A}_i\rho \hat{A}_i}{\mathrm{Tr}(\hat{A}_i^{\dagger}\hat{A}_i\rho)}$.

Building upon the standard von Neumann measurement model, Aharonov et al. have conducted research from the perspective of measurement coupling, introducing another measurement form, known as weak measurement theory. This theory is commonly employed to investigate the properties of quantum systems, particularly the relationships between states before and after measurement\cite{Aharonov_1988_PhysRevLett.60.1351}.

In the aforementioned von Neumann measurements, the outcomes are denoted as $\lambda_i$. Consequently, the quantum state of the system undergoes an inevitable collapse to the associated eigenstate of $\hat{P}_i$. The collapse states corresponding to distinct measurement outcomes are non-degenerate, expressed as $\hat{P}_i\hat{P}_j=\delta _{ij}$. At this juncture, the position of the reading pointer yields complete information about the measured physical quantity.

Von Neumann designed a model for implementing projective measurements, conceptualizing measurement as the interaction between a system and measurement apparatus. In essence, he interconnected the quantum measurement uncertainty with the macroscopic uncertainty of the measuring device
(\citealp{Neumann_1955_mathematical};
\citealp{Home_2007_PVM};
\citealp{von.Neumann_mathematical_2018}). The Aharonov model extends this concept, differing from von Neumann's measurement model in two key aspects
(\citealp{Aharonov_1988_PhysRevLett.60.1351};
\citealp{Aharonov_PhysRevA.41.11_1990};
\citealp{Aharonov_2005_Quantum_Paradoxes}). 
Firstly, in weak measurements, the interaction between the measuring device and the system is typically weak, and the collapsed states corresponding to different measurement outcomes are degenerate, minimizing significant interference with the quantum state of the system. This weak coupling property allows for a more nuanced exploration of the relationship between states before and after measurement. Secondly, Aharonov requires weak measurements to project selectively onto final states, preserving the time-reversal symmetry of the measurement process. Weak values are commonly used to represent such measurement scenarios.


Typically, the weak value denotes measurements on an ensemble of preselected and postselected systems
\cite{Aharonov_1964_PhysRev.134.B1410}. The weak value is defined as $M^w=\frac{\langle\psi_{f}|\hat{M}|\psi_{i}\rangle}{\langle\psi_{f}|\psi_{i}\rangle}$
\cite{Aharonov_1988_PhysRevLett.60.1351}, where $|\psi_{i}\rangle$ and $|\psi_{f}\rangle$ represent the initial state and the postselected state of the measurement, and $M$ is the measured observable.
In essence, weak measurement serves as a means for an observer to extract partial information from a system while causing minimal disturbance. 
There exists a delicate trade-off between the amount of information obtained from the system and the level of interference introduced to it
(\citealp{Sciarrino_2006_PhysRevLett.96.020408};
\citealp{Banaszek_2006_OpenSyst.Inf.Dyn.13.1};
\citealp{nagali_2012_SciRep_testing};
\citealp{Gisin.arXiv.1407.8122v1._2016}): 
on one hand, weaker measurements extract less information but cause less disturbance to the system; on the other hand, stronger measurements provide more information but result in greater system disturbance. We project the measurements onto the measurement operator's eigenstate space, which requires weak coupling of the system to the measuring device (often called the auxiliary device). The acquired information will be available through the measurement of the auxiliary device.

In the overall system, it is assumed that the initial quantum state of the system is $|\psi\rangle$. The quantum state of the measuring device is $|D\rangle$. The state of the measuring device on the projection of its pointer is represented by the probability density of its pointer reading, which has many possible distributions.

The initial state of the entire system is expressed as $|\Psi_{\mathrm{int}}\rangle=|\psi\rangle\otimes|D(q_0)\rangle$.
In the standard von Neumann measurement model, the interaction between the measuring device and the quantum system is
$
\mathrm{H}=\mathrm{H}_{\mathrm{int}}={\hat{M}}\otimes \hat{p}
$. Where $\hat{p}$ is the degree of freedom coupled with the pointer. The pointer reading $q$ falls on the eigenvalue $\lambda_i$ with probability $|\langle i|\psi\rangle|^2$. By projecting the pointer state onto $\lambda_i$, the final state of the whole system can be given as,
\begin{eqnarray}
	|\Psi_{\mathrm{fin}} \rangle=|\psi\rangle\otimes |D(q)\rangle=\sum_i\langle i|\psi\rangle |D(q-\lambda_i)\rangle|i\rangle.
\end{eqnarray}
During the measurement process, the movement of the pointer is proportional to the eigenvalue of the measurement.
For von Neumann measurements, the initial state of the pointer is narrower than the distance between the eigenvalues, $\langle D(q-\lambda_i)|D(q-\lambda_j) \rangle =\delta_{i,j}$,
hence reading the position of the pointer provides all information about the measured physical quantity. For weak measurements, their weak coupling properties result in the initial state of the pointer being wider than the eigenvalues, reading the position of the pointer cannot provide all information. If the pointer is very large and covers the entire eigenvalue spectrum, reading the position of the pointer provides no information due to $\langle D(q-\lambda_i)|D(q-\lambda_j) \rangle\approx 1$. The system will be not affected by interference,
\begin{align}
	{| \psi_{\mathrm{fin}} \rangle}_{q_0}=\sum_{i} \langle i |\psi \rangle \langle q_0|D(q-\lambda_i)\rangle|i\rangle \nonumber\\
	\approx \langle q_0|D(q)\rangle \sum_{i} \langle i|\psi\rangle|i\rangle\nonumber\nonumber\\
 =\langle q_0|D(q)\rangle|\psi\rangle=|\psi\rangle.
\end{align}
This is an extremely weak measurement.

Taking spin-$\frac{1}{2}$ particles as an example, the initial state is $|\psi\rangle=a|0\rangle+b|1\rangle$, where $(a^2+b^2=1)$.
After the interaction between the pointer and the system, the entire system state changes to $a|0\rangle\otimes D(q-1)+b|1\rangle\otimes D(q+1)$. The weak measurement can be characterized by two parameters, namely the quality $F$ and the precision factor $G$. The quality factor can be given as $F=\int^\infty _{-\infty}\langle D(q+1)|D(q-1)\rangle dq$. It quantifies the degree to which the system is not disturbed after measurement. The precision factor can be given as $G=\int^{+1}_{-1}D^2(q)dq$. It quantifies the information about the state obtained by the measurement.
The greater the value of $G$, the smaller the corresponding F, indicating greater information extraction ability of the measurement and more disturbances in the system. Strong measurement performance is achieved when $G=1$ and $F=0$. The trade-off between two factors is defined as a pointer. Through tracking the state of the pointer, the system can be represented as,
\begin{align}
	\rho'=F\rho+(1-F)(\Pi^+\rho\Pi^++\Pi^-\rho\Pi^-),
\end{align}
where $\rho=|\psi\rangle\langle\psi|$, and $\Pi^\pm$ is the projection of the results $|0\rangle$ and $|1\rangle$. The probability distribution can be given as $p(\pm)=G\langle \psi|\Pi^\pm|\psi\rangle+(1-G)\frac12$.
\citeauthor{Silva.Ralph_PhysRevLett.114.250401_2015} found that the square pointer, where $G+F=1$, is not optimal. It can not provide the optimal trade-off for the measurement. Specifically, for a given quality factor $F$, it does not offer the maximum accuracy $G$. Subsequently, \citeauthor{Silva.Ralph_PhysRevLett.114.250401_2015} investigated the optimal trade-off condition, $G^2+F^2=1$.

Weak measurements are a classification of POVM that can be expressed as a set of positive operators, denoted as $\hat{Q}\equiv \{\hat{Q}_i|\sum_i\hat{Q}_i=\mathbb{I}, 0<\hat{Q}_i\le \mathbb{I} \}$. They can also be represented as a projective measurement with white noise, given by $\hat{Q}^\gamma_i=\gamma \hat{P}_i+(1-\gamma)\frac{\mathbb{I}}{d}$, where $\gamma$ is the unsharpness parameter in the range of $[0, 1]$. It is possible to construct the Kraus operator, $\hat{A}_i^\gamma=\sqrt{\hat{Q}_i^\gamma}$, to describe the post-measurement state,  $\rho'=\sum_i{\hat{A}^\gamma_i}\rho{\hat{A}^\gamma_i}$.
Continuing with the example of a spin-$\frac12$ system, let's define $i\in\{+,-\}$. The measurement operator can be given as $ \hat{Q}^\gamma_\pm=\gamma \hat{P}_\pm+(1-\gamma)\frac{\mathbb{I}}{2}=\frac12(\mathbb{I}\pm\gamma \vec{n}.\hat\sigma)$, and the corresponding Kraus operator can be formed as $\hat{A}_\pm^\gamma=\sqrt{\hat{Q}_\pm^\gamma}=\frac{1}{\sqrt2}(\sqrt{1+\gamma}\hat{P}_++\sqrt{1-\gamma}\hat{P}_-)$. The post-measurement state can be represented as $\rho'={\hat{A}^\gamma_+}\rho{\hat{A}^\gamma_+}+{\hat{A}^\gamma_-}\rho{\hat{A}^\gamma_-}$, extending as,
\begin{align}
	\rho'=\sqrt{1-\gamma^2}\rho+(1-\sqrt{1-\gamma^2})(\hat{P}_+\rho \hat{P}_++\hat{P}_-\rho \hat{P}_-).
\end{align}
The probability distribution remain as $p(\pm)=\mathrm{Tr}[{\hat{Q}^\gamma_\pm}\rho]=\gamma \mathrm{Tr}[\hat{P}_\pm\rho]+\frac{1-\gamma}{2}$.
Compared with the previous form of weak measurement, it can be found that $\gamma$ corresponds to the precision factor $G$, and $\sqrt{1-\gamma^2}$ corresponds to the quality parameter $F$ under Kraus decomposition.
It is worth noting that constructing a general operator under the Kraus decomposition inherently results in the current POVM being an optimal weak measurement that satisfies the optimal trade-off, $G^2+F^2=1$. This provides additional support for the optimality of the pointer proposed by \cite{Silva.Ralph_PhysRevLett.114.250401_2015}.
}

\section{Nonlocality Sharing}\label{nonlocality_sharing_section}


In quantum theory, quantum correlation obeys the principle of monogamy \cite{Masanes.PhysRevA.73.012112,Toner_2009_monogamy}. It refers to the restriction on the distribution of quantum correlations between multiple particles. For example, if particle A is highly entangled with particle B (i.e. violate Bell inequality), then particle A cannot be entangled with particle C to the same degree. In other words, quantum correlation, including quantum entanglement, quantum steering, and Bell nonlocality, cannot be freely shared or distributed among all particles in a system.

{\color{black}In 2015, \citeauthor{Silva.Ralph_PhysRevLett.114.250401_2015} achieved the sharing of nonlocal correlations among multiple observers through weak measurements. Using the CHSH inequality criterion, they demonstrated that at most two Bobs can share nonlocality with Alice, challenging the inherent exclusivity of one observer per measurement pair \cite{Mal.Shiladitya_math4030048_2016}. This conclusion challenges intuition, as it disrupts the inherent monogamy observed in quantum correlations. Immediately, nonlocal sharing has garnered considerable attention. 

Subsequently,
\citet{Ren.Changliang_PhysRevA.100.052121_2019} and \citet{Feng.Tianfeng_PhysRevA.102.032220_2020} analyzed the ``weak'' nature of the ``measurements'' involved in nonlocal sharing, proposing the viewpoint that the motivation behind the measurements performed by the intermediate observer, Bob, in weak measurements influences the sharing of nonlocality. Based on this motivation, nonlocal sharing in the CHSH scenario was defined as active nonlocal sharing and passive nonlocal sharing.

\citet{Brown.Peter.J_PhysRevLett.125.090401_2020}  delved into the understanding of ``measurements'' in the context of Bell nonlocality sharing, attempting to increase the number of Bobs capable of sequentially sharing nonlocality with Alice. They employed an asymmetric POVM taken by an intermediate observer between Bobs to prove that any amount of independent Bobs can violate the CHSH inequality with a single Alice, thereby establishing the possibility of multiple rounds of nonlocal sharing.

Against this backdrop, \citet{Anna._PhysRevLett.129.230402_2022} explored the sharing of nonlocality without resorting to weak measurements, instead opting for projective measurements only. Furthermore, a series of studies have investigated nonlocal sharing in bilateral and trilateral scenarios, with some conclusions already experimentally validated.

Furthermore, several studies have investigated nonlocality sharing in bilateral
(\citealp{Cheng.Shuming_PhysRevA.104.L060201_2021};
\citealp{Zhu.Jie_PhysRevA.105.032211_2022}) and trilateral scenarios \cite{Ren.Changliang_PhysRevA.105.052221_2022} and some of these schemes have been experimentally validated
(\citealp{Schiavon._Quantum.Sci.Technol.2_015010_2017};
\citealp{Hu.Meng.Jun_Quantuminf.s41534-018-0115-x_2018};
\citealp{Feng.Tianfeng_PhysRevA.102.032220_2020};
\citealp{Foletto.Giulio_PHYS.REV.APPL.13.044008_2020};
\citealp{Xiao.Ya_arXiv.2212.03815_2022}).
}

\subsection{The Nonlocality Sharing Via Weak Measurement}

Here we provide an overview of the study conducted by \citeauthor{Silva.Ralph_PhysRevLett.114.250401_2015} focused on achieving nonlocality sharing through optimal weak measurements. \citeauthor{Silva.Ralph_PhysRevLett.114.250401_2015} initially investigated the delicate balance between information gain and the degree of perturbation in the measurements of spin particles. This exploration led to the derivation of pointer states that effectively reached the optimal trade-off. The findings were then applied to the nonlocality-sharing of a single pair of maximally entangled states, illustrating that nonlocality can be observed among multiple observers.

\begin{figure}[htbp]
\centering
	\includegraphics[width=0.45\textwidth]{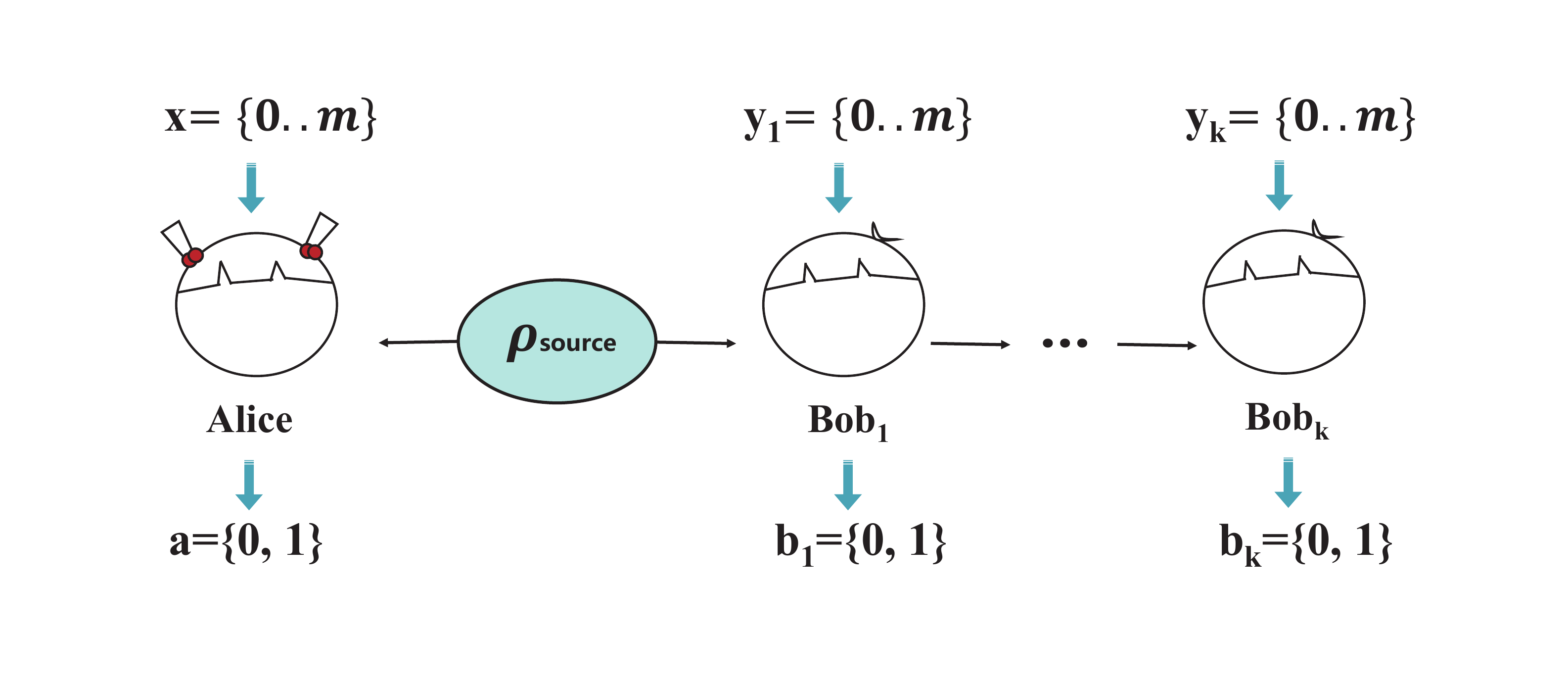}
	\caption{\justifying{\small{The scenario of nonlocality sharing}: the source $\rho_{\text{source}}$ distributes to distant observers, namely Alice and Bob$_1$, after which Bob$_1$ conveys it to subsequent observers Bob$_s$. Upon receiving the quantum state, each observer is equipped with measurement inputs, denoted as $x, y_1, \ldots, y_k \in {0, \ldots, m}$, and produces measurement outcomes $a, b_1, \ldots, b_k$.}}\label{bell nonlocality sharing}
\end{figure}

The nonlocality sharing scenario is illustrated in Fig. \ref{bell nonlocality sharing}. The initial state of the entire system can be expressed as $\left|\psi\right\rangle = \frac{1}{\sqrt{2}}\left|\uparrow\downarrow\right\rangle - \left|\downarrow\uparrow\right\rangle$. This state is distributed among multiple observers, including Alice and sequential observers known as Bobs.
Alice selects a measurement from a set of m possible measurements, denoted by dichotomic strong measurements $\hat{A}_x$, where $x\in {0, ..., m}$. The outcomes of Alice's measurements are marked as $a={0,1}$.
On the Bob side, there are multiple observers labeled as Bob$_k$, each implementing dichotomic measurements $\hat{B}_{y_k}$ corresponding to m inputs $y_k \in {0, ..., m}$. The resulting outputs are also marked as $b_k={0,1}$. Importantly, these observers operate independently, particularly in scenarios where measurement choices remain unbiased.



In order to realize nonlocality sharing, the observer Bob$_{s}$ (where $s \le k-1$) may conduct weak measurements on the received qubit, while only Bob$_{k}$ performs strong measurements. The trade-off between information gain and the degree of disturbance is evaluated through quality factor $F_{s}$ and precision factor $G_{s}$. $F_{s}$ signifies the undisturbed degree of the quantum system after measurement, and $G_{s}$ quantifies the information gain as measured by $\hat{B}_{y{s}}$.

\begin{figure}[htbp]
\centering
\includegraphics[width=0.45\textwidth]{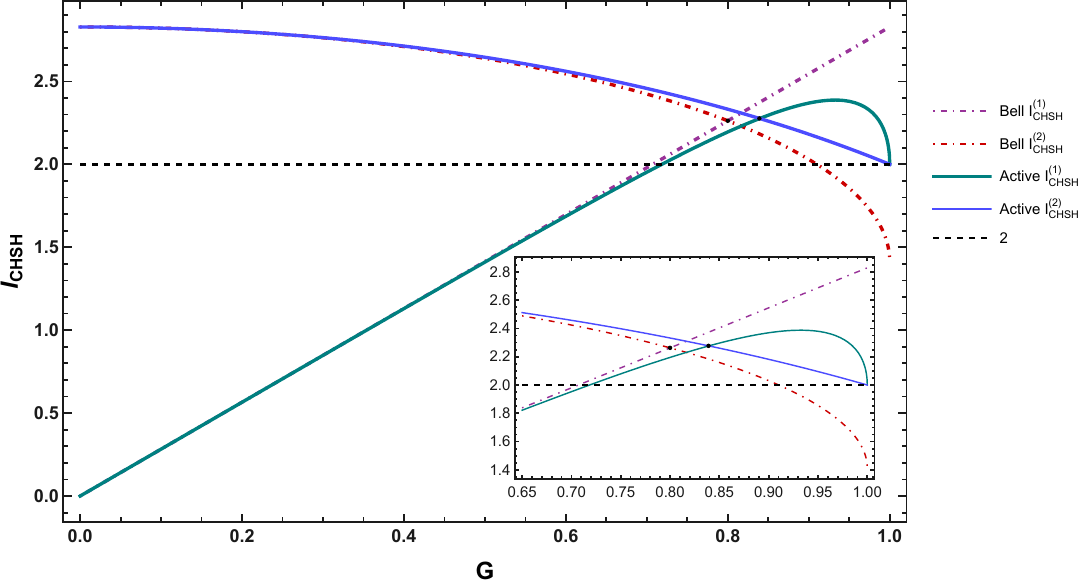}
\caption{\justifying{Plot of the CHSH value as a function of the precision factor G of Bob$_1$ for optimal pointer types in the sequential CHSH scenario, with Bell-$I^{(k)}_{\mathrm{CHSH}}$ for sharing in the bell test, and Active-$I^{(k)}_{\mathrm{CHSH}}$ for active sharing type. This figure is adapted from \cite{Ren.Changliang_PhysRevA.100.052121_2019}}}\label{chshvalue_figure}.
\end{figure}

Now let's review the nonlocality sharing between Alice and two sequential Bobs, i.e. Bob$ _1$ and Bob$ _2$. As shown in Fig. \ref{bell nonlocality sharing},
Alice randomly performs the measurements of $\hat{A}_0=\sigma_z$ or $\hat{A}_1=\sigma_x$ on one of qubit of $\rho_{source}$.
The post-measurement state with the measurement outcome of Alice $a=\{0,1\}$ is described as,
\begin{eqnarray}
	\rho_{\hat{A}_x }^{a}=(U_{\hat{A}_{x} }^{a})\rho_{source}(U_{\hat{A}_{x} }^{a})^{\dagger },
\end{eqnarray}
where $U_{\hat{A}_{x} }^{a}=\frac{\mathbb{I}+(-1)^{a}\hat{A}_x }{2} \otimes \mathbb{I}$.
Subsequently, Bob$ _1$ and Bob$ _2$ measures the another qubit of $\rho_{source}$ sequentially. Their measurement observables can be given as,
$\hat{B}_{0}=\frac{-(\sigma_z+\sigma_x)}{\sqrt{2} }$, $\hat{B}_{1}=\frac{(-\sigma_z+\sigma_x)}{\sqrt{2} }$. Here Bob$_1$ chooses the optimal pointer to implement weak measurement, satisfying $F^{2}+G^{2}=1 $. 
The post-measurement state of Bob$ _1$ with the outcome $b_{1}=\{0,1\}$ is given as,
\begin{align}
	\rho _{\hat{B}_{y_1}}^{b_{1}}=\frac{F}{2}\rho_{\hat{A}_x}^{a}
	+\frac{1+\left (-1\right )^{b_{1}}G-F}{2}
	\left [U_{\hat{B}_{y_1} }^{1}	\rho_{\hat{A}_x}^{a}
	\left (U_{\hat{B}_{y_1} }^{1}\right )^{\dagger }\right ]
	\nonumber\\+\frac{1-\left (-1\right )^{b_{1}}G-F}{2}
	\left [U_{\hat{B}_{y_1} }^{0}	\rho_{\hat{A}_x}^{a}
	\left (U_{\hat{B}_{y_1} }^{0}\right )^{\dagger }\right],
\end{align}
where $U_{\hat{B}_{y_n} }^{b_n}=(\mathbb{I}\otimes \frac{\mathbb{I}+(-1)^{b_n}\hat{B}_{y_n} }{2} )$. The post-measurement state of Bob$ _2$ with the output $b_2\in\{0,1\}$ becomes,
\begin{eqnarray}
	\rho_{\hat{B}_{y_2}}^{b_2}=U_{\hat{B}_{y_2} }^{b_{2}}
	\rho _{\hat{B}_{y_1}}^{b_{1}}
	\left (U_{\hat{B}_{y_2}}^{b_{2}}\right )^{\dagger }.
\end{eqnarray}
Ultimately, one may derive the  joint probability distribution from $\rho_{\hat{B}_{y_2}}^{b_2}$, 
\begin{align}
	P(a,b_1,b_2|x,y_1,y_2)=\text{Tr}(\rho_{\hat{B}_{y_2}}^{b_2}).
\end{align}
Specifically, this joint probability is 
\begin{eqnarray}
	&P(a,b_{1},b_{2}|x,y_{1},y_{2})=\frac{b_{1}G}{4}(\frac{a\hat{A}_{x}\cdot \hat{B}_{y_{1}}+b_{2}\hat{B}_{y_{1}}\cdot \hat{B}_{y_{2}}  }{2}  )\nonumber\\
	&+\frac{F}{4}(\frac{1+ab_{2}\hat{A}_{x}\cdot \hat{B}_{y_{2}} }{2} ) 
	+(\frac{1-F}{4}) (\frac{1+ab_{2}\hat{A}_{x}\cdot \hat{B}_{y1}\cdot \hat{B}_{y_{2}} }{2} ).
\end{eqnarray}
The marginal probability of the combination of Alice-Bob$_k (k\in\left\{1,2\right\})$ can be obtained from the whole joint probability distribution,
\begin{eqnarray}\label{marginal-pro}
P(a,b_{k}|x,y_k)=
\sum_{b_{k'}}^{k\ne k'}
P(a,b_{1},b_{2}|x,y_1,y_2),
\end{eqnarray}
and its average value is
\begin{align}\label{average_value}
	E_{xy_k}^{k}=\sum_{a,b_k=0,1}(-1)^{a+b_k}[P(a,b_k|x,y_k)].
\end{align}
One may calculate the quantum correlation between Alice-Bob$_1$ and Alice-Bob$_2$, $I_{\mathrm{CHSH}}^{(1)}=2\sqrt{2} G$ and $I_{\mathrm{CHSH}}^{(2)}=\sqrt{2}(1+F) $ respectively using CHSH inequality,
\begin{align}
I_{\mathrm{CHSH}}^{(n)}=E_{00}^{n}+E_{01}^{n}+E_{10}^{n}-E_{11}^{n}.
\end{align} 
Clearly, the quantum correlation between Alice and Bob$_2$ is contingent upon both the precision and quantity of measurements carried out by Bob$_1$.
As depicted in Fig. \ref{chshvalue_figure}, the distinct measured strengths of Bob$_1$ play a pivotal role in determining both the quantum correlation of Alice with itself and the quantum correlation between Alice and Bob$_2$. In the scenario of an optimal pointer for weak measurements ($F^2+G^2=1$), the precision factor G within the range of $(\frac{1}{\sqrt2},\sqrt{2(\sqrt2-1)})$ contributes to the violation of the two CHSH criteria. Notably, a double violation of CHSH inequalities with the same correlation is achievable when $G=0.8$, resulting in $I_{\mathrm{CHSH}}^{(1)}=I_{\mathrm{CHSH}}^{(2)}=\frac{8\sqrt2}{5}\approx2.26$.
These results underscore that both Alice-Bob$_1$ and Alice-Bob$_2$ can simultaneously reveal Bell nonlocality. Likewise, when examining other types of weak measurement cases, such as square pointer states, the observation of double CHSH violations remains possible, albeit with a reduced degree of violation compared to the optimal case at the same measurement strength.

In a more general scenario, the presence of multiple observers on the Bob side is considered, denoted as {Bob$_1$, Bob$_2$,..., Bob$_k$}, as illustrated in Fig. \ref{bell nonlocality sharing}. Silva et al. addressed this configuration in their original work, specifically examining the case of biased measurements by Bobs. 
In this scenario, Bob$_1$ randomly selects two measurement settings and subsequently transmits the measured quantum states to Bob$_2$ with different probabilities.
Additionally, now the measurement operators for Bob are defined as $\hat{B}_0^{k}=\sigma_z,\hat{B}_1^{k}=\cos\theta_{k}\sigma_z+\sin\theta_{k}\sigma_x$. 
Building upon the aforementioned analysis, it is crucial to recognize that the post-measurement state of Alice and Bob$_k$ depends on the inputs from all preceding observers on Bob's side. This dependency is denoted as,
\begin{align}
	\rho _{y_{k-1}}=F_{k-1}\rho _{y_{k-2}}+(1-F_{k-1})D_{B_{y_{k-1}}}(\rho _{y_{k-2}}),
\end{align}
where $D_{v}(\rho )=\pi _{v}^{+}\rho\pi _{v}^{+}+\pi _{v}^{-}\rho\pi _{v}^{-}$, and $\pi _{v}^{\pm}$ represents the projection in the direction $\pm v$. 
Given the independence of each Bob from the previous one, it becomes necessary to consider the summation over all possible choices of inputs and their corresponding probabilities. 
Thus, the post-measurement state of Alice and Bob$ _k$  is expressed as,
\begin{align}
	\rho_k=\sum_{i=1}^{k-1}\rho _{y_{i}}\prod^{k-1}_{i=1}P(y_i),
\end{align}
where $P(y_i)$ represents the probability of input $y_i$ received by Bob$ _i$.
This allows us to derive the complete joint probability distribution $P(a,b_1,...b_n|x,y_1,....b_k)$.
Likewise, the average value $E_{xy_k}^{k}$ can be determined using Eq. (\ref{marginal-pro}) and Eq. (\ref{average_value}), enabling the calculation of the quantum correlation $I_{\mathrm{CHSH}}^{(k)}$.
Silva et al. established the presence of arbitrarily longer sequences of CHSH violations between Alice and Bobs in the biased case, considering two measurement settings, as the number of Bobs approaches infinity ($k\to \infty$, $k\in \mathbb{N}$).



Subsequently, in the unbiased case, \citet{Mal.Shiladitya_math4030048_2016} concluded that achieving a violation of the Clauser-Horne-Shimony-Holt (CHSH) inequality by more than two Bobs using two unbiased input settings with Alice is impossible.
\citet{Matteo.Schiavon_Quantum.Sci.Technol.2.015010_2017} experimentally demonstrate the violation of double CHSH inequalities by exploiting a two-photon polarisation maximally entangled state. 
Additionally, \citet{Hu.Meng.Jun_Quantuminf.s41534-018-0115-x_2018} reported the observation of double CHSH inequality violations using a continuous-tunable optimal weak measurement strength in a photonic system.
\citet{Feng.Tianfeng_PhysRevA.102.032220_2020} introduced a balanced type of nonlocality sharing that enables the observation of double CHSH violations using no-so-weak measurements. 
Moreover, \citet{Das.Debarshi_PhysRevA.99.022305_2019} explored the sequential demonstration of bipartite nonlocality with a single Alice, investigating the number of Bobs that can exhibit nonlocality as the number of measurement settings per observer increased.
Various applications of sequential scenarios have been explored in studies such as \cite{Curchod.F.J_PhysRevA.95.020102_2017, Karthik.Mohan_NEW.J.PHYS.10.1088.1367-2630.ab3773_2019}. Furthermore, experimental demonstrations have been conducted in (\citealp{Schiavon._Quantum.Sci.Technol.2_015010_2017};
\citealp{Hu.Meng.Jun_Quantuminf.s41534-018-0115-x_2018}). These investigations contribute to the understanding and practical implementation of sequential quantum scenarios.

\subsection{Passive And Active Nonlocality Sharing}


In the earlier discussion, the attainment of a double CHSH violation for Alice-Bob$_1$ and Alice-Bob$_2$ through Bob$_1$'s optimal weak measurements in the sequential CHSH scenario showcased the nonlocality sharing. However, this phenomenon was not fully comprehended at that time. In 2019, \citet{Ren.Changliang_PhysRevA.100.052121_2019} classified this phenomenon into two types—active nonlocality sharing and passive nonlocality sharing, depending on Bob$_1$'s motivation for measurement. They demonstrated that it is still possible to observe nonlocality sharing as long as the intermediate observers refrain from making ideal strong measurements.  This phenomenon is counter-intuitive since it allows for Bob$_1$ to conduct measurements with intensities close to 1, while enabling nonlocality sharing—the situations that were not permitted in the original version \cite{Silva.Ralph_PhysRevLett.114.250401_2015}. 


In the sequential CHSH scenario (refer to Fig. \ref{bell nonlocality sharing}), the nature of the double violation observed in Alice-Bob$_1$ and Alice-Bob$_2$ depends on Bob$_1$'s intentions. If Bob$_1$ is solely focused on achieving a maximum violation for himself without considering nonlocality sharing with Bob$_2$, this double violation is identified as passive nonlocality sharing. On the other hand, if Bob$_1$ is motivated to assist Bob$_2$ in achieving the maximum CHSH violation, the observed double violation in Alice-Bob$_1$ and Alice-Bob$_2$ is categorized as active nonlocality sharing.
As we all know, an arbitrary 2-qubit state $\rho$ can be decomposed as,
\begin{small}
\begin{align*}
	\rho=\frac14\left(\mathbb{I}\otimes \mathbb{I}+\sum_i^3\lambda_i\sigma_i\otimes \mathbb{I}+\sum_j^3\eta_j \mathbb{I}\otimes\sigma_j+\sum_k^3\tau_k\sigma_k\otimes\sigma_k\right).
\end{align*}
\end{small}
Here, $\sigma_{i}$ represents the Pauli matrix, where $i, j, k \in \{1,2,3\}$, and $\left | \tau _{1} \right | \ge \left | \tau _{2} \right |\ge \left | \tau_{3} \right |$. Assuming each observer employs binary inputs (with $m=2$) denoted as $n\cdot \hat{\sigma}$, and Alice's strong measurement is denoted as $\hat{\omega}_x$, while the measurements executed by Bob$_1$ (Bob$_2$) are labeled as $\hat{\mu}_{y_1}$ ($\hat{\nu}_{y_2}$), where $x, y{_1}, y{_2} \in \{1,2\}$.
The CHSH values of Alice-Bob$ _1$ and Alice-Bob$ _2$ can be expressed as,
\begin{eqnarray}
I_{\mathrm{CHSH}}^{(1)}=\sum_{i, j=0}^{1} (-1)^{ij}\langle\omega_i\otimes\mu_j\rangle, \\
I_{\mathrm{CHSH}}^{(2)}=\sum_{i,j=0}^{1}(-1)^{ij}\langle\omega_i\otimes\nu_j\rangle.
\end{eqnarray}

Indeed, the CHSH value of Alice-Bob$_2$ is influenced by several factors, including the motivation of Bob$_1$ (whether active or passive), the probability of his measurement choice (whether biased or unbiased), and the pointer distribution of the weak measurement. These elements collectively contribute to shaping the quantum correlations observed in the sequential CHSH scenario.

In the scenario of passive nonlocality sharing where Bob$_1$ aims to achieve a maximum violation for himself in a biased case, the expressions for the maximum CHSH values of Alice-Bob$_1$ and Alice-Bob$_2$ are given by \cite{Ren.Changliang_PhysRevA.100.052121_2019},
\begin{eqnarray}
&I_{\mathrm{CHSH}}^{(1)}=2\sqrt{\tau_1^2+\tau_2^2}G\nonumber,\\
&I_{\mathrm{CHSH}}^{(2)}=\sqrt{\tau_1^2+\tau_2^2}+\sqrt{\frac{\tau_1^4+2(-1+2F^2)\tau_1^2\tau_2^2+\tau_2^4}{\tau_1^2+\tau_2^2}}.
\end{eqnarray}
While the maximum CHSH values for the unbiased case can be expressed as,
\begin{eqnarray}
&I_{\mathrm{CHSH}}^{(1)}=2\sqrt{\tau_1^2+\tau_2^2}G\nonumber,\\
	&I_{\mathrm{CHSH}}^{(2)}=2\sqrt{\frac{\tau_1^4+[-1+F(2+F)]\tau_1^2\tau_2^2+\tau_2^4}{\tau_1^2+\tau_2^2}} .
\end{eqnarray} 
Since the Bell state satisfies $\tau_{1}=1$, $\tau_{2}=-1$, and $\tau_{3}=1$, the maximal CHSH values in both the biased and unbiased cases are exactly the same, characterized by $I_{\mathrm{CHSH}}^{(1)}=2\sqrt{2} G$ and $I_{\mathrm{CHSH}}^{(2)}=\sqrt2(1+F)$ which is coincident with the result in \cite{Silva.Ralph_PhysRevLett.114.250401_2015}. 
Hence, the original instance of nonlocality sharing can be categorized as a distinctive illustration of passive non-locality sharing in Bell states. It is important to note that when Bob$_1$ employs a weak measurement with the square pointer, leading to the condition $F+G=1$, a clear outcome emerges: both $I_{\mathrm{CHSH}}^{(1)}$ and $I_{\mathrm{CHSH}}^{(2)}$ cannot simultaneously exceed 2. However, if Bob$_1$ opts for a weak measurement using the optimal pointer, satisfying $F^2+G^2=1$, the observation of nonlocality sharing is possible. Specifically, as shown in Fig. \ref{chshvalue_figure}, within the parameter range of $G\in[\frac{1}{\sqrt2},\sqrt{2(\sqrt2-1)}]$, it is found that both $I_{\mathrm{CHSH}}^{(1)}$ and $I_{\mathrm{CHSH}}^{(2)}$ can exceed two simultaneously. This result indicates the presence of passive nonlocality sharing in the system.

In the context of active nonlocality sharing, where Bob$_1$ assists Bob$_2$ in achieving the maximum CHSH violation under biased conditions, the maximum CHSH values can be expressed as follows,
\begin{eqnarray}
	&I_{\mathrm{CHSH}}^{(1)}=G(\frac{2F\tau_1^2+\tau_2^2}{\sqrt{F^2{\tau_1}^2+{\tau_2}^2}}+\frac{\tau_2^2}{\sqrt{\tau_1^2+F^2\tau_2^2}})	\nonumber,\\
	&I_{\mathrm{CHSH}}^{(2)}=\frac{2F^2\tau_1^2+\tau_2^2}{\sqrt{F^2\tau_1^2+\tau_2^2}}+\frac{\tau_2^2}{\sqrt{\tau_1^2+F^2\tau_2^2}}.
\end{eqnarray}
Active nonlocality sharing can be observed with appropriate measurement settings.
For a Bell state, $\tau _{1}=1,\tau _{2}=-1,\tau _{3}=1$, the CHSH values are turn to,
\begin{eqnarray}
&I_{\mathrm{CHSH}}^{(1)}=\frac{2G(1+F)}{\sqrt{1+F^{2}}},\nonumber\\
&I_{\mathrm{CHSH}}^{(2)}=2\sqrt{1+F^{2}}.
\end{eqnarray}
When Bob$_1$ performs weak measurement with the optimal pointer, $I_{\mathrm{CHSH}}^{(1)} $ and $I_{\mathrm{CHSH}}^{(2)} $ can exceed two simultaneously in the range of $G\in\left ( 0.718,1 \right ) $. Even as $G$ approaches one, double CHSH violations still can be observed.
On the other hand, when Bob$_1$ performs weak measurement with the square pointer, the CHSH value turns to $I_{\mathrm{CHSH}}^{(1)}=(1+\sqrt{2} ) G,I_{\mathrm{CHSH}}^{(2)}=2\sqrt{1+F^2} $ under simple measurement settings \cite{Ren.Changliang_PhysRevA.100.052121_2019}. 
Certainly, the presence of double CHSH violation becomes evident in the range of $G\in\left[\frac{2}{1+\sqrt{2}},1\right]$. This phenomenon is unattainable in the case of passive nonlocality sharing.

While for the unbiased case of active nonlocality sharing, an analytical result for the Bell state can be achieved. 
when $G\in(0,0.8)$, the CHSH values attain their optimum as follows: $I_{\mathrm{CHSH}}^{(1)} = 2\sqrt{2}G$ and $I_{\mathrm{CHSH}}^{(2)} = \sqrt{2}(1+F)$.
When $G\in(0.8,1)$ the optimal CHSH values transform into,
\begin{eqnarray}
&I_{\mathrm{CHSH}}^{(1)}=\frac{(-3+F)\mathcal{J}\mathcal{L}}{3\sqrt6(-1+F)F},\nonumber\\ &I_{\mathrm{CHSH}}^{(2)}=\frac{\mathcal{J}\mathcal{L}}{\sqrt6(-1+F)F},
\end{eqnarray}
where
\begin{eqnarray}
	&\mathcal{J} =\sqrt{3+\frac{(-3+F)(-1+F)^2(-3+5F)}{(-1+F)^2}},\nonumber\\
	&\mathcal{L}=3+(-4+F)F-\sqrt{(-3+F)(-1+F)^2(-3+5F)}.\nonumber
\end{eqnarray}
When Bob$_1$ utilizes weak measurement with the optimal pointer, it is notable that $I_{\mathrm{CHSH}}^{(1)}$ and $I_{\mathrm{CHSH}}^{(2)}$ can simultaneously surpass two within the parameter range of $G\in\left(\frac{1}{\sqrt{2}},1\right)$, see Fig. \ref{chshvalue_figure}.
However, in the case where Bob$_1$ opts for weak measurement with the square pointer, achieving double CHSH violation is impossible.

These results suggest that implementing nonlocality sharing becomes more feasible when Bob is cooperative in assisting subsequent observers to attain the maximum CHSH violation. This observation was experimentally validated by \citet{Feng.Tianfeng_PhysRevA.102.032220_2020}.
Subsequently, \citet{Hou.Wenlin_PhysRevA.105.042436_2022} delved into network nonlocality sharing within the extended bilocal scenario through the lens of weak measurements. Their study revealed that network nonlocality sharing can be dissected into passive and active network nonlocality sharing by considering the divergent motivations of intermediate observers. Notably, these motivations lack counterparts in standard Bell scenarios.
In 2022, \citet{caizinuo_arXiv} extensively explored two types of potential full network nonlocality sharing (FNN)——passive FNN sharing and active FNN sharing. The findings underscored that passive FNN sharing is unattainable, while active FNN sharing can be accomplished through suitable measurements. This indicates that achieving FNN sharing in this scenario necessitates greater cooperation from intermediate observers compared to Bell nonlocality sharing and network nonlocality sharing.


\subsection{Asymmetric POVM For Boundless Nonlocality Sharing}

The earlier research on nonlocality sharing by \citet{Silva.Ralph_PhysRevLett.114.250401_2015} established a significant result. Specifically, it was found that, under the assumption of unbiased binary choices and the condition that all observers are independent, no more than two observers on Bob's side can simultaneously violate the CHSH inequalities with Alice.
Subsequently, \citet{Mal.Shiladitya_math4030048_2016} and \citet{Zhu.Jie_PhysRevA.105.032211_2022} further proved this result. 
However, the question of whether nonlocality can be shared among more observers on Bob's side remains an open and unresolved inquiry.
In 2020, \citet{Brown.Peter.J_PhysRevLett.125.090401_2020} delved into this question and provided a positive answer.
\citeauthor{Brown.Peter.J_PhysRevLett.125.090401_2020} proposed a scheme where the observers in the middle of sequential Bobs perform measurements with unequal sharpness. 
In this scheme, Bob$_s$ ($1 \le s \le k-1$) independently selects two different dichotomic measurements: one being a strong measurement and the other a generalized POVM. This configuration enables an arbitrarily longer sequence of Bobs to simultaneously violate the CHSH inequality with a single Alice, showcasing boundless nonlocality sharing. 

In the scenario of boundless nonlocality sharing, Alice still performs strong measurements in the x-z plane of the Bloch sphere, defined as $\hat{A}_x=n \cdot\hat{\sigma}$.
For Bob's side, one measurement of Bob$_k$ can be defined as $\hat{B}_0^{(k)}=\sigma_z$; the other is generalized POVM, which can be defined as a pair of positive semidefinite matrices $(\hat{B}^{(k)}_{01},\hat{B}^{(k)}_{11})$ with binary outcomes, $\hat{B}^{(k)}_{01}+\hat{B}^{(k)}_{11}=I$.
Without loss of generality, $\hat{B}^{(k)}_{01}=\frac{1}{2}(I+\gamma_k\sigma_z)$, where $\gamma_k$ represents the sharpness parameters of measurements. Obviously, $B^{(k)}_{01}$ returned to strong measurement when $\gamma_k=1$ and the measurement disregards the state and is equivalent to an unbiased coin toss when $\gamma_k=0$.
If the initial state is a singlet state, the CHSH value corresponding to Alice and Bob$ _k$ can be expressed as \cite{Brown.Peter.J_PhysRevLett.125.090401_2020},
\begin{align}
	I^{(k)}_{\mathrm{CHSH}}=2^{2-k}(\gamma_k\sin\theta+\cos\theta\prod_{j=1}^{k-1}(1+\sqrt{1-\gamma^{2}_{j}})).\label{roger_chshk}
\end{align}
Through Eq. (\ref{roger_chshk}), one may identify suitable values for $\gamma_k$ and measurement directions that lead to multiple instances where $I_{\mathrm{CHSH}}^{(k)}$ can surpass the classical bound of the CHSH inequality simultaneously.
A result is clearly demonstrated in the following. If $\gamma_k$ satisfies the conditions provided as,
\begin{align}
	\gamma_k> \frac{2^{k-1}\cos\theta {\textstyle \prod_{j=1}^{k-1}(1+\sqrt{1-\gamma^{2}_{j}}} )}{\sin\theta },
\end{align}
arbitrary violations of $I_{\mathrm{CHSH}}^{(k)}$ can occur, implying that any number of independent Bobs can simultaneously violate the CHSH inequality with a single Alice.
Furthermore,  Brown and Colbeck introduced a function of $\gamma_1$ with $\theta$, $\gamma_1(\theta)=(1+\varepsilon)\frac{1-\cos\theta}{\sin\theta}$, where 
$\varepsilon$ is an infinitesimal quantity greater than zero. Subsequently, $\gamma_k(\theta)$ is defined recursively,
\begin{equation}
	\gamma_k(\theta)=\begin{cases}
		(1+\varepsilon ) \frac{2^{k-1}-\cos\theta P_k}{\sin\theta },& \text{ if } \gamma_{k-1}(\theta)\in(0,1),\\
		\infty, & \text{ otherwise},
	\end{cases}
\end{equation}
where $P_{k}= {\textstyle \prod_{j=1}^{k-1}} (1+\sqrt{1-\gamma^{2}_{j}(\theta )})$.
They demonstrated the existence of $\left \{ \gamma _{k}(\theta ) \right \} $, which always let $\{I^{(1)}_{\mathrm{CHSH}}, ..., I^{(k)}_{\mathrm{CHSH}}\}$ exceed 2 for $k\to \infty$, $k\in \mathbb{N}$.
This result unveils the phenomenon of boundless nonlocality sharing.

Furthermore, Roger discovered the maximum limit of violation size through further analysis of this strategy. As the number of observers on Bob's side decreases, the violation is reduced exponentially at a double rate. Additionally, this strategy can achieve an unlimited number of violations in the same process as described above. 
However, this measurement strategy is not appropriate for device-independent tasks, such as randomness expansion, where the number of provable randomness increases with the size of CHSH violations.

The above results arise from a comprehensive understanding of ``measurement'' in the Bell nonlocality test, 
contributing valuable insights to unresolved inquiries within the field. Specifically, the question arises as to whether a set of 2-qubit states exists that can be sufficiently characterized to permit an arbitrary number of CHSH violations. Although \citet{Brown.Peter.J_PhysRevLett.125.090401_2020} proposed a condition deemed sufficient for achieving arbitrary violations in a 2-qubit state, the necessity of this condition remains uncertain. Moreover, the diminishing exponential trend in CHSH inequality violation as the number of independent Bobs increases poses limitations on its future applicability. The intriguing prospect of achieving a larger violation still raises a significant question.

The investigation of related extensions is ongoing. In \cite{Zhang.Tinggui_PhysRevA.103.032216_2021}, it has been constructively demonstrated that a class of pure entangled states allows an arbitrary number of independent Bobs can share nonlocality with a single Alice. 
Building on this insight, \citet{Cheng.Shuming_PhysRevA.105.022411_2022} derive corresponding one-sided monogamy relations that rule out two-sided nonlocality sharing for a wide range of parameters, based on a general trade-off relation between the strengths and maximum reversibilities of qubit measurements.
In another study, \citet{Srivastava.Chirag_PhysRevA.103.032408_2021} delve into sequential measurement-device-independent entanglement detection by multiple observers, exploring scenarios with both equal and unequal sharpness parameters. Additionally, \citet{Qiao-Qiao.Lv_Phys.AMath.Theor.2307.09928_2023} investigate the bilateral sharing of EPR steering through weak measurement with unequal sharpness parameters in a generalized case. The findings demonstrate that an unbounded number of sequential Alice-Bob pairs can share EPR steering.

\subsection{Nonlocality Sharing Through Projective Measurements}
As established in the aforementioned studies, breaking the symmetry of weak measurements allows for two or more independent Bobs, leading to CHSH inequality violations by Alice. In a recent development, \citet{Anna._PhysRevLett.129.230402_2022} explored a strategy that achieves nonlocality sharing exclusively through the implementation of  projective measurements in the case of a single Alice and two Bobs.
In this specific scenario, all observers execute strong measurements, followed by Bobs performing arbitrary unitaries based on the input and output of the measurements. A Bell test was then conducted using three distinct projective measurement strategies, determined by Bob's two measurement choices:
(i) both measurements are projective, (ii) both measurements are identity, (iii) one measurement is projective while the other is identity. These three strategies are labeled as $\lambda \in \left \{ 1,2,3 \right \} $. 
Note that all Bob$ _k$'s measurements are independent, and the state shared by Alice and Bob$_k$ can be expressed as,
\begin{align}
	\rho^{(k)}=\frac{1}{2}\sum_{b_k,y_k={k=0,1}}\left(\mathbb{I}\otimes \hat{K}_{b_k|y_k}^k\right).\rho^{(k-1)}.\left(\mathbb{I}\otimes \hat{K}_{b_k|y_k}^k\right)^\dagger .
\end{align}
where $\hat{K}_{b_{k}|y_{k}}^{k}$ the Kraus operators. $\hat{B}_{y_k}^{k}$ are Bob$_k$'s projective measurements, which can be decomposed as $(\hat{K}_{b_{k}|y_{k}}^{k})^{\dagger}(\hat{K}_{b_{k}|y_{k}}^{k})$.
The Kraus operators can be given as,
$
	\hat{K}_{b_{k}|y_{k}}^{k}=\sqrt{\hat{B}_{y_k}^{k}}=\hat{U}_{b_{k}|y_{k}}^{k}\hat{B}_{y_k}^{k}
$,
where $\hat{U}_{b_{k}|y_{k}}^{k}$ are arbitrary unitary.

Suppose Alice, Bob$_1$, and Bob$_2$ carry out independent measurements $\hat{A}_{x}^{(\lambda) }$, $\hat{B}_{y}^{(\lambda) }$ and $\hat{C}_{z}^{(\lambda )}$ respectively. 
The CHSH value for Alice and Bob$_1$ can be achieved as,
\begin{align}
	S_{AB}^{(\lambda)}=\sum_{x,y}(-1)^{x+y}\mathrm{Tr}[\left(\hat{A}_{x}^{(\lambda)}\otimes \hat{B}_{y}^{(\lambda)}\right)\rho^{(1)}],
\end{align}
and for Alice and Bob$_2$ can be given as,
\begin{align}
	S_{AC}^{(\lambda)}=\sum_{x,z}{(-1)^{x+z}\text{Tr}[\left(\hat{A}_{x}^{(\lambda)}\otimes \hat{C}_{z}^{(\lambda)}\right)\rho^{(2)}]}.
\end{align}
The optimal measurement settings for the three measurement strategies are as follows.

\emph{When $\lambda=1$}: 
Bob$_1$'s two measurement choices are projective, represented as $\hat{B}_{0}^{(1)}=\cos\phi\sigma_{x}+\sin\phi\sigma_{z}$ and $\hat{B}_{1}^{(1)}=\sin\phi\sigma_{x}+\cos\phi\sigma_{z}$. The corresponding unitary operators for Bob$_1$ are defined as $\hat{U}_{0}^{(1)}=\mathbb{I}$ and $\hat{U}_{1}^{(1)}=e^{i(\phi-\frac{\pi}{4})\sigma_{y}}$.
For Alice's measurements, the choices are $\hat{A}_{0}^{(1)}=\frac{\sigma_{x}+\sigma_{z}}{\sqrt{2}}$ and $\hat{A}_{1}^{(1)}=\frac{\sigma_{x}-\sigma_{z}}{\sqrt{2}}$. On the other hand, for Bob$_2$, both $\hat{C}_{0}^{(1)}$ and $\hat{C}_{1}^{(1)}$ are given by $\cos\phi\sigma_{x}+\sin\phi\sigma_{z}$.
These choices lead to the following CHSH values: $S_{AB}^{(1)}=2\sqrt{2}\cos\phi$ and $S_{AC}^{(1)}=\sqrt{2}(\cos\phi+\sin\phi)$.
In this case, the trade-off of Alice-Bob$_1$ and Alice-Bob$_2$ becomes,
\begin{eqnarray}
	&S_{AC}^{(1)}=\frac{1}{2}\bigg(S_{AB}^{(1)}+\sqrt{8-\left(S_{AB}^{(1)}\right)^2}\bigg).
\end{eqnarray}

\emph{When $\lambda=2$}: 
Bob$_1$ has two measurement choices, both being identity operators, i.e., $\hat{B}_{0}^{(2)}=\hat{B}_{1}^{(2)}=\mathbb{I}$. The unitary operator for Bob$_1$ is also identity, denoted as $\hat{U}_{b_y}^{(2)}=\mathbb{I}$.
The operators for Alice are given by $\hat{A}_0^{(2)}=\cos\theta\sigma_x+\sin\theta\sigma_z$, $\hat{A}_1^{(2)}=\cos\theta\sigma_x-\sin\theta\sigma_z$. While for Bob$_2$, $\hat{C}_{0}^{(2)}=\sigma_{x},\hat{C}_{1}^{(2)}=\sigma_{z}$.
The CHSH values are immediately determined as $S_{AB}^{(2)}=0$, while $S_{AC}^{(2)}$ can reach a maximum value of $2\sqrt{2}$.

\emph{When $\lambda=3$}: 
Bob$_1$ has two measurement choices: a projective measurement denoted as $\hat{B}_{0}^{(3)}=\mathbb{I}$ and an identity measurement as $\hat{B}_{1}^{(3)}=\sigma_{z}$. The unitary operator for Bob$1$ is also the identity, $\hat{U}_{b_y}^{(3)}=\mathbb{I}$. Alice's measurements are set as $\hat{A}_0^{(3)}=\cos\theta\sigma_x+\sin\theta\sigma_z$ and $\hat{A}_1^{(3)}=\cos\theta\sigma_x-\sin\theta\sigma_z$. For Bob$_2$, $\hat{C}_{0}^{(3)}=\sigma_{x}$ and $\hat{C}_{1}^{(3)}=\sigma_{z}$.
The CHSH values are given by $S_{AB}^{(3)}=2\sin\theta$ and $S_{AC}^{(3)}=\cos\theta+2\sin\theta$.
Their trade-off can be given as,
\begin{eqnarray}
	S_{AC}^{(3)}=S_{AB}^{(3)}+\frac{1}{2}\sqrt{4-\left(S_{AB}^{(3)}\right)^{2}}.
\end{eqnarray}

\begin{figure}[htbp]
\centering
\includegraphics[width=0.45\textwidth]{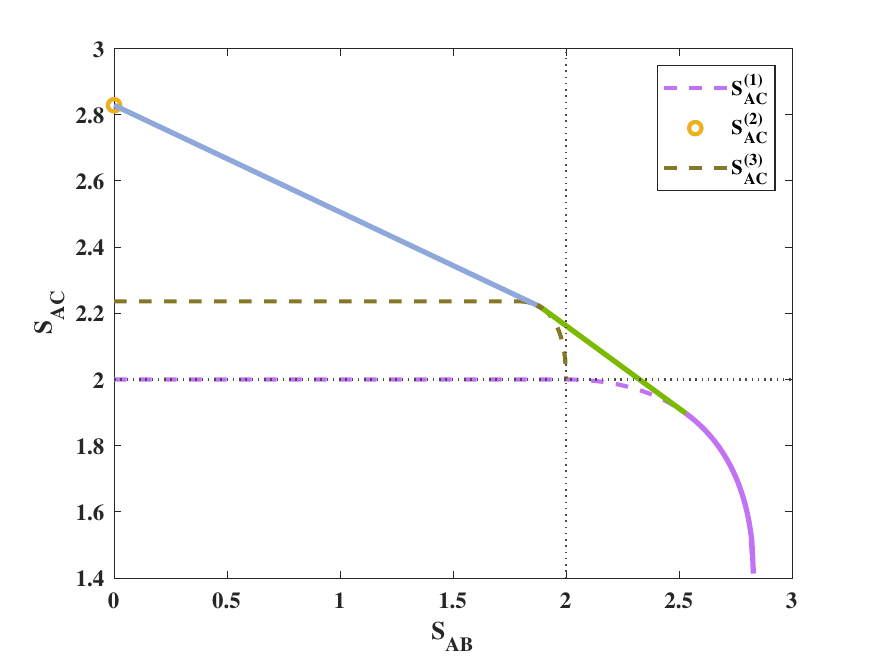}
\caption{\justifying{Plot of the optimal trade-off between $S_{AB}$ and $S_{AC}$ for a maximally entangled qubit pair under projective measurements. This figure is adapted from \cite{Anna._PhysRevLett.129.230402_2022}.}}\label{chshvalue_projective_measure_figure}
\end{figure}

The above results reveal that the violation occurs exclusively between Alice and Bob$_1$ when $\lambda=1$ and exclusively between Alice and Bob$_2$ when $\lambda=2$ and $\lambda=3$. Achieving a double violation necessitates employing a random combination of these three cases. The boundary for {$S_{AB}^{(\lambda)},S_{AC}^{(\lambda)}$} is divided into four segments: (a) a combination of the cases $\lambda=2$ and $\lambda=3$; (b) the case $\lambda=3$; (c) a combination of the cases $\lambda=1$ and $\lambda=3$; and (d) the case $\lambda=1$.
Certainly, the maximal CHSH values, $S_{AB}=S_{AC}=\frac{2\sqrt{10}}{3}=2.108$, can be achieved.
The optimal trade-off is piecewise illustrated with solid
lines in Fig. \ref{chshvalue_projective_measure_figure}.

The aforementioned analysis pertains to the scenario where a maximally entangled state is shared. However, if the shared state is only partially entangled, denoted as $\left | \psi \right \rangle =\cos \alpha\left | 00 \right \rangle +\sin \alpha\left | 11 \right \rangle $ with $\alpha \in [0,\frac{\pi}{4}]$, noteworthy observations emerge.
A crucial point to highlight is that, in the conventional CHSH scenario, achieving the most robust quantum nonlocality necessitates the sharing of maximally entangled states. Conversely, in this sequential CHSH scenario, superior double violations can be attained by sharing partially entangled states rather than maximally entangled ones. For further details, refer to \cite{Anna._PhysRevLett.129.230402_2022}. Subsequently, \citet{Xiao.Ya_arXiv.2212.03815_2022} experimentally demonstrated the nonlocality sharing with projective measurements, extending beyond unsharp measurement scenarios. Similarly, Dong et al. [****] extended this method to the steering sharing.


Recently, \citet{Sasmal_2023_arXiv_unbounded} introduced a novel local randomness-assisted projective measurement protocol (PPM). This protocol enables an unrestricted number of sequential observers (referred to as Bobs) to establish nonlocality with a single Alice simultaneously. The authors endeavor to devise an elegant set of Kraus operators designed to minimally disturb the system. In essence, this strategy circumvents the limitations encountered in the earlier work \cite{Anna._PhysRevLett.129.230402_2022} marking a notable advancement in the field.

\subsection{Nonlocality Sharing In Multilateral And Multi-Qubit Cases }
\subsubsection{2-Qubit System}\label{nonlocality_2qubit}

\begin{figure}[htbp]
	\centering
	\includegraphics[width=0.45\textwidth]{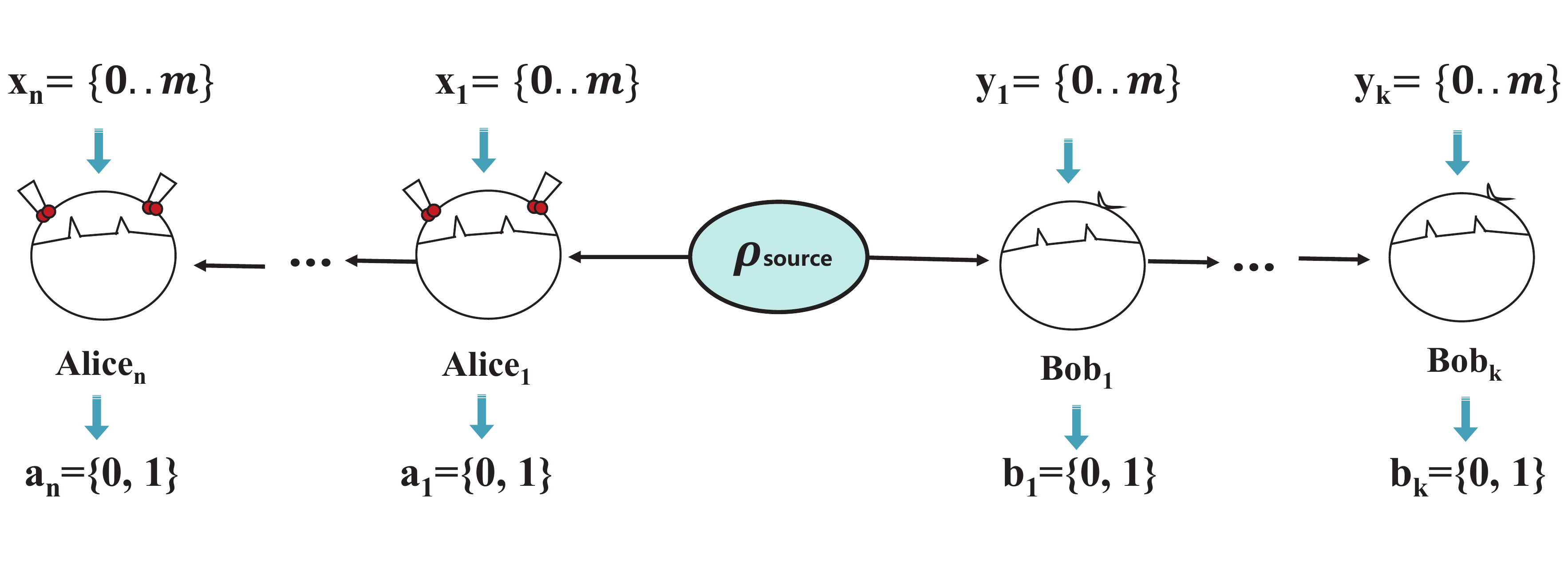}
	\caption{\justifying{\small{The bilateral sequential scenario of nonlocality sharing}.
 The source $\rho_{source}$ distributes to distant observers, namely Alice$_1$ and Bob$_1$, which transmit the received particle to the following observers after measuring. Alices and Bobs have m measurement choices, denoted as $x_n=\{0,...,m\}$ and $y_k=\{ 0,...,m\}$ respectively. $a_n=\{0,1\}$, $b_k=\{0,1\}$ represent measurement outcomes of Alices and Bobs.}}\label{bilateral nonlocality sharing}.
\end{figure}

To date, the majority of research has focused on the unilateral scenario of nonlocality sharing, involving the sharing of an entangled state with Alice and multiple Bobs.
Drawing on previous work, the possibility of generating nonlocality sharing in the bilateral scenario (refer to Fig. \ref{bilateral nonlocality sharing}) has attracted attention. \citet{Zhu.Jie_PhysRevA.105.032211_2022} investigated this matter, determining that two CHSH inequality violations cannot be attained simultaneously in the bilateral scenario.

In the bilateral scenario, involving multiple Alices and Bobs, we assume the shared state to be the singlet state $|\psi\rangle=\frac{1}{\sqrt{2}}\left(\left|01\right\rangle-\left|10\right\rangle\right)$.
Alice$_1$ performs the optimal weak measurement with her two unbiased measurement choices being $\sigma_{x}$ and $\sigma_{z}$. Subsequent Alices perform the same measurement as Alice$_1$ until the last one performs a strong measurement.
Similarly, Bob$_1$ performs the optimal weak measurement with unbiased measurement choices set to $\frac{\sigma{x}-\sigma_{z}}{\sqrt{2}}$ and $\frac{-\sigma_{z}-\sigma_{x}}{\sqrt{2}}$. Subsequent Bobs perform the same measurement until the last one performs a strong measurement. The two parameters for Alice$_k$'s weak measurement are defined as $G_{A_k}$ and $F_{A_k}$, and for Bob$_k$ are $G_{B_k}$ and $F_{B_k}$.

Consider the scenario with two Alices and two Bobs, the CHSH values can be presented as,
\begin{eqnarray}\label{two-side-chshvalue}
	&I_{A_{1}B_{1}}=2\sqrt{2}G_{A_1}G_{B_1},\nonumber\\
	&I_{A_{1}-B_{2}}=\sqrt{2}G_{A_1}(1+F_{B_1})G_{B_2},\nonumber\\
	&I_{A_{2}-B_{1}}=\sqrt{2}G_{B_1}(1+F_{A_1})G_{A_2},\nonumber\\
	&I_{A_{2}B_{2}}=\frac{\sqrt{2}}{2}(1+F_{A_1})G_{A_2}(1+F_{B_1})G_{B_2}.
\end{eqnarray}
When both $G_{A_2}$ and $G_{B_2}$ are set to 1, Alice$_2$ and Bob$_2$ perform a strong measurement. In the event that $F_{A_1}=1$ and $G_{A_1}=0$, this scenario transforms into the sequential CHSH scenario (refer to Fig. \ref{bell nonlocality sharing}).
From Eq. (\ref{two-side-chshvalue}), it becomes evident that Alice$_1$-Bob$_1$ and Alice$_2$-Bob$_2$ cannot simultaneously violate the CHSH inequality.

For the bilateral scenario with $n$ Alices and $k$ Bobs, the CHSH value of Alice$_r$-Bob$_s$ satisfies,
\begin{align*}
	I_{A_{r}-B_{s}}=\frac{2\sqrt{2}}{2^{r-1}\times2^{s-1}}(1+F_{A_{1}})\cdots(1+F_{A_{r-1}})G_{A_{r}}\\
	\times(1+F_{B_{1}})\cdots(1+F_{B_{s-1}})G_{B_{s}},
\end{align*}
where $r\in\{1,..,n \}$, $s\in\{1,..,k\}$, and Alice$_n$ and Bob$_k$ perform a strong measurement with $G_{A_n}=G_{B_k}=1$.
Thus, it remains evident that the bilateral scenario cannot simultaneously achieve two CHSH inequality violations. As a result, sequential observers from both sides are unable to share nonlocality. Nonlocality sharing is only demonstrable between any number of Bobs and a single Alice.

Along with that, \citet{Cheng.Shuming_PhysRevA.104.L060201_2021} proposed a conjecture asserting that in scenarios where observers are independent with unbiased case, nonlocality sharing is feasible in one side extended of the bell scenario. However, this conjecture does not apply to scenarios with multiple observers on each side, which aligns with \cite{Zhu.Jie_PhysRevA.105.032211_2022}.


\citet{Cheng.Shuming_PhysRevA.104.L060201_2021} considered a scenario where the shared state is partially entangled, and all measurement operators are expressed as POVM. In an effort to validate the conjecture in the bilateral scenario, Cheng et al. systematically analyzed 16 parameter sets of measurement directions and one parameter set for the initial state. This analysis encompassed CHSH values, specifically $\mathrm{S(A_1, B_1)}$, $\mathrm{S(A_1, B_2)}$, $\mathrm{S(A_2, B_1)}$, and $\mathrm{S(A_2, B_2)}$. Numerical evidence was derived from the optimization of $\mathrm{S(A_2, B_2)}$ and $\mathrm{S(A_2, B_1)}$ under the constraint that $\mathrm{S(A_1, B_1)}$ and $\mathrm{S(A_1, B_2)}$ reach minimum fixed values. Employing a constrained differential evolution algorithm implemented in the SCIPY library, the maximal CHSH values within the 17-dimensional parameter space were calculated. The optimization process revealed that both CHSH values remain below 2, signifying the absence of double violations of the CHSH inequality.
This result furnishes robust analytical and numerical evidence supporting the conjecture that double CHSH inequality violations do not occur in the bilateral scenario.


\citet{cabello_2021_bell} has shown that it is possible to have arbitrarily long sequences of Alices and Bobs such that every (Alice, Bob) pair violates a Bell inequality in a four-dimensional quantum system. However, the accuracy of this result needs further verification.
Subsequently, \citet{Zhang.Tinggui_Quantum.Inf.Process.s11128.022.03699.z_2022} investigated the nonlocality sharing under bilateral measurements in such a scenario, and found Bell nonlocality cannot be shared for a limited number of times under a specific class of projection operators. Such a result can
be generalized to higher-dimensional cases.

\citet{Cheng.Shuming_PhysRevA.105.022411_2022} generalized a scheme for generating Bell nonlocality between arbitrarily many independent observers on each side, via the two-sided recycling of multiqubit states.
Subsequently, a series of works \cite{Lijian_2022EPR_steering,Han.Xin-Hong_PhysRevA.106.042416_2022,Qiao-Qiao.Lv_Phys.AMath.Theor.2307.09928_2023} present a steering scenario where both subsystems are accessible by multiple observers. \citet{Tong.Jun.Liu_OPT.OE.470229_2022} observe multi-observer steering on both sides simultaneously via weak measurements.
\citet{Hou.Wenlin_PhysRevA.105.042436_2022} observed network nonlocality sharing in the extended bilocal scenario. Additionally, works by Hu et al. (\citealp{Pandit.Mahasweta_PhysRevA.106.032419_2022};
\citealp{HuMingLiang_PhysRevA.108.012423}) considered scenarios involving two-sided sequential measurements where the entangled pair is distributed to multiple Alices and Bobs for entanglement sharing.

\subsubsection{Nonlocality Shairng In 3-Qubit System}

Compared with the case of two qubits, nonlocality sharing in three qubits systems exhibits richer properties.

\begin{figure}[htbp]
	\centering
	\includegraphics[width=0.45\textwidth]{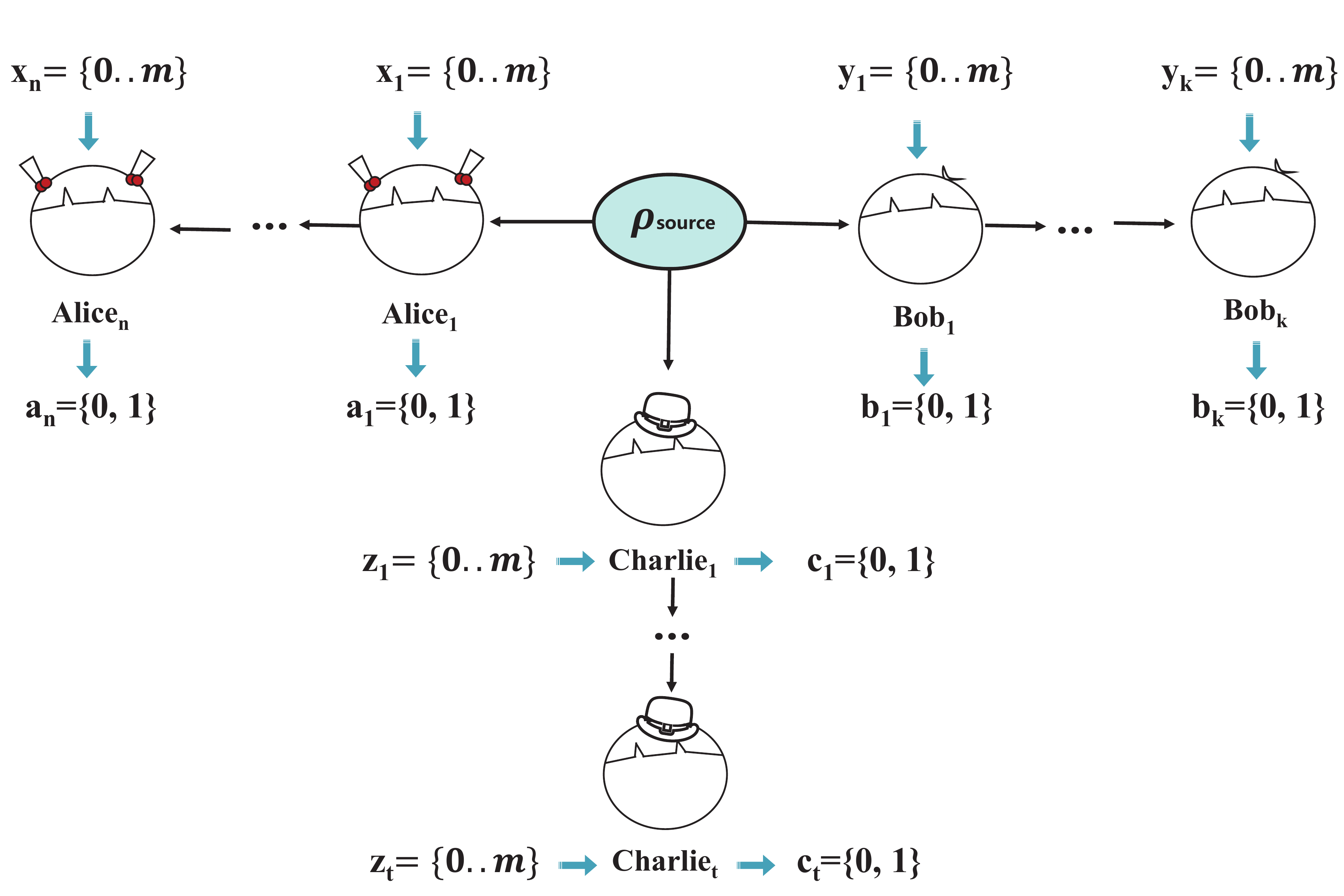}
	\caption{\justifying{\small{The trilateral scenario of nonlocality sharing}. The source $\rho_{source}$ distributes to distant observers, namely Alice$_1$, Bob$_1$ Charlie$_1$, which transmits the received particle to the following observers after measuring. The observer has measurement inputs $x_n$, $y_k$, $z_t\in\{0,...,m\}$ respectively. Measurement outputs after receiving the quantum state are labeled as $a_n, b_k,c_t\in\{0,1\}$ respectively. }}\label{trilateral nonlocality sharing}
\end{figure}

\citet{Ren.Changliang_PhysRevA.105.052221_2022} have extensively explored nonlocality sharing in a 3-qubit system by leveraging multiple violations of the Mermin-Ardehali-Belinskii-Klyshko (MABK) inequality. Their investigation reveals complete nonlocality sharing in the context of trilateral sequential scenarios, particularly when employing Greenberger-Horne-Zerlinger (GHZ) states, as illustrated in Fig. \ref{trilateral nonlocality sharing}. In comparison to 2-qubit scenarios, the phenomenon of nonlocality sharing in 3-qubit systems exhibits more intricate and diverse properties.

For the trilateral sequential scenario considered, they specifically set $n=k=t\in\{1,2\}$, where each observer has two measurement directions to choose from denoted as $n\cdot \hat{\sigma}$. The GHZ state is represented as $\left | \psi \right \rangle = \frac{1}{\sqrt{2}} (\left|000\right\rangle+\left|111\right\rangle)$.
In this framework, Alice$_n$'s $i$-th measurement choices are labeled as $\hat{A}_{n,i}$, Bob$_k$'s $j$-th measurement choices are denoted as $\hat{B}_{k,j}$, and Charlie$_t$'s $t$-th measurement choices are represented as $\hat{C}_{t,l}$.
The MABK inequality for the scenario can be expressed as,
\begin{eqnarray}
	B=|-M_a+M_b+M_c+M_d|\le2.
\end{eqnarray}
Where
\begin{align}
	&M_a=-E\left(A_{n,1},B_{\mathrm{k,1}},C_{\mathrm{t,1}}\right),
 &&M_b=E\left(A_{\mathrm{n,2}},B_{\mathrm{k,1}},C_{\mathrm{t,2}}\right),\nonumber\\
 &M_c=E\left(A_{\mathrm{n,2}},B_{\mathrm{k,2}},C_{\mathrm{t,1}}\right),&&M_d=E\left(A_{\mathrm{n,1}},B_{\mathrm{k,2}},C_{\mathrm{t,2}}\right).\nonumber
\end{align}
The list of quantum correlations of Alice$_n$-Bob$_k$-Charlie$_t$ corresponds to eight MABK inequalities. The measurement direction of each observer is always in the X-Y plane. Let Alice$_1$, Bob$_1$, and Charlie$_1$ perform a weak measurement with the same parameter, $G=G_1=G_2=G_3$.
In this scenario, the MABK values of Alice$_1$-Bob$_1$-Charlie$_1$ and  Alice$_2$-Bob$_2$-Charlie$_2$ can be expressed as $B_1=4G_1^3$;
$B_8=\frac{1}{2} (1+\sqrt{1-G^2})^{3}$.
It is easy to find that $B_1$ and $B_8$ can exceed 2 simultaneously when $G\in(\sqrt{2(2^{\frac{2}{3}}-2^{-\frac{1}{3}})},2^{\frac{1}{3}})$.
For the other list of Alice$_n$-Bob$_k$-Charlie$_t$, the corresponding MABK value, $B_2-B_7$, can also exceed 2 simultaneously when $G\in(\sqrt{\frac{\sqrt{5}-1}{2}},0.839)$.
Certainly, for the whole list of Alice$_n$-Bob$_k$-Charlie$_t$, the MABK value $B_1-B_8$ can exceed 2 simultaneously when $G\in(\sqrt{2(2^{\frac{2}{3}}-2^{-\frac{1}{3}})},2^{\frac{1}{3}})$, and they can achieve the maximal violation when $G=0.8$.
The findings indicate that trilateral sequential scenarios can exhibit complete nonlocality sharing. 



Besides,
\citet{Saha.Sutapa_Quantum.Inf.Process.s11128.018.2161.x_2019} and \citet{Ya.Xi_arXiv.2207.00296v1_2022} have explored the feasibility of tripartite nonlocality sharing involving more than three observers. \citet{Shashank._PhysRevA.103.022421_2021}  delved into the potential for multiple uses of a single copy of a 3-qubit state to detect genuine tripartite EPR steering. \citet{Wang.Jian.Hui_PhysRevA.106.052412_2022} observed nonlocality sharing in a star network scenario, demonstrating its applicability with any number of branches. Additionally, \citet{Maity.Ananda.G_PhysRevA.101.042340_2020} investigated the sequential detection of genuinely multipartite entanglement in quantum systems, considering an arbitrary number of qubits. The sequential detection of genuinely multipartite entanglement in quantum systems with an arbitrary number of qubits has also been explored in the work by \cite{Srivastava_2022sequential}.

\subsection{Experimental Implementation Of Nonlocality Sharing}


In this section, we review the recent advancements in experimental studies on nonlocality sharing.

The first experimental demonstration of nonlocality sharing is  presented by \citet{Schiavon._Quantum.Sci.Technol.2_015010_2017}. Specifically, \citeauthor{Schiavon._Quantum.Sci.Technol.2_015010_2017} observed double violations of the CHSH inequality among three observers who shared a 2-particle maximally entangled state through weak measurements. At the same time, \citet{Hu.Meng.Jun_Quantuminf.s41534-018-0115-x_2018} delved deeper into the continuously tunable optimal weak measurements for nonlocality sharing.
Subsequently, \citet{Feng.Tianfeng_PhysRevA.102.032220_2020} maximized the violation of Bell's inequality under optimal weak measurements of varying strengths and developed an optimal protocol for nonlocality sharing among three observers. Prior to this advancement, nonlocality sharing in experiments was constrained to three observers, until \citet{Foletto.Giulio_PHYS.REV.APPL.13.044008_2020,Foletto.Giulio_PhysRevResearch.2.033205_2020} observed triple violations of the CHSH inequality among four observers, thereby further validating the multiple sharing of nonlocality in the experiment.
Additionally, \citet{Xiao.Ya_arXiv.2212.03815_2022} diverged from a weak measurement strategy in the nonlocal sharing experiment, opting for projective measurements to share nonlocality. They observed that three independent observers could effectively share Bell nonlocality.

\subsubsection{Double Violations Of CHSH Inequality}

\begin{figure*}[htbp]
	\centering
\includegraphics[width=0.8\textwidth]{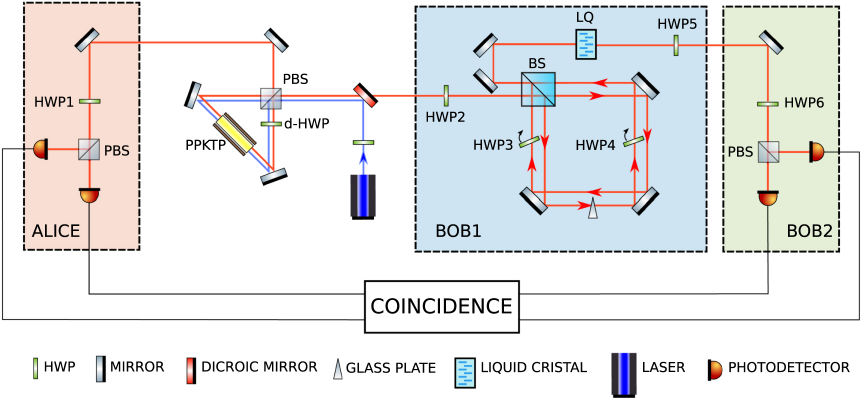}
\caption{\justifying{\small{Scheme of the experimental setup. This figure is adapted from \cite{Schiavon._Quantum.Sci.Technol.2_015010_2017}}}}\label{Schiavon_scheme_figure}
\end{figure*}

\citet{Schiavon._Quantum.Sci.Technol.2_015010_2017} demonstrated experimentally that nonlocality sharing can be observed between three observers \citet{Silva.Ralph_PhysRevLett.114.250401_2015}.
As depicted in Fig. \ref{Schiavon_scheme_figure}, the nonlocality-sharing scenario involves three observers: Alice, Bob$_1$, and Bob$_2$. They share pairs of a 2-particle entangled state, denoted as $\left|\Psi^-\right\rangle=\frac{1}{\sqrt{2}}(|H\rangle|V\rangle-|V\rangle|H\rangle)$, and choose the number of measurement options as $m=2$. Alice performs a strong measurement on her system using the basis ${\left|u_x\right\rangle, \left|u_x^\perp\right\rangle}$. Subsequent to her measurement, the state on Bob's side is projected onto $\left|u_x^{\left(-a\right)}\right\rangle$, where $\left|u_x^+\right\rangle\equiv\left|u_x\right\rangle$ and $\left|u_x^-\right\rangle\equiv\left|u_x^\perp\right\rangle$.

Specifically, Bob$_1$ employs a weak measurement utilizing a controlled phase gate, denoted as $CP\epsilon=\left|H\right\rangle\left\langle H\right|\otimes I+\left|V\right\rangle\left\langle V\right|\otimes e^{i\epsilon\sigma_z}$, to establish entanglement between an auxiliary qubit and the system. The ancillary qubit is initialized in the state $\left|+\right\rangle$ and subsequently measured in the basis ${\left|+\right\rangle, \left|-\right\rangle}$. By adjusting the rotation angle in the controlled phase gate, the strength of the measurement can be modulated.
To extend the measurement to an arbitrary basis ${\left|\omega_{y_1}\right\rangle, \left|\omega_{y_1}^\perp\right\rangle}$, the state requires rotation using a rotation matrix $\hat{R}_{y_1}$. This rotation ensures that $\hat{R}{y_1}\left|\omega_{y_1}\right\rangle=\left|H\right\rangle$ and $\hat{R}{y_1}\left|\omega_{y_1}^\perp\right\rangle=\left|V\right\rangle$, thereby generalizing the measurement to accommodate different bases.

\begin{figure}[htbp]
    \centering
 \includegraphics[width=0.4\textwidth]{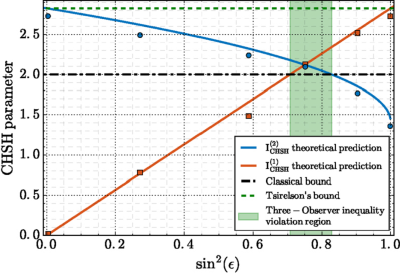}
 \caption{\justifying{\small{Measurement of $I_{\mathrm{CHSH}}^{(1)}$ (squares) and (diamonds) for several values of $\epsilon$. This figure is adapted from  \cite{Schiavon._Quantum.Sci.Technol.2_015010_2017}}}}
 \label{Schiavon_result_figure}
\end{figure}

The qubit transmitted to Bob$_1$ system $\left|\psi_{a|x}\right\rangle=\left|u_x^{\left(-a\right)}\right\rangle$ can be rewritten in his measurement basis as $\alpha\left|\omega_{y_1}\right\rangle+\beta\left|\omega_{y_1}^\perp\right\rangle$, where $\alpha=\left\langle\omega_{y_{1}}|u_x^{\left(-a\right)}\right\rangle,\beta=\left\langle\omega_{y_{1}}^\perp|u_x^{\left(-a\right)}\right\rangle$.
Inside Bob$_1$'s measurement device, the joint state $\left|\psi_{a|x}\right\rangle\otimes\left|+\right\rangle$ becomes $(\alpha\left|H\right\rangle+\beta\left|V\right\rangle)\left|+\right\rangle$ by the rotation.
The controlled phase gate applies the unitary operator $e^{i\epsilon\sigma_z}$ to the ancilla qubit, subject to the condition that the system qubit is in the state $\left|V\right\rangle$. This transforms the state into $\alpha\left|H\right\rangle\left|+\right\rangle+\beta\left|V\right\rangle(\cos\epsilon\left|+\right\rangle+i\sin\epsilon\left|-\right\rangle)$.
After the inverse rotation $\hat{R}_{y_1}^\dagger$, the joint state changes to $\left|\psi_{a|xy_1}\right\rangle=\alpha\left|\omega_{y_1}\right\rangle\left|+\right\rangle+\beta\left|\omega_{y_1}^\perp\right\rangle(\cos\epsilon\left|+\right\rangle+i\sin\epsilon\left|-\right\rangle)$.

Bob$_1$ measures the ancillary qubit in the basis $\{\left|+\right\rangle,\left|-\right\rangle\}$, while Bob$_2$ measures the polarization qubit in the basis $\{\left|v_{y_2}\right\rangle,\left|v_{y_2}^\perp\right\rangle\}$. The whole joint probability distribution corresponding to Alice-Bob$_1$ and Alice-Bob$_2$ can be given as,
\begin{align}
	&p\left(a,b_1|x,y_1\right)=\mathrm{Tr}\left[\left(\mathbb{I}\otimes \hat{\Pi}_{b1}\right)\rho_{a|xy_1}\right]p\left(a|x\right),\nonumber\\&p\left(a,b_2|x,y_2\right)=\sum_{y_1}p\left(y_1\right)Tr\left[\left(\hat{\Pi}_{b1}^{y_2}\otimes \mathbb{I}\right)\rho_{a|xy_1}\right]p\left(a|x\right),\nonumber
\end{align}
where $\hat{\Pi}_{b1},\hat{\Pi}_{b2}^{y_2}$ are the projectors for
Bob$_1$ and Bob$_2$ respectively.

If Alice randomly chooses two measurement settings  $-\frac{\sigma_z+\sigma_x}{\sqrt{2}}$ and $\frac{-\sigma_z+\sigma_x}{\sqrt{2}}$ while Bobs randomly choose two measurement settings $\sigma_z$ and $\sigma_x$,
the CHSH values of Alice-Bob$_1$ and Alice-Bob$_2$ are  $I_{\mathrm{CHSH}}^{\left(1\right)}=2\sqrt{2}\sin^2\epsilon$ and $I_{\mathrm{CHSH}}^{\left(2\right)}=\sqrt{2}\left(1+\cos\epsilon\right)$ respectively. 
In the experimental setup, Alice utilizes two measurement bases determined by the orientation of a half-wave plate set at angles of $11.25^{\circ}$ and $33.75^{\circ}$. On the other hand, Bob's measurement bases are defined by rotations of $0^{\circ}$ and $22.5^{\circ}$.

The experimental results of double violations of CHSH inequality are shown in
Fig. \ref{Schiavon_result_figure}.
For $\epsilon=0$, there is no interaction between the polarization and the ancillary state, $I_{\mathrm{CHSH}}^{\left(1\right)}$ approaches 0, whereas $I_{\mathrm{CHSH}}^{\left(2\right)}$ is close to the Tsirelson's bound.
With an increase in $\epsilon$, the experimental findings revealed a rise in $I_{\mathrm{CHSH}}^{\left(1\right)}$ and a concurrent decrease in $I_{\mathrm{CHSH}}^{\left(2\right)}$. To achieve the double violation, researchers conducted sequential measurements using two distinct values of $\epsilon$.
The results of 8 sequntial measurements with $\epsilon=1.049\pm0.002$ clearly demonstrate that both $I_{\mathrm{CHSH}}^{\left(1\right)}$ and $I_{\mathrm{CHSH}}^{\left(2\right)}$ surpass the classical bound, registering values of $I_{\mathrm{CHSH}}^{\left(1\right)}=2.125\pm0.003$ and $I_{\mathrm{CHSH}}^{\left(2\right)}=2.096\pm0.003$, respectively. Similarly, in a second series of trials with $\epsilon=1.053\pm0.002$, both $I_{\mathrm{CHSH}}^{\left(1\right)}$ and $I_{\mathrm{CHSH}}^{\left(2\right)}$ exceed the classical bound, maintaining values of $I_{\mathrm{CHSH}}^{\left(1\right)}=2.114\pm0.003$ and $I_{\mathrm{CHSH}}^{\left(2\right)}=2.064\pm0.003$.
The experimental results demonstrate the possibility of achieving double violations of CHSH inequalities, i.e. nonlocaltiy sharing, even if weak measurements are not optimal.



Almost the same period, \citet{Hu.Meng.Jun_Quantuminf.s41534-018-0115-x_2018} reported the experimental observation of the double violation of CHSH-Bell inequality for a single pair of entangled photons, utilizing a photonic system with continuous-tunable optimal weak measurement.
In their experiments, 
the pointer states are labeled as $|\phi_H\rangle (|\phi_V\rangle)$. The states $\{\left|\phi_{+1}\right\rangle,\left|\phi_{-1}\right\rangle\}$ chosen as reading states in the implementation represent two separate paths. These paths are denoted by $\left|0\right\rangle$ and $\left|1\right\rangle$.
By introducing a rotation of the half-wave plate, the pointer states transform into $\left|\phi_H\right\rangle=\cos\theta\left|0\right\rangle+\sin\theta\left|1\right\rangle$ and $\left|\phi_V\right\rangle=\sin\theta\left|0\right\rangle+\cos\theta\left|1\right\rangle$, where $0\le\theta\le\frac{\pi}{2}$.
In this context, the quality factor and information gain are expressed as $F=\sin2\theta$ and $G=\cos2\theta$, respectively. It is noteworthy that the condition for optimal weak measurement, given by $F^2+G^2=1$, is satisfied in this scenario.

\begin{figure}[htbp]
    \centering
\includegraphics[width=0.38\textwidth]{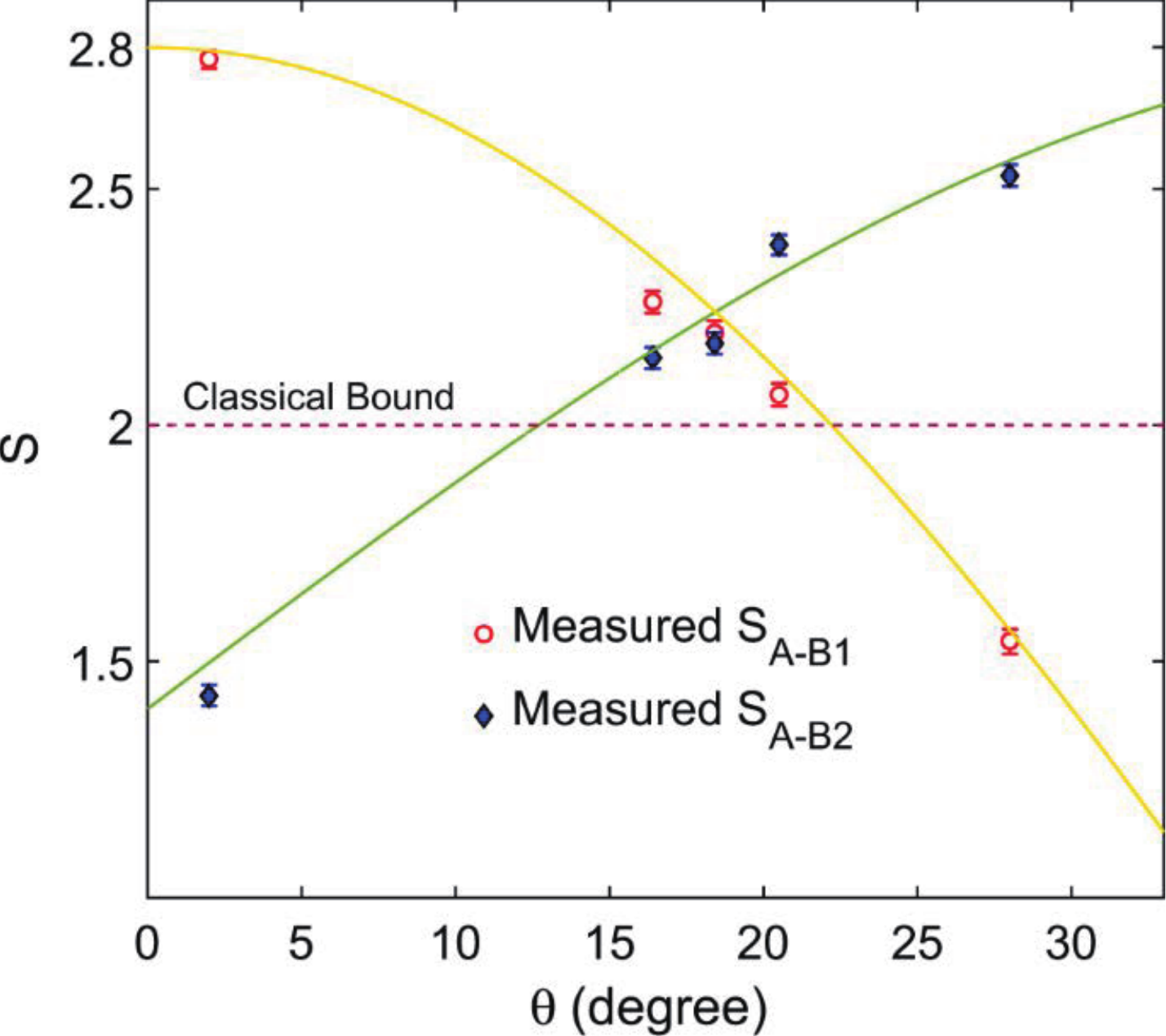}
\caption{\justifying{\small{Experimental results. This figure is adapted from \cite{Hu.Meng.Jun_Quantuminf.s41534-018-0115-x_2018}.}}}
\label{humengjun_results}
\end{figure}

To attain the maximal CHSH bound of $2\sqrt{2}$, Alice's measurement directions are chosen as ${\sigma_z, \sigma_x}$, while Bob's measurement directions are chosen as ${ \frac{-\sigma_z+\sigma_x}{2}, \frac{-(\sigma_z+\sigma_x)}{2} }$. In the experiment, the angles for Alice's half-wave plate can be set to either $(0^{\circ},45^{\circ})$ or $(22.5^{\circ},67.5^{\circ})$, and similarly, for Bob$_1$ and Bob$_2$ to $(-11.25^{\circ},33.75^{\circ})$ or $(11.25^{\circ},56.25^{\circ})$.
Five different angles $\theta={4^{\circ},16.4^{\circ},18.4^{\circ},20.5^{\circ},28^{\circ}}$ were selected, among which $\theta={16.4^{\circ},18.4^{\circ},20.5^{\circ}}$ lie in the region where double violations are predicted to be observed. The theoretical case of $F=0.6$, corresponding to $\theta=18.4^{\circ}$, yields a balanced double violation with $I_{\mathrm{CHSH}}^{\left(1\right)}=I_{\mathrm{CHSH}}^{\left(2\right)}=2.26$ under optimal weak measurements.
The experimental results, depicted in Fig. \ref{humengjun_results}, show a double violation at $\theta={16.4^{\circ},18.4^{\circ},20.5^{\circ}}$ with $I_{\mathrm{CHSH}}^{\left(1\right)}=2.20\pm0.02$ and $I_{\mathrm{CHSH}}^{\left(2\right)}=2.17\pm0.02$. This observed double violation at $\theta={16.4^{\circ},18.4^{\circ},20.5^{\circ}}$ is significant, with a confidence level of 10 standard deviations.

\begin{figure*}[htbp]
    \centering
\includegraphics[width=0.9\textwidth]{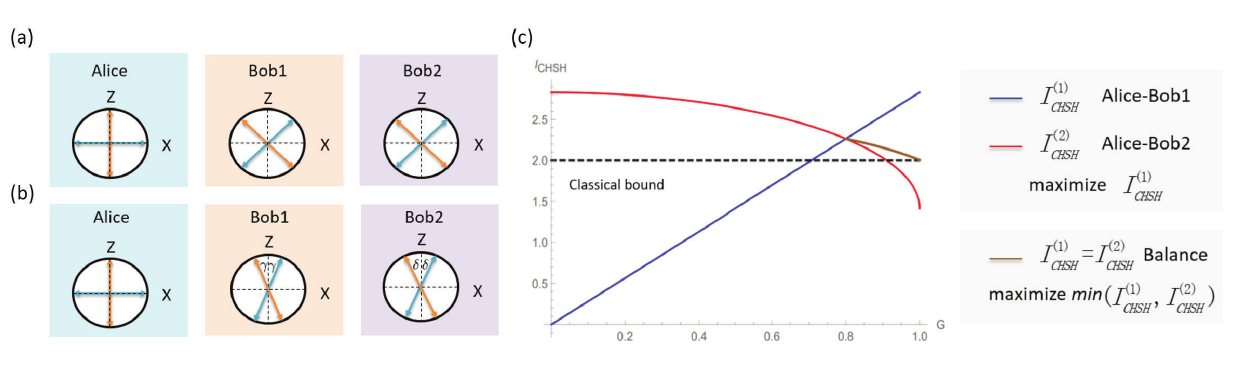}
\caption{\justifying{\small{The measurement setting and the corresponding CHSH values. This figure is adapted from \cite{Feng.Tianfeng_PhysRevA.102.032220_2020}}}}
\label{fengtianfeng_result}
\end{figure*}


\subsubsection{Observation Of Active Nonlocal Sharing}


While \citet{Silva.Ralph_PhysRevLett.114.250401_2015} successfully demonstrated nonlocality sharing among three observers, the optimal protocol for achieving this remains unclear. The existing demonstration establishes the feasibility of nonlocality sharing, yet a systematic and optimized protocol for implementing this phenomenon has yet to be fully elucidated. In 2020,
\citet{Feng.Tianfeng_PhysRevA.102.032220_2020} solved this problem and developed an optimal protocol for nonlocality sharing among three observers, allowing for Bob$_1$'s weak measurement strength to be nearly maximized. Counter-intuitively, the nonlocality sharing in this area is impossible in the previous scheme.


\citet{Feng.Tianfeng_PhysRevA.102.032220_2020} improved the measurement strategy based on the original work of \citet{Silva.Ralph_PhysRevLett.114.250401_2015}.
In the original work, Bob$_1$'s observables only consider the maximum value of the CHSH value $I_{\mathrm{CHSH}}^{(1)}$ corresponding to him and Alice. This inevitably lowers the upper bound for the CHSH value $I_{\mathrm{CHSH}}^{(2)}$ corresponding to Alice-Bob$_2$. To maximize the simultaneous violation range of both $I_{\mathrm{CHSH}}^{(1)}$, Feng et al. opted for a suitable measurement strategy to maximize $\min\{I_{\mathrm{CHSH}}^{(1)}, I_{\mathrm{CHSH}}^{(2)}\}$, as shown in Fig. \ref{fengtianfeng_result}.
They determined Alice's observables as $\sigma_x$ and $\sigma_z$, with the choices of observables for both Bob$_1$ and Bob$_2$ contingent upon the precision factor G of Bob1's measurement.
When $G<0.8$, $I_{\mathrm{CHSH}}^{(1)}$ is always less than or equal to $I_{\mathrm{CHSH}}^{(2)}$. Consequently, Bob$_1$ and Bob$_2$ can employ Silva's original scheme to maximize $\min\{I_{\mathrm{CHSH}}^{(1)},I_{\mathrm{CHSH}}^{(2)}\}$.
When $G>0.8$, $I_{\mathrm{CHSH}}^{(2)}$ may not necessarily exceed $I_{\mathrm{CHSH}}^{(1)}$. It can be observed that, under a given value of G, the smaller the difference between the two observables for Bob$_1$, the greater the quantum correlation between Alice and Bob$_2$. Therefore, enhancing the value of $\min\{I_{\mathrm{CHSH}}^{(1)}, I_{\mathrm{CHSH}}^{(2)}\}$ can be achieved by adjusting the similarity between the two observables for Bob$_1$,
\begin{align}
    \hat{\mu}&=\cos\gamma X+\sin\gamma Z;&&
\hat{\mu}'=\cos\gamma X-\sin\gamma Z;\nonumber\\
\hat{v}&= \cos\delta  X+\sin\delta Z ;&&
\hat{v}'= \cos\delta  X-\sin\delta Z;
\end{align}
where $\gamma,\delta\in\{0,\frac{\pi}{4}\}$.
For any given G greater than 0.8, they can always find the values of $\gamma$ and $\delta$ to achieve $I_{\mathrm{CHSH}}^{(1)}=I_{\mathrm{CHSH}}^{(2)}$, thereby obtaining the maximum value of $\min\{I_{\mathrm{CHSH}}^{(1)}, I_{\mathrm{CHSH}}^{(2)}\}$. 
This measurement strategy further extends the scheme proposed by Silva et al., enabling the attainment of nonlocality sharing when $0.910 < G < 1$, which is unattainable in the original scheme.


Similar to the setup proposed by \citet{Hu.Meng.Jun_Quantuminf.s41534-018-0115-x_2018}, \citet{Feng.Tianfeng_PhysRevA.102.032220_2020} made use of an ancillary qubit to realize a weak measurement with tunable strength. 
Specifically, defining any single qubit state as 
$|\psi\rangle=\alpha|0\rangle+\beta|1\rangle$, $\left(\alpha^2+\beta^2=1\right)$,
which will be conducted as a weak measurement.
The qubit $|\psi\rangle$ is coupled to an ancillary qubit prepared in the state $|0\rangle$ by a 2-qubit unitary $U$. The resulting 2-qubit state is expressed as $\alpha\left|0\right\rangle\otimes\left(\cos\theta\left|0\right\rangle+\sin\theta\left|1\right\rangle\right)+\beta\left|1\right\rangle\otimes\left(\sin\theta\left|0\right\rangle+\cos\theta\left|1\right\rangle\right)$. Then a projective measurement on the $0/1$ basis is conducted on the ancillary qubit, effectively resulting in a weak measurement on $|\psi\rangle$. The quality and precision factors satisfy the optimal pointer of the weak measurement since $G=\cos2\theta$ and $F=\sin2\theta$.


In the experiment, it is assumed that the qubit is $\left|\psi\right\rangle=\alpha\left|H\right\rangle+\beta\left|V\right\rangle$, exploiting the spatial degree of freedom as the ancillary qubit. The state outgoing will become $\alpha\left|H\right\rangle\otimes\left(\cos\theta\left|l\right\rangle+\sin\theta\left|u\right\rangle\right)+\beta\left|V\right\rangle\otimes\left(\sin\theta\left|l\right\rangle+\cos\theta\left|u\right\rangle\right)$. Where $\left|l\right\rangle$ and $\left|u\right\rangle$ denote the lower and upper spatial modes, respectively. The initial state is the maximum entangled state of two photons, which is shared with observers on both sides. They selected 9 different values of $G$ in the experiment and set the measurement directions. Theoretically, the maximum value of $\min\left(I_{\mathrm{CHSH}}^{\left(1\right)},I_{\mathrm{CHSH}}^{\left(2\right)}\right)$ can reach $2.263$ when $G = 0.8$. The experimental value is $I_{\mathrm{CHSH}}^{\left(1\right)}=2.214\pm0.011,I_{\mathrm{CHSH}}^{\left(2\right)}=2.168\pm0.007$. Both values exceed the classical bound by 10 standard deviations.
Even if Bob$_1$'s measurement strength reaches a very high level, such as $G=0.96$, the double violation of CHSH inequality can still be observed. Specifically,  $I_{\mathrm{CHSH}}^{\left(1\right)}=2.028\pm0.024,I_{\mathrm{CHSH}}^{\left(2\right)}=2.047\pm0.020$, refer to Fig. \ref{fengtianfeng_result}.

Overall, the authors devised an optimal approach for achieving nonlocality sharing among three observers and demonstrated in experiments that it uses weak measurements for a broad range of strengths. Previous experiments revealed that double violation of the CHSH inequality could only be observed using weak measurements of moderate strength. \citet{Feng.Tianfeng_PhysRevA.102.032220_2020} experimentally proved that double violation of CHSH inequality can also be achieved using not-so-weak measurements.

\subsubsection{Certification Of Sustained Entanglement And Nonlocality}


In the work by \citet{Foletto.Giulio_PHYS.REV.APPL.13.044008_2020,Foletto_PhysRevApplied.13.069902}, it is observed that pairs of entangled photons in polarization maintain their entanglement even when one particle undergoes three sequential measurements. Remarkably, each of these measurements has the potential to violate a CHSH inequality. The study specifically explores a sequential scenario, as illustrated in Figure \ref{bell nonlocality sharing}, where 
$k$ is defined as 3, and the number of measurement choices is set to 2.

In their experiments, Alice and Bob share a 2-qubit maximally entangled state $\left|\psi_1\right\rangle=\frac{1}{\sqrt{2}}\left(\left|00\right\rangle+\left|11\right\rangle\right)$. Bobs perform sequential measurements, $\{\hat{B}_0,\hat{B}_1\}=\{\sigma_z,\sigma_x\}$, on his photon. At the generic step $k$, Bob$_k$ applies $\hat{U}_{B,k}^\dagger$ to his photon ($k\ge2$); he performs either measurement $\hat{B}_0,\hat{B}_1$ with strength parameter $\mu_k$ and the state takes the form of $\left|\psi_k\right\rangle=\hat{U}_{A,k}\otimes \hat{U}_{B,k}\left[\cos\left(\eta_k\right)\left|00\right\rangle+\sin\left(\eta_k\right)\left|11\right\rangle\right]$. For the step $k+1$, if Bobs choose $\mu_j>0,\forall j\le k$, then $\eta_{k+1}>0$, it means that $\left|\psi_{k+1}\right\rangle$ is still entangled. And if Bobs perform measurement $\hat{B}_0$ with strength parameter $\mu_k>\arctan\left[\tan^2\left(\eta_k\right)\right]$ and the outcome is $-1$, then the new entanglement parameter is $\eta_{k+1}=\arctan\left[\frac{\tan\left(\mu_k\right)}{\tan\left(\eta_k\right)}\right]>\eta_k$, which amplifies the entanglement.
Alice's side must apply $\hat{U}_{A,k}^\dagger$, and her projective measurement can be given as $\hat{A}_{0,k}=\cos\left(\theta_k\right)\sigma_x+\sin\left(\theta_k\right)\sigma_z,\hat{A}_{1,k}=-\cos\left(\theta_k\right)\sigma_x+\sin\left(\theta_k\right)\sigma_z$, where $\theta_k=\operatorname{arccot}\left[\sin\left(2\eta_k\right)\right]$. The CHSH value corresponding to Alice-Bob$_k$ can be obtained as $I_{\mathrm{CHSH},k}=2\cos\left(2\mu_k\right)\sqrt{1+\sin^2\left(2\eta_k\right)}$. For any choice of k, if Bobs choose that $\mu_k<\mu_{k,\max}=\frac{1}{2}\arctan\left[\sin\left(2\eta_k\right)\right]$, the CHSH value can exceed 2.

\citet{Foletto_PhysRevApplied.13.069902} verified that Bob's side makes at most three sequential measurements and the protocol can be stopped at steps 1, 2, or 3. They choose that $\mu_1=0.34,\mu_2=0.19,\mu_3=0$.
Overall, they measured a total of 9 independent $I_{\mathrm{CHSH}}$ (one stopped in the first step of the protocol, four stopped in the second step, and four stopped in the third step).
\citeauthor{Foletto.Giulio_PHYS.REV.APPL.13.044008_2020} encode two qubits in the polarization degree of freedom of two separated photons. Polarization-entangled photon pairs are generated by a custom-built source based on a Sagnac interferometer. For each step of Bobs's measurement setup in the experiment: two half-wave plates represent the application $\hat{U}_B^\dagger$, the other half-wave plate represents the measurement choices between $\hat{B}_0$ and $\hat{B}_1$, a polarization-based Mach-Zehnder interferometer (MZI) implements the unsharp measurement. It entangles the polarization with the path degree of freedom, while the sharpness parameter is set by the angles $\{-\frac{\mu}{2},\frac{\pi}{4}-\frac{\mu}{2}\}$ of the internal half-wave plate.

The experimental results reveal that all nine independent $I_{\mathrm{CHSHs}}$ values surpass 2, providing evidence that sequential measurements do not disrupt entanglement. Notably, the $I_{\mathrm{CHSH}}$ value obtained in step 3 exceeds those in steps 1 and 2. This observation aligns with expectations based on the sharpness parameters employed in the experiment, illustrating the feasibility of applying the protocol for entanglement amplification. It's important to note that this amplification effect is observed only for a subset of measurement choices and outcomes.
The reported results demonstrate robust violations of the CHSH inequality, supported by a statistical significance exceeding 10 standard deviations, even at the third step of the sequential process.


\subsubsection{Realization Of Double Violation By Projective Measurements Strategy}


Recent studies by \citet{Anna._PhysRevLett.129.230402_2022} have demonstrated that projective measurements alone are sufficient for recycling nonlocality. Additionally, the work of \citet{Xiao.Ya_arXiv.2212.03815_2022} has employed projective measurements and validated the CHSH inequality to show that, for initial states that are either maximally entangled or partially entangled, it is experimentally observable that three independent observers can share the Bell nonlocality of binary qubit states.

In their experiment, the initial state is chosen as a two-photon entangled state given by $\left|\psi_{\varphi}\right\rangle=\cos\varphi\left|HH\right\rangle+\sin\varphi\left|VV\right\rangle$, where $\varphi\in{0^{\circ},45^{\circ}}$. The parameter $\varphi$ is manipulated using a half-wave plate. One of the entangled photons is directly transmitted to Alice, who performs a projective polarization measurement using a half-wave plate and a polarization beam splitter. Simultaneously, the second photon undergoes a distinct path, traversing an unbalanced interferometer before reaching Bob and Charlie for their respective projective measurements. Bob employs linear polarizers, while Charlie utilizes half-wave plates and polarization beam splitters.
The unitary operation setup involves a sequence of optical elements, including a quarter-wave plate, a half-wave plate, and another quarter-wave plate on both paths. This configuration allows Bob to implement arbitrary single-qubit unitary transformations. Furthermore, variable neutral density filters, easily rotatable, provide control over the relative probability of combining the two paths in the interferometer setup.

\begin{figure}
    \centering
\includegraphics[width=0.45\textwidth]{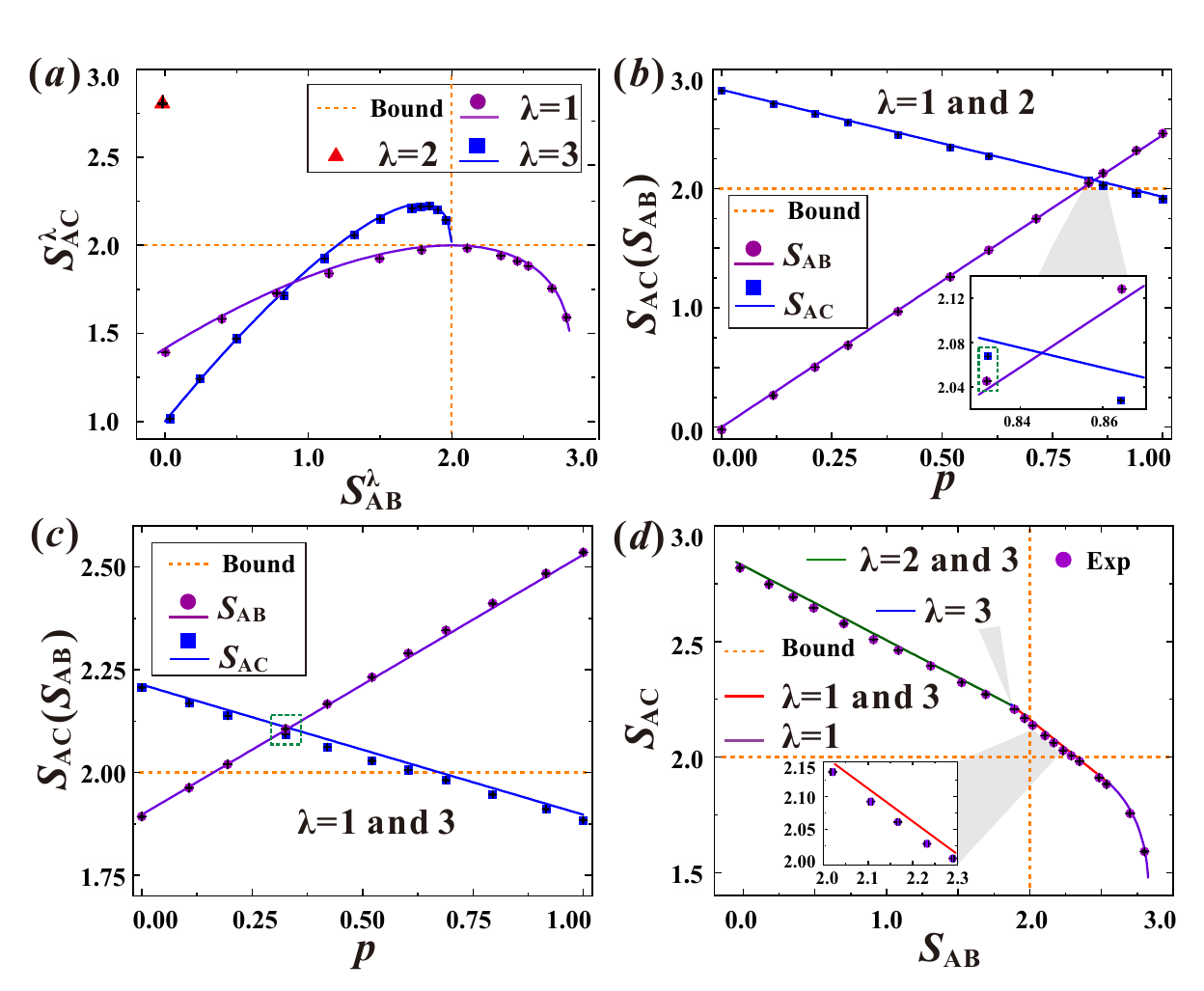}
\caption{\justifying{\small{Experimental results of double violation by projective measurements.} This figure is adapted from \cite{Xiao.Ya_arXiv.2212.03815_2022}.}}
\label{xiaoya_results}
\end{figure}

\citet{Xiao.Ya_arXiv.2212.03815_2022} first investigate the Bell nonlocality recycling for a maximally entangled state. In order to obtain the optimal double violation, the first and second cases are randomly combined \cite{Anna._PhysRevLett.129.230402_2022}, and the optimal settings for the first and second cases are, $\phi=75^{\circ},\chi=45^{\circ}$. As $p$ increases, $I_{AB}$ increases and $I_{AC}$ decreases, where $p$ is the probability of the first scenario. The double violation of CHSH inequality can be observed in the range of $p\in\left[\frac{2}{\sqrt{6}},\frac{4-2\sqrt{2}}{3-\sqrt{3}}\right]$ (see Fig. \ref{xiaoya_results}). In theory, $I_{AB}$ and $I_{AC}$ reach their maximum value when $p=\frac{6-2\sqrt{3}}{3}$. In the experiment, $I_{AB}=2.0451\pm0.0015,I_{AC}=2.0676\pm0.0014$ can be obtained when $p=0.83207\pm0.0015$, both of which are 30 standard deviations higher than the classical bound. 

Furthermore, the first and third cases are randomly combined \cite{Anna._PhysRevLett.129.230402_2022}, with the optimal measurement setting being $\phi=71.57^{\circ}$ and $\theta=18.43^{\circ}$. The maximum values of the two CHSH expressions are $I_{AB}=2.1056\pm0.0013$, and $I_{AC}=2.0927\pm0.0013$, respectively,  when $p=0.3256\pm0.0012$. Both values are 71 standard deviations higher than the classical bound.
Obviously, these results are stronger than the combination of the first and second cases.
They further studied the Bell nonlocality of partially entangled states. By varying the probability $p$ of the first scenario, they discovered that if the proper $\phi$ is chosen, partially entangled states can achieve a stronger double violation than the maximum entangled state, regardless of whether it is a blend of the first and second cases or the first and third cases. When examining the second case, where $\phi$ takes on values of $\{45^\circ,39.23^\circ,34.08^\circ,28.32^\circ,21.77^\circ\}$, all results exceed the bound established by the maximum entangled state. The average fidelity of the state is approximately $0.9853\pm0.0012$. 

In conclusion, the researchers successfully conducted an experimental demonstration of nonlocality sharing involving a 2-qubit entangled state among three observers through the utilization of projective measurements. Through the random combination of basis projective and identity measurements, they observed a double violation of the CHSH inequality for the maximally entangled state. Interestingly, the violations became more pronounced when employing random combinations of basis projective and mixed measurements. Notably, certain partially entangled states exhibited even stronger double violations, with the maximum violation surpassing that achievable with the maximally entangled state by up to 11 standard deviations.

{\color{black}
\section{Steering Sharing}\label{steering_sharing_section}

Einstein-Podolaky-Rosen(EPR) steering is a unique quantum correlation of two-particle quantum states between Bell nonlocality and entanglement \cite{EPR1935PhysRev.47.777}. The steerable states are a subset of entangled states and a superset of Bell states. Especially, steerability is an asymmetrical correlation resulting in a more intricate monogamous relationship than bell nonlocality
(\citealp{J.S.Bell.PhysicsPhysiqueFizika.1.195};
\citealp{Werner.PhysRevA.40.4277};
\citealp{Wiseman2007PhysRevLett.98.140402};
\citealp{Jones.PhysRevA.76.052116};
\citealp{Cavalcanti.Rep.Prog.Phys.80.024001_2017};
\citealp{Uola.RevModPhys.92.015001}). 
\citet{Reid.PhysRevA.88.062108} established monogamy constraints contingent and demonstrated that the steerable states have a directional property. Subsequently, \citet{Bowles.PhysRevLett.112.200402} also demonstrated such directional property, i.e., one can steer the other but not the other way around \cite{Mal.arXiv.1711.00872,Adesso_2016_IOPPublishing473001,Lami.PhysRevLett.117.220502}.

Inspired by the investigation of nonlocality sharing, it is straightforward to extend to quantum steering. \citet{Sasmal.Souradeep_PhysRevA.98.012305_2018} investigated the maximum number of multiple observers that could steer a single observer in the sequential scenario. Conversely, \citet{YaoDan.PhysRevA.103.052207_2021} examined the maximum number of multiple observers that a single observer could steer. Subsequently, a series of works concentrated on steering sharing in the one-sided sequential case has been explored (\citealp{Sasmal.Souradeep_PhysRevA.98.012305_2018};
\citealp{Shenoy.H._PhysRevA.99.022317_2019};
\citealp{Yeon.Ho.Choi._optica.394667_2020};
\citealp{Shashank._PhysRevA.103.022421_2021};
\citealp{YaoDan.PhysRevA.103.052207_2021};
\citealp{Han.XinHong_arXiv.2303.05954_2023};
\citealp{Chen.Yuyu.Chin.Phys.B.32.040309_2023}). \citet{Zhu.Jie_PhysRevA.105.032211_2022} further explored steering sharing in the two-sided case and observed double EPR steering.
In addition, in a tripartite sequential scenario (where sequential measurements performs only on one side), \citet{Shashank._PhysRevA.103.022421_2021} investigated all types of (1→2) and (2→1) genuine tripartite EPR steering. Recently, steering sharing with unequal sharpness measurement has been investigated in \cite{Qiao-Qiao.Lv_Phys.AMath.Theor.2307.09928_2023,Han.XinHong_arXiv.2303.05954_2023}.





\subsection{Sequential Steering By Multiple Observers}

The phenomenon of sharing nonlocality offers a different perspective for delving into quantum correlations. Subsequent studies have expanded this concept to encompass the sharing of steering phenomena through sequential measurements. \citet{Sasmal.Souradeep_PhysRevA.98.012305_2018} explored a limited number of Bobs that Alice can steer in the sequential scenario. They choose the $\rho_{\mathrm{source}}$ with the singlet state. To explore the sequential steering by multiple observers, they employ CHSH-like inequality as criteria, which is also known as Cavalcanti Foster Fuwa Wiseman(CFFW). Suppose that each observer has a choice between two dichotomic measurements, the CHSH-like inequality remains, 
\begin{align}
	S_{BA}=&\sqrt{\left \langle  (B_1+B_2)A_1\right  \rangle ^2+\left \langle  (B_1+B_2)A_2\right  \rangle ^2}\nonumber\\
	&+\sqrt{\left \langle  (B_1-B_2)A_1\right  \rangle ^2+\left \langle  (B_1-B_2)A_2\right  \rangle ^2}
	\le 2. 
\end{align}
And when each observer has a choice between m dichotomic measurements, they employ linear steering inequality (see Eq. (\ref{linear_inequality})) as criteria.

In the sequential scenario, both Alice and Bob$_k$ have two independent dichotomic observables, $\hat{x}^j=\hat{y}^i_k=\vec{n}.\hat{\sigma}$, $i,j\in\{0,1\}$. Since all Bobs steer Alice’s system, the measurement settings of Alice are mutually unbiased and satisfy $\hat{x}^0.\hat{x}^1=0$. The intermediate Bob$_k$ carry out weak measurements with precise parameter $\lambda_k$. While the last observer on Bob's side will perform strong measurements ($\lambda_k=1$). They discussed the simplest case, $k=2$, it is shown that the double violation of the CHSH-like inequalities can be observed in the range of $\lambda_1\in(0.71,0.91)$. So Bob$_1$ and Bob$_2$ can steer Alice simultaneously. Besides, they determined that no more than two Bobs can steer a single Alice based on the CHSH-like inequality criterion.

Subsequently, they increased the number of measurement settings of all the observers in the sequential scenario. Assuming that Alice chooses between three binary measurements $\hat{x}^i$, and Bob$_k$ chooses between three binary measurements $\hat{y}^i_k$, where $i\in\{0,1,2\}$. The steering verification that Bob$_k$ steers Alice’s system can be determined by the violation of the three-setting linear inequality, which can expressed as $F^3_m=\frac{1}{\sqrt{3}}\sum_{i=1}^{3}\bar{C}^{ii}_m$. Since all Bobs steer Alice's system, the measurement settings of Alice are orthogonal to each other, $\hat{x}^0.\hat{x}^1=0, \hat{x}^0.\hat{x}^2=0, \hat{x}^1.\hat{x}^2=0$. In this scenario, they show that no more than three Bobs simultaneously can steer Alice based on the violation of the three-setting linear inequality.

Since then, a logical proposition has emerged, suggesting that the capacity for multiple Bobs to steer Alice escalates with the increasing number of measurement choices each observer possesses. The hypothesis suspected that an ensemble of n-Bobs can steer the Alice system when probing steering through the n-settings CJWR inequality, which remains unproven or unrefuted.

The other related work includes,
\citet{Shenoy.H._PhysRevA.99.022317_2019} has found that with isotropic entangled states of local dimension d,
the number of Bobs that can steer Alice is found to be $N_{Bob}\sim \frac{d}{logd}$ thus leading to an arbitrarily large number of successive instances of steering with independently chosen and unbiased inputs.
\citet{Yeon.Ho.Choi._optica.394667_2020} experimentally demonstrate multiple-observer quantum steering by exploiting sequential weak measurements.
\citet{Han.XinHong_arXiv.2303.05954_2023} have proposed an effective method for constructing optimal nonlocal measurements using quantum ellipsoids to share quantum steering in a 3-qubit systems. It has further investigated the (n→1) steering sharing in 3-qubit system (\citealp{Shashank._PhysRevA.103.022421_2021};
\citealp{Chen.Yuyu.Chin.Phys.B.32.040309_2023}).

\subsection{Alice Steers Two Bobs}

Conversely, \citet{YaoDan.PhysRevA.103.052207_2021} investigated the case where a single Alice steers multiple Bobs by sharing an arbitrary 2-qubit state in the sequential scenario. 
The arbitrary 2-qubit state can be expressed as 
$\rho _{AB}=\frac{1}{4}(\mathbb{I}\otimes \mathbb{I}+\sum^3_i\lambda_i\sigma_i\otimes \mathbb{I}+\sum^3_j\eta _j\mathbb{I}\otimes\sigma_j+\sum^3_k\gamma _k\sigma_k\otimes\sigma_k)$, where $i,j,k\in\left\{ 1,2,3 \right\}$, $|\gamma_1|\le|\gamma_2|\le|\gamma_3|$.
Alice performs a strong measurement $\hat{\sigma}_{w_x}=\sum^3_{i=1}w^i_x\sigma_i$, where $\vec{w}_x=\{w^1_x,w^2_x,w^3_x\}$ represents the measurement directions. 
Bob$_1$ performs weak measurement on the received qubit along the $\vec{\mu}_{y_1}$, the remaining observers on Bob's side can be defined similarly.

To explore the phenomenon of steering sharing, they have chosen the CHSH-like inequality and the three-setting linear steering inequality as the benchmarks for evaluating steering. Under the CHSH-like inequality, each observer is provided with two measurement options. While, in the case of the three-setting linear inequality, each observer has three measurement choices.

They started by exploring whether Alice can steer two Bobs. In the scenario, Bob$_2$ performs a strong measurement, where $F_2=0$ and $G_2=1$. 
Under the optimal measurement setting, the CHSH-like average value corresponding Alice-Bob$_n$ (n=1,2) can be expressed as
$I^{(1)}_{\mathrm{CHSH-like}}=2\sqrt{\gamma^2_2+\gamma_3^2 }G_1$,
$I^{(2)}_{\mathrm{CHSH-like}}=\sqrt{2(1+F^2_1)(\gamma^2_2+\gamma_3^2) }$.
Defined $|\gamma_i|=1$, the CHSH-like average value was given as $I^{(1)}_{\mathrm{CHSH-like}}=2\sqrt{2}G_1$, $I^{(2)}_{\mathrm{CHSH-like}}=2\sqrt{1+F^2_1}$.
No matter whether the pointer distribution of Bob$_1$'s weak measurement is square or optimal, it is evident that a double violation of two inequalities exists within the appropriate range. Hence, it demonstrates Alice can steer two Bobs.

Furthermore, they show that three Bobs can be steered by a single Alice when the triple violation of the three-setting linear steering inequality is observed. In this case, each observer has three measurement choices $(m=3)$.
According to \cite{Costa.PhysRevA.93.020103_2016}, the measurement settings of Bobs also need to meet the following two conditions,
\begin{eqnarray}
	\sum _{y_1}(\mu^1_{y_1})^2=\sum _{y_1}(\mu^2_{y_1})^2=\sum _{y_1}(\mu^3_{y_1})^2=1,\nonumber\\
	\sum_{y_1}\mu^1_{y_1}\mu^2_{y_1}=\sum_{y_1}\mu^1_{y_1}\mu^3_{y_1}=\sum_{y_1}\mu^2_{y_1}\mu^3_{y_1}=0.\nonumber
\end{eqnarray}
The three-setting linear inequalities corresponding to Alice-Bob$_n$ can be simplified as $I^{(1)}_{\mathrm{linear}}=3G_1, I^{(2)}_{\mathrm{linear}}=1+2F_1$ under the optimal measurement settings. It is evident that achieving double violations of three-setting linear inequalities remains easily attainable, signifying Alice's capability to steer two Bobs.

While the more interesting question is how many multiple observers can share the correlation in this scenario.  Assuming the observer has three inputs $x,y_1,y_2\in\{0,1\}$, and employing the linear steering inequality, when the measurement settings of Bob$_2$ and Bob$_3$ match those of Bob$_1$, the average value of linear steering inequality $I^{(n)}_{\mathrm{linear}},(n\in\left\{1,2,3\right\})$  can attain its maximum value. Under the optimal measurement settings, three linear inequalities can be given as,
$I^{(1)}_{\mathrm{linear}}=3G_1,
I^{(2)}_{\mathrm{linear}}=3\gamma_1 G_2,
I^{(3)}_{\mathrm{linear}}=3\gamma_1\gamma_2$.
Obviously, the triple violation of the linear inequalities can be obtained under the optimal pointer of weak measurement, but not under the square pointer. This result proves that Alice can steer three Bobs. 

Subsequently, they investigated the potential for quadruple violations, encompassing Alice and four Bobs. In cases where the measurement configurations of Bob$_2$, Bob$_3$, and Bob$_4$ align with those of Bob$_1$ in the preceding measurement settings, the linear inequalities can reach their utmost value. However, irrespective of whether the distribution of pointer measurements is optimal or square, no observable quadruple violation occurs. Thus, in this sequential scenario, the number of Bobs steered by Alice aligns with the number of measurement settings for each observer. Currently, the ongoing debate regarding multiple violations of inequality revolves around the prioritization of maximized violation by Alice and Bob$_1$. It remains an open question whether there exist multiple violations simultaneously when the violation of the inequality associated with Alice and Bob$_1$ is not maximal. In addition, it has further investigated ($1\to n$) steering sharing in a 3-qubit system (\citealp{Shashank._PhysRevA.103.022421_2021};
\citealp{Chen.Yuyu.Chin.Phys.B.32.040309_2023}).

\subsection{Steering Sharing In The Multipartite Systems And Multilateral Cases}
\subsubsection{Genuine Tripartite EPR Steering}
Multipartite quantum steering has demonstrated its importance as a vital resource for diverse quantum protocols, including quantum key distribution and quantum teleportation. Delving into the subtleties of multipartite quantum steering offers valuable insights into the quantum correlations present in intricate quantum systems with more than two parties. Specifically, comprehending the effective distribution and utilization of multipartite quantum steering among diverse parties represents a noteworthy focus. This exploration stands to contribute significantly to the advancement of sophisticated quantum communication protocols and technologies. \citet{Shashank._PhysRevA.103.022421_2021} explored the sharing of genuine tripartite EPR steering among multiple sequential observers. Within the trilateral sequential scenario, a pure 3-qubit state of either the GHZ type or W type is distributed among three observers (Alice, Bob, Charlie). On one side, arbitrary sequential observers conduct unsharp measurements. Due to the inherent asymmetry of EPR steering, two distinct types of genuine tripartite EPR steering scenarios: one where Alice endeavors to genuinely steer the particles of both Bob and Charlie (1→2), and another where Alice and Bob collaborate to genuinely steer Charlie's particles (2→1), as illustrated in Fig. \ref{genuine steering_1} and Fig. \ref{genuine steering_2}.

\begin{figure}[htbp]
	\centering
	\includegraphics[width=0.45\textwidth]{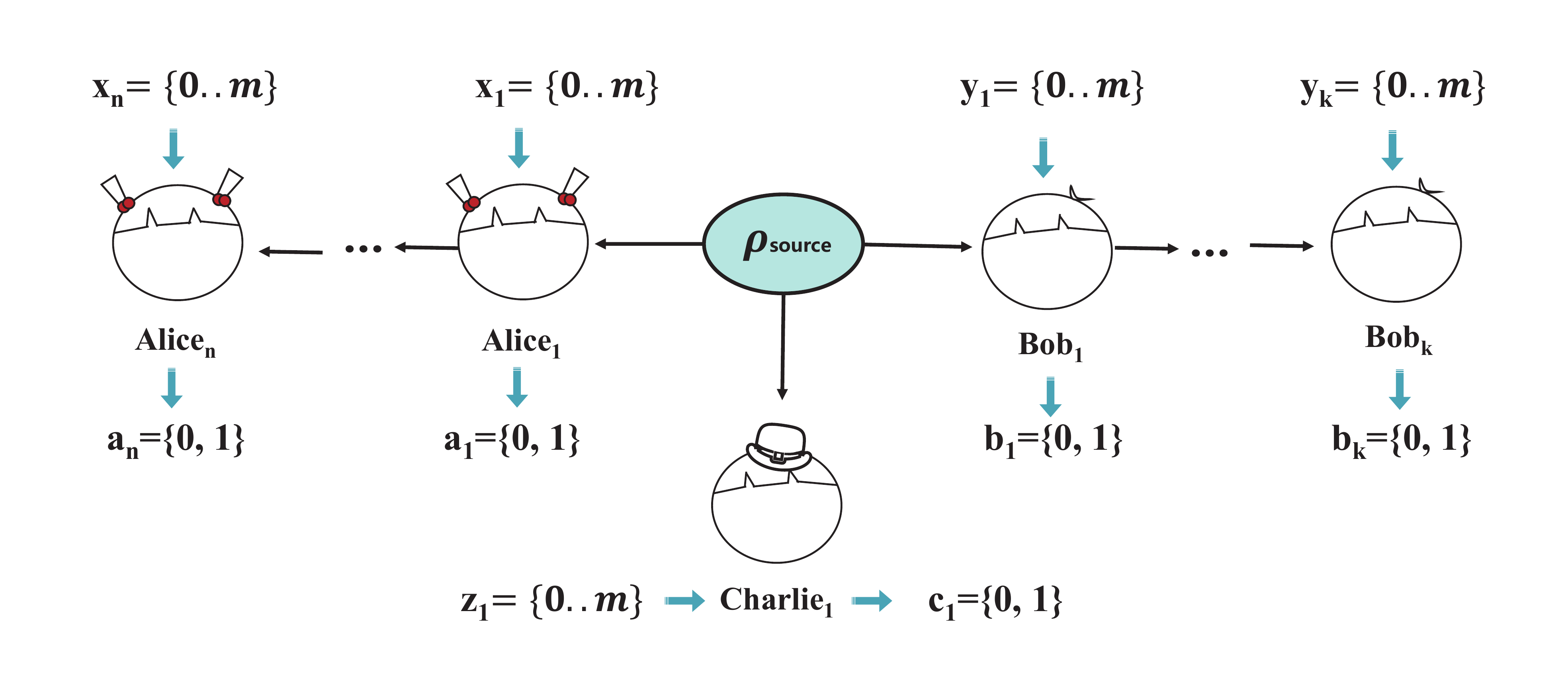}
	\caption{\justifying{\small{A schematic diagram (1→2): Alice attempting to genuine steer Bob and Charlie's particles. }}}\label{genuine steering_1}
 \includegraphics[width=0.45\textwidth]{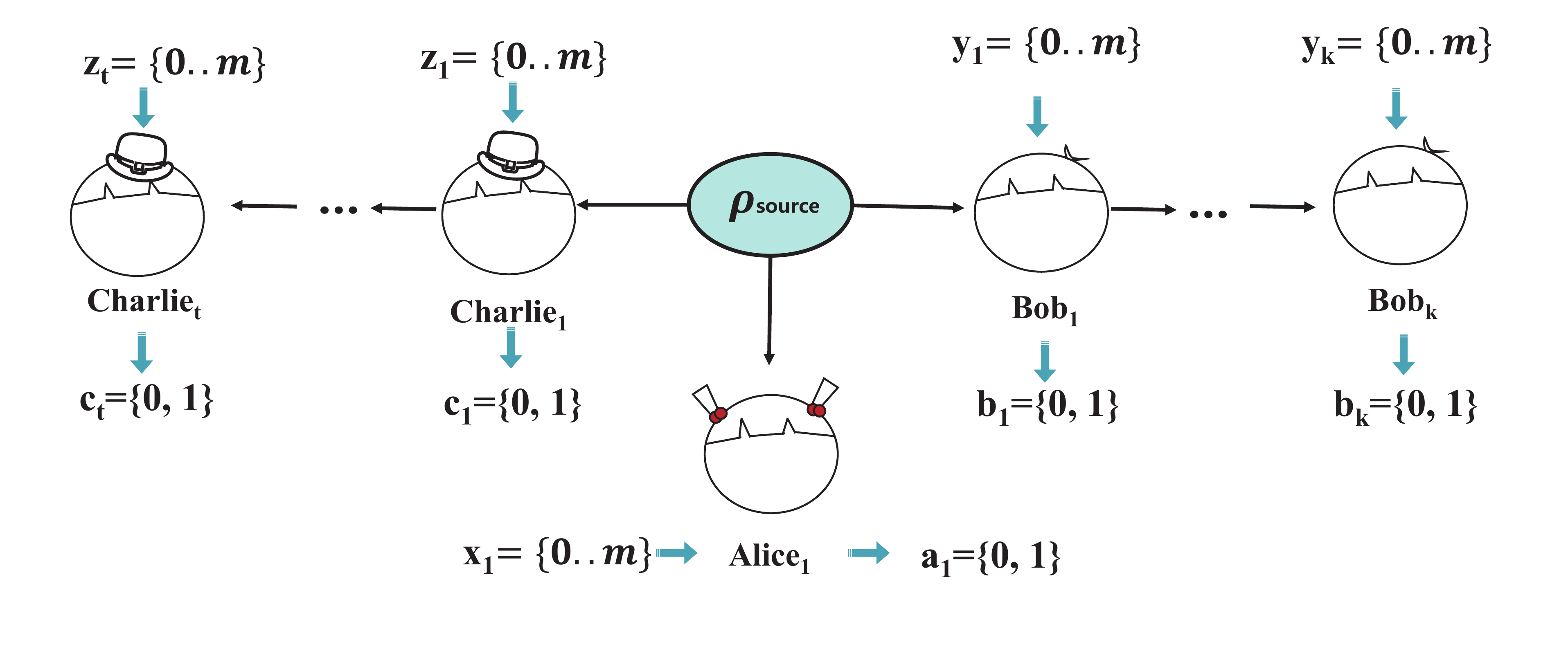}
	\caption{\justifying{\small{A schematic diagram (2→1): Alice and Bob attempting to genuine steer Charlie's particles.}}}\label{genuine steering_2}
\end{figure}

In (1→2) case, Supposed that Alice's measurement operator is $\hat{A}_{a|x}$ which is the projective measurement along the $\hat{A}$-direction with measurement outcome $a$. The unnormalized conditional states on Bob-Charlie’s side can be expressed as $\left\{ \sigma^{\mathrm{BC}}_{a|x} \right\}_{a|x}$, where every element can be given by as $\sigma^{\mathrm{BC}}_{a|x}=\mathrm{Tr}[(\hat{A}_{a|x}\otimes \mathbb{I}_B \otimes \mathbb{I}_C )\rho]$. When the initial state $\rho$ is not genuinely entangled but separable, it always can be decomposed as,
\begin{align}
	\rho=&\sum_\lambda p^{\mathrm{A:BC}}_\lambda\rho^{\mathrm{A}}_\lambda\otimes \rho^{\mathrm{BC}}_\lambda+p^{\mathrm{B:AC}}_\mu\rho^{\mathrm{B}}_\mu\otimes \rho^{AC}_\mu\nonumber\\
	&+p^{\mathrm{AB:C}}_v \rho^{\mathrm{AB}}_v \otimes \rho^{\mathrm{C}}_v.
\end{align}
Where $p^{\mathrm{A:BC}}_\lambda,p^{\mathrm{B:AC}}_\mu,p^{\mathrm{AB:C}}_v$ are probability distributions. $\mathrm{A:BC}$, $\mathrm{B:AC}$, and $\mathrm{AB:C} $ represent the different types of bi-partitions. It is worth noting that, there is no difference between $\mathrm{AB: C}$ and $\mathrm{C: AB}$. 
Once each element of $\left\{ \sigma^{\mathrm{BC}}_{a|x} \right\}_{a|x}$ cannot be described as,
\begin{align}
	\sigma^{\mathrm{BC}}_{a|x}=&\sum_\lambda p^{\mathrm{A:BC}}_{\lambda}p_\lambda(a|x)\rho^{\mathrm{BC}}_\lambda +\sum_\mu p^{\mathrm{B:AC}}_{\mu}\rho^{\mathrm{B}}_\mu\otimes\sigma^{\mathrm{C}}_{a|x\mu}\nonumber\\
	&+\sum_v p^{\mathrm{AB:C}}_v \sigma^{\mathrm{B}}_{a|xv} \otimes \rho^{\mathrm{C}}_v, \quad\quad \forall a,x,
\end{align}
which means the genuine tripartite EPR steering from Alice to Bob-Charlie. $\sigma^{\mathrm{C}}_{a|x\mu}$ is the unnormalized conditional state on Charlie’s side when Alice performs the measurement $x$ with the outcome `$a$'. Similarly, $\sigma^{\mathrm{B}}_{a|xv}$ is the unnormalized conditional state on Bob’s side when Alice performs the measurement $x$ with the outcome `$a$'.

In the (2→1) case, assuming that Alice's measurement operator is represented by $\hat{A}_{a|x}$ and Bob's measurement operator is represented by $\hat{B}_{b|y}$, the unnormalized conditional state prepared on Charlie's side can be given as $\sigma^{\mathrm{C}}_{ab|xy}=\mathrm{Tr}[(\hat{A}_{a|x}\otimes \hat{B}_{b|y} \otimes \mathbb{I}_C )\rho]$.
Similarly, once each element of $\left\{ \sigma^{\mathrm{C}}_{ab|xy} \right\}_{a,b,x,y}$ cannot be written as,
\begin{align}
	\sigma^{\mathrm{C}}_{ab|xy}=&\sum_\lambda p^{\mathrm{A:BC}}_{\lambda}p_\lambda(a|x)\sigma^{\mathrm{C} }_{b|y\lambda}
	+\sum_\mu p^{\mathrm{B:AC}}_{\mu}p_\mu(b|y)\sigma^{\mathrm{C}}_{a|x\mu}\nonumber\\&+
	\sum_v p^{\mathrm{AB:C}}_v p_v(ab|xy) \rho^{\mathrm{C}}_v,
 \quad\quad\forall a,b,x,y,
\end{align}
which means the genuine tripartite EPR steer from Alice-Bob to Charlie.

Unfortunately, the above definition is not practical, and few criteria for determining genuine multipartite quantum steering were constructed. Nevertheless, Cavalcanti et al. designed several inequalities to detect the genuine quantum steering of GHZ and W states in these two cases, which can be expressed as,
\begin{widetext}
	\begin{equation}
 \left\{
		\begin{aligned}
			\quad G_1=&1+g_\alpha\langle Z_BZ_C\rangle-\frac{1}{3}(\langle A_3Z_B\rangle+\langle A_3Z_C\rangle+\langle A_1X_BX_C\rangle-\langle A_1Y_BY_C\rangle-\langle A_2X_BY_C\rangle-\langle A_2Y_BX_C\rangle)\ge0,\\
\\
	\quad G_2=&1-\alpha(\langle A_3B_3\rangle+\langle A_3Z\rangle+\langle B_3Z\rangle)-\beta(\langle A_1B_1X\rangle-\langle A_1B_2Y\rangle-\langle A_2B_1Y\rangle-\langle A_2B_2X\rangle)\ge0,\\
\\
			\quad W_{1}=& 1+w_\alpha(\langle Z_B\rangle+\langle Z_C\rangle)-w_\beta\langle Z_BZ_C\rangle  -w_\gamma(\langle X_BX_C\rangle+\langle Y_BY_C\rangle+\langle A_3X_BX_C\rangle+\langle A_3Y_BY_C\rangle) \\
			&+w_{\delta}(\langle A_{3}\rangle+\langle A_{3}Z_{B}Z_{C}\rangle)+w_{\epsilon}(\langle A_{3}Z_{B}\rangle+\langle A_{3}Z_{C}\rangle) -w_\phi(\langle A_1X_B\rangle+\langle A_1X_C\rangle+\langle A_2Y_B\rangle+\langle A_2Y_C\rangle  \\
			&+\langle A_1X_BZ_C\rangle+\langle A_1Z_BX_C\rangle)+\langle A_2Y_BZ_C\rangle+\langle A_2Z_BY_C\rangle) \ge0,\\
\\
		\quad W_{2}=& 1+w_{\kappa}(\langle A_{3}\rangle+\langle B_{3}\rangle)+w_{\lambda}\langle Z\rangle-w_{\eta}(\langle A_{1}X\rangle   +\langle A_{2}Y\rangle+\langle B_{1}X\rangle+\langle B_{2}Y\rangle)+w_{\mu}(\langle A_{3}Z\rangle+\langle B_{3}Z\rangle) \\
		&-w_{\nu}(\langle A_{1}B_{1}\rangle+\langle A_{2}B_{2}\rangle)+w_{\omega}\langle A_{3}B_{3}\rangle-w_\pi(\langle A_1B_1Z\rangle+\langle A_2B_2Z\rangle)+w_\theta\langle A_3B_3Z\rangle  \\
		&-w_\xi(\langle A_1B_3X\rangle+\langle A_2B_3Y\rangle+\langle A_3B_1X\rangle+\langle A_3B_2Y\rangle)\ge0,
	\end{aligned}
 \right.
\end{equation}
\end{widetext}
where the two inequalities $G_1$ and $W_1$ are used to determine EPR steering from Alice to Bob-Charlie, while $G_2$ and $W_2$ are used to determine EPR steering from Alice-Bob to Charlie.
Obviously, the genuine tripartite EPR steering sharing can be investigated in two trilateral sequential scenarios based on these four inequalities.

In the first scenario, where multiple Alices (Alice$_1$, Alice$_2$,..., Alice$_n$) perform sequential measurements, exploring the amount of Alices capable of effectively steering Bob-Charlie, as well as the amount of Alices capable of effectively steering Charlie when cooperated with a single Bob, are interesting. Without loss of generality, supposed that Alice$_n$'s weak measurement is $\hat{x}^n_i$, while the Bob and Charlie's sharp measurement are $\hat{y}_j$ and $\hat{z}_k$ respectively, where $i,j,k\in\{0,1,2\}$, which can be expressed as,
\begin{align}
&\hat{y}_j =\sin\theta^y_j\cos\phi^y_j\hat{X}+\sin\theta^y_j\sin\phi^y_j\hat{Y}+\cos\theta^y_j\hat{Z},\nonumber\\
&\hat{z}_k =\sin\theta^z_k\cos\phi^z_k\hat{X}+\sin\theta^z_k\sin\phi^z_k\hat{Y}+\cos\theta^z_k\hat{Z},\nonumber\\
&\hat{x}^{n}_i =\sin\theta^{x_n}_i\cos\phi^{x_n}_i\hat{X}+\sin\theta^{x_n}_i\sin\phi^z_k\hat{Y}+\cos\theta^{x_n}_i\hat{Z}.\nonumber
\end{align}
When the shared state is the GHZ state, either in Alice$_n$ genuine tripartite steer Bob-Charlie(1→2) or in Alice$_n$-Bob genuine tripartite steer Charlie(2→1), under the optimal measurement setting for all observers, the maximum amount of Alices is 3 by analyzing the multiple violations of $G_1$ and $G_2$.
Especially, the allowable range of sharpness parameters is relatively large under 1→2 EPR steering case.
When the shared state is the $W$ state, the maximum amount of Alices is 2 in both two EPR steering cases according to the existence of the multiple violations of $W_1$ and $W_2$. Similarly, the allowable range of sharpness parameters is relatively large under 1→2 EPR steering case.

In the other scenario, where multiple Charlies (Charlie$_1$, Charlie$_2$, ..., Charlie$_n$) perform sequential measurements, they also explored the amount of Alices capable of effectively steering Bob-Charlie, as well as the amount of Alices capable of effectively steering Charlie when cooperating with a single Bob. Similar discussions with the above, when the shared state is GHZ state, it is shown that up to six sequential Charlies and a single Bob can be genuinely steered by a single Alice (1→2) by analyzing multiple violations of the $G_1$ inequalities. In addition, up to three Charlie can be genuinely steered by Alice-Bob (2→1). Compared to the 2→1 steering case, the amount of sequential observers and range of sharpness parameters in the 1→2 steering case are higher. when the shared state is W state, at most four Charlies and Bob can genuinely be steered by Alice (1→2) based on the multiple violations of $W_1$ inequalities. On the other hand, Alice-Bob can truly steer up to three Charlies. The amount of sequential observers and the sharpness range for the 1→2 steering case are also higher than those for the 2→1 steering case.

The result shows that the GHZ state is more powerful than the W state, allowing for more observers in the sequential steering scenario. 
Additionally, the 1→2 EPR steering case is more effective across a wider range of allowable sharpness parameters than that of the 2→1 EPR steering case. 
They also explain that these steering sharing phenomena can be verified in quantum secret sharing protocol, and determining the precise relationship between the security of such sequential quantum secret-sharing protocols and genuine tripartite EPR-steering sequential detection is worthy of further investigation. 

Subsequently, \citet{Chen.Yuyu.Chin.Phys.B.32.040309_2023} also investigated steering sharing in three-qubit systems.
\citet{Han.XinHong_arXiv.2303.05954_2023} has found that each pair of Alice$_1$-Bob$_1$ remotely steers the quantum state of Charlie simultaneously and independently.

\subsubsection{EPR Steering In The Two-sided Sequential Measurement Scenario}

The above discussion is the simplest one-sided sequential measurement scenario. In a more general case,  it is interesting to extend it to the bilateral scenario. As introduced in Sec. \ref{nonlocality_2qubit}, \citet{Zhu.Jie_PhysRevA.105.032211_2022} indicated that nonlocal sharing of 2-qubit systems in the bilateral scenario is impossible. While it is possible to observe the EPR steering sharing in the two-sided sequential measurement scenario.

In their work, they investigated whether Alice$_1$-Bob$_1$ and Alice$_2$-Bob$_2$ can observe EPR steering simultaneously in the simplest bilateral scenario with two Alices and two Bobs.

Assumed that the entangled pair in a singlet state, are assigned to multiple Alices and Bobs on both sides. The measurement operators of Alice$_i$ and Bob$_j$ are defined by  $\hat{x}_i$ and $\hat{y}_j$
with the binary outcomes $a_i$ and $b_j$ respectively. The linear steering inequality (\ref{linear_inequality}), $S_m$, was used as the criteria for verifying quantum steering, where $m$ is the amount of measurement settings for each observer. The average value $S_{A_1B_1}$ for Alice$_1$-Bob$_1$ is only determined by $G_{A_1B_1}$, and the average value $S_{A_2B_2}$ for Alice$_2$-Bob$_2$ is only determined by $F_{A_1B_1}$ when the measurement settings are determined. When the amount of measurement settings for each observers $m=3$,
where the measurement settings are chosen as $\{\sigma_x,\sigma_y,\sigma_z\}$, and the quality factor G is same for Alice$_1$ and Bob$_1$, the maximal average values can be given as $S_3^{A_1B_1}=G^2,S_3^{A_2B_2}=1-2\frac{G^2}{3}$. It is shown that the EPR steering sharing can be observed between Alice$ _1$-Bob$ _1$ and Alice$ _2$-Bob$ _2$ when $G\in(0.7598, 0.796)$.

In practical terms, obtaining a higher quantity of measurement settings (m) is essential to observe a greater number of instances of violation. Notably, in a unilateral sequential scenario where multiple Bobs aim to steer a single Alice, all Bobs must select the same measurement to steer Alice synchronously. Conversely, in a bilateral sequential scenario, each Bob endeavors to independently steer several Alices. Consequently, the measurement choices of different Bobs are independent of each other. It would be intriguing to explore the potential for multiple pairs of Alice$_i$-Bob$_j$ to simultaneously demonstrate EPR steering in a bilateral sequential scenario.

Similarly, the steering sharing in the two-sided sequential measurement scenario has also been investigated in  (\citealp{Kun.Liu_arXiv.2102.12166_2021};
\citealp{Han.Xinhong_Quantum.Inf.Process.s11128-021-03211-z_2021};
\citealp{Lijian_2022EPR_steering};
\citealp{Tong.Jun.Liu_OPT.OE.470229_2022};
\citealp{Han.Xin-Hong_PhysRevA.106.042416_2022};
\citealp{Qiao-Qiao.Lv_Phys.AMath.Theor.2307.09928_2023}).
Especially, 
\citet{Han.Xin-Hong_PhysRevA.106.042416_2022} showed that how to manipulate that steering direction among multiple Alices and Bobs through sequential unsharp measurement. It is indicated that the change of steering direction depends on the number of measurement settings and the sharpness of measurements. \citet{Tong.Jun.Liu_OPT.OE.470229_2022} demonstrated this steering scenario experimentally.

\subsubsection{Steering Sharing Using Nonlocal Measurements}
The investigation of quantum steering sharing among multiple observers is presently constrained to imprecise local measurements.

Recently, \citet{Han.XinHong_arXiv.2303.05954_2023} showcases that imprecise nonlocal product measurements can trigger steering sharing, surpassing the efficacy of local measurements. Furthermore, the optimal nonlocal measurements can be obtained through the use of steering ellipses. Experimental findings indicate that, in the sharing of GHZ states with two measurement settings, the activation of steering sharing is more pronounced when employing imprecise nonlocal product measurements compared to imprecise local measurements. Notably, the activation of steering sharing through nonlocal measurements, particularly those with unequal strength measurements, exceeds that observed with equal strength measurements.

As illustrated in Fig. \ref{genuine steering_1}, assumed that three-qubit state $\rho_\mathrm{{ABC}}$ initially shared between the spatially separated Alice$_1$, Bob$_1$, and Charlie. Alice$_1$ and Bob$_1$'s qubits will be sequentially handed down. It is worth exploring whether each pair of A$_i$ and B$_i$ can steer Charlie's quantum state simultaneously and independently. Different from the previous discussion, they relax the measurement restrictions, Alice$_1$ and Bob$_1$  can adopt nonlocal measurements. The k-th nonlocal measurement of A$_i$-B$_i$ is set to $\hat\Pi ^{(i)}_k$, which strength is labeled as $\lambda ^{(i)}_k$. And k-th local sharpness measurement of Charlie is represented as $\hat\Lambda _k$. They adopted n-setting Linear Steering inequality to verify steering.

For each group of observers, the post-measurement state shared between A$_i$-B$_i$ and Charlie can be expressed as,
\begin{align}
	\rho_{ABC}^{(i)}=\frac{1}{n}\sum_{k=1}^{n}\sum_{o_k=-1,1}[(K_{o_k}^{(i)}\otimes \mathbb{I}_C)\rho_{ABC}^{(i-1)}(K_{o_k}^{(i)\dagger}\otimes \mathbb{I}_C)],
\end{align}
where $K^{(i)}_{o_k}{K^{(i)}_{o_k}}^{\dagger}=\hat{\Pi}_{k}^{(i)}$, $o_k\in\{+1,-1\}$ is of $\hat{O}_i$.
The average value is given as $\langle\hat{\Pi}_{k}^{(i)}\otimes\hat{\Lambda}_{k}\rangle=\mathrm{Tr}[\hat{\Pi}_{k}^{(i)}\otimes\hat{\Lambda}_{k}\rho_{ABC}^{(i)}]$. Through the steering ellipsoids, the optimal nonlocal measurements can be obtained as $\hat\Pi ^{(i)}_1=\lambda_1^{(i)}(\sigma_y\otimes\sigma_y),
\hat\Lambda _1=\sigma_x $, $\hat\Pi ^{(i)}_2=\lambda_2^{(i)}(\sigma_y\otimes\sigma_x),
\hat\Lambda _x=\sigma_y$. So the average value of the corresponding linear steer inequality can be given as,
\begin{align}
	S_{2}^{(i)}\!=\frac{1}{2^{i}}\Big[\lambda_{2}^{(i)}\!\prod_{1\!\le\! j\le \! i-1\!}(\!1\!+\!F_{\lambda_{1}^{(j)}}\!)+\lambda_{1}^{(i)}\!\prod_{1\!\le\! j\le\! i-1\!}(\!1\!+\! F_{\lambda_{2}^{(j)}}\!)\Big]\!\le \! C_{2},
\end{align}
where $F_{\lambda_{1}^{(j)}}=\sqrt{1-(\lambda_{1}^{(j)})^{2}}$, $ F_{\lambda_{2}^{(j)}}=\sqrt{1-(\lambda_{2}^{(j)})^{2}}$, $ C_{2}=\frac{1}{\sqrt{2}}$. 
While if the measurements are constrained local, then the optimal measurement is, $\{\hat{M}_{1}^{A_{i}}=\eta_{1}^{(i)}\sigma_{y},\hat{M}_{1}^{B_{i}}=\gamma_{1}^{(i)}\sigma_{y},\hat{M}_{1}^{C}=\sigma_{x}\},\{\hat{M}_{2}^{A_{i}}=\eta_{2}^{(i)}\sigma_{y},\hat{M}_{2}^{B_{i}}=\gamma_{2}^{(i)}\sigma_{x},\hat{M}_{2}^{C}=\sigma_{y}\}$.
The average value of the corresponding linear steer inequality can be given as,
\begin{align}
	\widetilde{S}_{2}^{(i)}\!=\frac{1}{2^{i}}\!\Big[\lambda_{2}^{(i)}\!\prod_{\!1\!\le\! j\!\le\! i-1\!}(1\!+F_{\gamma_{1}^{(j)}}\!)+\!\lambda_{1}^{(i)}\!\prod_{1\!\le\! j\le\! i-1}(\!1\!+F_{\gamma_{2}^{(j)}}\!)\Big]\!\le\! C_{2},
\end{align}
where $F_{\gamma_{1}^{(j)}}~=~\sqrt{1-(\gamma_{1}^{(j)})^{2}}, F_{\gamma_{2}^{(j)}}~=~\sqrt{1-(\gamma_{2}^{(j)})^{2}}$.
The parameters of the weak measurement should satisfy
$\lambda_k^{(i)}=\eta_k^{(i)}*\gamma_k^{(i)},k\in\{1,2\}$.

Assumed that the strengths of two-setting measurements
used by each observer are equal,
which is defined as $\lambda^{(i)}=\lambda^{(i)}_1=\lambda^{(i)}_2$, $\eta^{(i)}=\eta^{(i)}_1=\eta^{(i)}_2$, $\gamma^{(i)}=\gamma^{(i)}_1=\gamma^{(i)}_2$. The criterion inequalities in both cases can be given as
$S_{2}^{(i)}~=~\frac{1}{2^{i-1}}\biggl[\lambda^{(i)}\prod_{1\leq j\leq i-1}(1+\sqrt{1-(\lambda^{(j)})^{2}})\biggr],
\widetilde S_{2}^{(i)}=\frac{1}{2^{i-1}}\bigg[\lambda^{(i)}\prod_{1\leq j\leq i-1}(1+\sqrt{1-(\gamma^{(j)})^{2}})\bigg]$.
Due to the requirement $\lambda^{(i)}=\eta_k^{(i)}*\gamma_k^{(i)}$, which means $\lambda^{(i)}<\gamma_k^{(i)}$, hence $S^{(i)}_2>\tilde{S}^{(i)}_2$.
This result indicated that using the nonlocal product measurement can activate steering sharing where it is unattainable through local measurement. Additionally, it can be observed that a maximum of two pairs of $A_i$ and $B_i$ can steer Charlie simultaneously. 
When the strength of two-setting measurements used by A$_i$ and B$_i$ are unequal,
$\lambda^{(i)}_1\ne\lambda^{(i)}_2$, $\eta^{(i)}_1\ne\eta^{(i)}_2,\gamma^{(i)}_1\ne\gamma^{(i)}_2$,
$\eta_k^{(i)}=\gamma_k^{(i)},k\in\{1,2\}=\sqrt{\lambda_k^{(i)}}$, 
the maximum number of A$_i$-B$_i$ pairs that can simultaneously steer Charlie does not increase.

In the case of nonlocal measurements and unequal strength of the two-settings measurements, the steering ellipsoid of $A_i$ and $B_i$, as well as Charlie's turning ellipsoid, vary with $\lambda_1^{(i)}$ and $\lambda_2^{(i)}$. The volume of Charlie's steering ellipsoids generated by the measurements of A$_i$ and B$_i$ can be given as $V_C^{(i)}=\frac{|\det(T-\widetilde{m}\vec{n}^\top)|/(1-|\widetilde{m}\vec{n}|^2)^2}{4\pi/3}$.
It is shown that the volume reduction rate of the steering ellipsoid is relatively slow compared to using measurements of equal intensity.
Since the volume of the steering ellipsoid characterizes the steering capacity, nonlocal measurements with unequal strength are relatively more favorable for activating quantum steering than with equal strength. \citet{Qiao-Qiao.Lv_Phys.AMath.Theor.2307.09928_2023} also investigated the sharing of EPR steering via weak measurement with unequal sharpness parameters in a generalized case, showing that an unbounded number of sequential Alice-Bob pairs can share the EPR steering.

\subsection{Experiments Of Steering Sharing}
\subsubsection{Detection Steering Sharing}

\begin{figure}[t!]
    \centering
\includegraphics[width=0.45\textwidth]{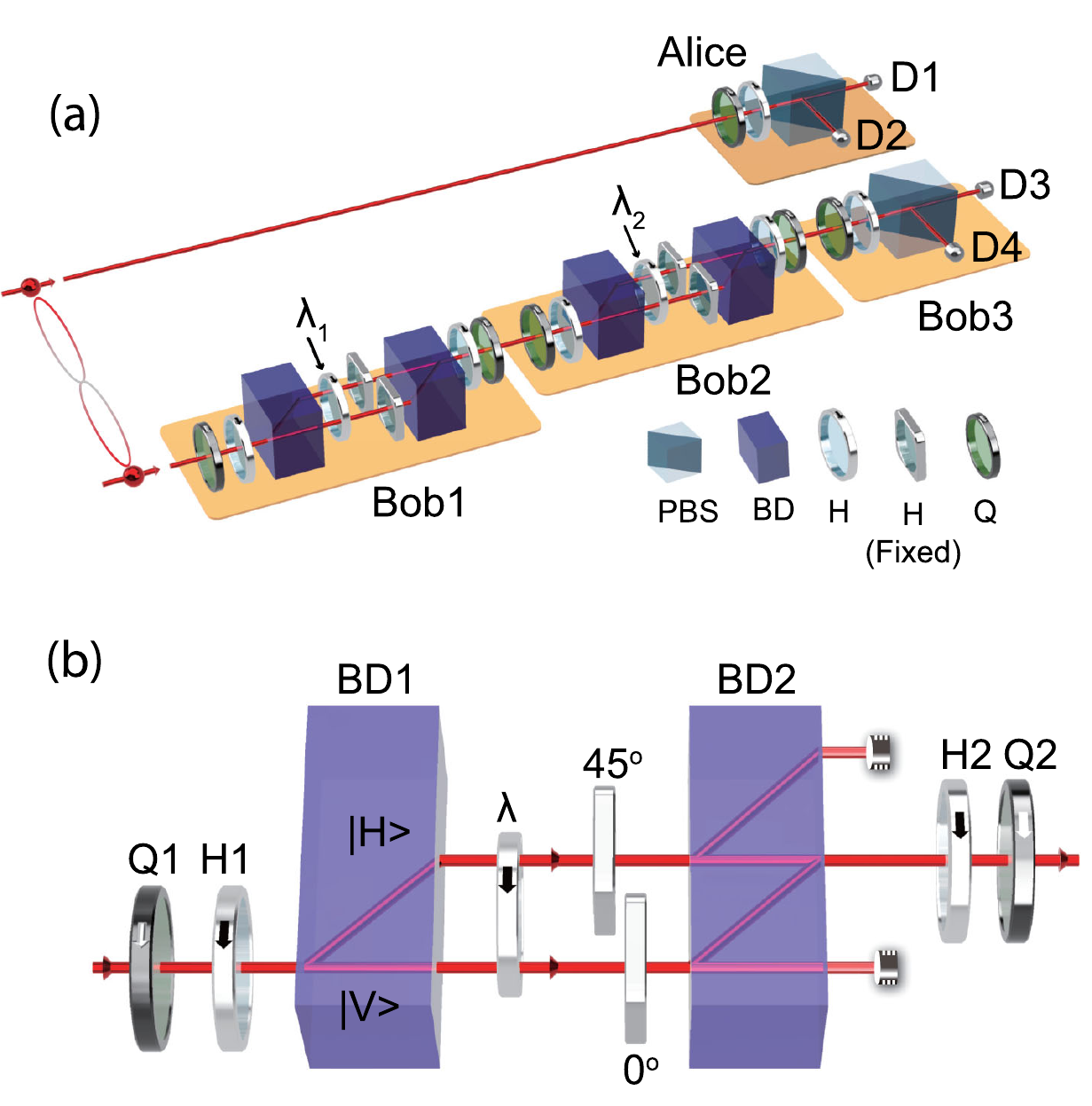}
\caption{\justifying{\small{Experimental setup and schematic of the weak measurement
with variable strength $\lambda$. This figure is adapted from \cite{Yeon.Ho.Choi._optica.394667_2020}.}}}
\label{choi_cricuit_figure}
\end{figure}

\begin{figure*}[t!]
    \centering
\includegraphics[width=0.7\textwidth]{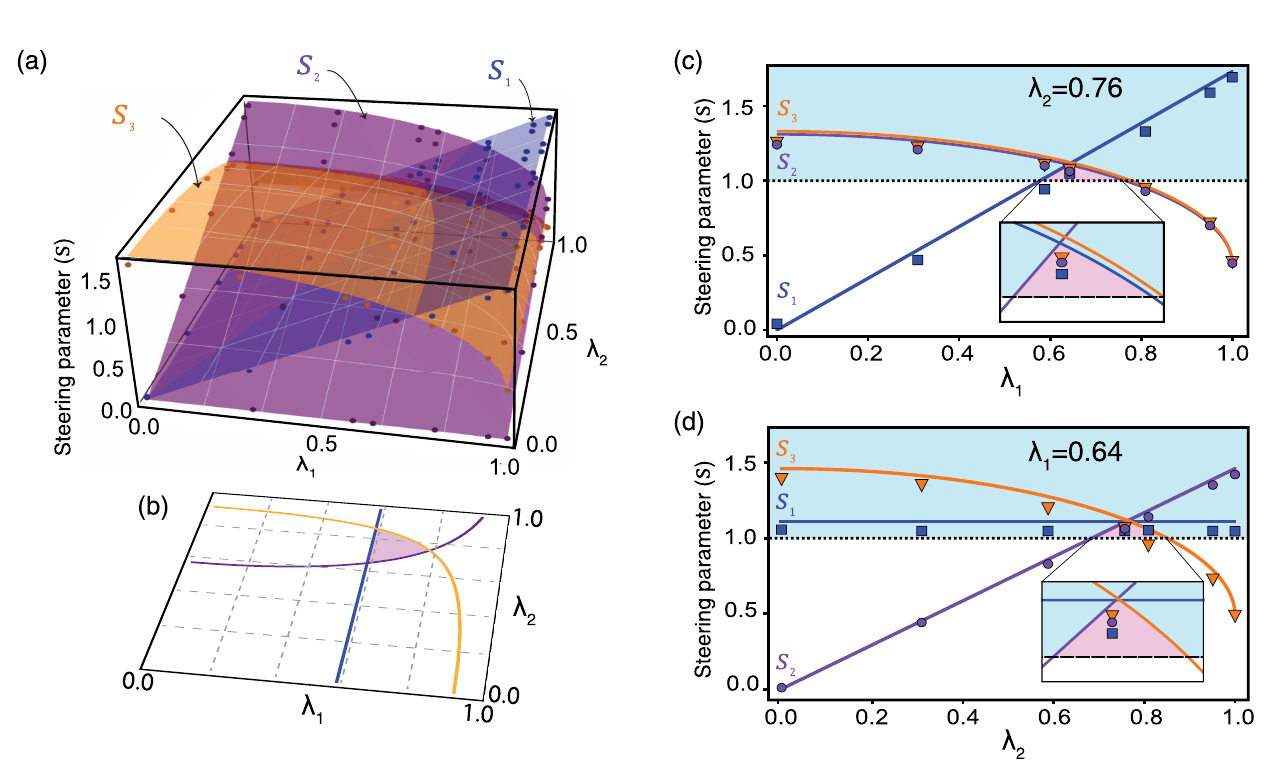}
\caption{\justifying{\small{Demonstrating the triple violations of the steering inequalities, indicating simultaneous quantum steering by three observers. This figure is adapted from \cite{Yeon.Ho.Choi._optica.394667_2020}.}}}
\label{choi_result_figure}
\end{figure*}

\citet{Yeon.Ho.Choi._optica.394667_2020} has experimentally showcased multi-observer quantum steering by leveraging sequential weak measurements, providing direct confirmation that quantum steering can be shared among more than four observers.

In their experiment (see Fig. \ref{choi_cricuit_figure}), considering a unilateral sequential scenario, the initial state is a singlet state $\left|\Psi^-\right\rangle\equiv\frac{1}{\sqrt{2}}(\left|HV\right\rangle-\left|VH\right\rangle)$. To achieve weak measurement, it is necessary to suppose an arbitrary pure state $\left|\psi\right\rangle_{in}=\alpha\left|\phi\right\rangle+\beta\left|\phi_\perp\right\rangle$ with $\left\langle\phi|\phi_\perp\right\rangle=0$ given as input. A quarter-wave plate and a half-wave plate rotate the polarization state in such a way that $\hat{R}_+\left|\psi\right\rangle_{in}\to\alpha\left|H\right\rangle+\beta\left|V\right\rangle$. The photon is then entered into the interferometer with polarizing beam displacers. Inside the interferometer, both the horizontal and vertical polarization components at each path evolve to $\sin2\theta_\lambda\left|H\right\rangle+\cos2\theta_\lambda\left|V\right\rangle$. Then, the unnormalized output state after polarizing beam displacers is given as $\alpha\cos2\theta_\lambda\left|H\right\rangle+\beta\sin2\theta_\lambda\left|V\right\rangle$. Finally, $\hat{R}_+^{-1}$ operation is applied by using a quarter-wave plate and a half-wave plate, giving the final (unnormalized) output state as $\left|\psi\right\rangle_{out}=\alpha\cos2\theta_\lambda\left|\phi\right\rangle +\beta\sin2\theta_\lambda\left|\phi_\perp\right\rangle$, which has the same effect as the evolution of the Kraus operator. The measurement strength $\lambda$ can be expressed as $\lambda\equiv2\cos^2(2\theta_\lambda)-1$, where $\theta_\lambda\in\left[0,\frac{\pi}{8}\right]$. The half-wave plate at $\theta_\lambda$ determines the measured strength. This completes the weak measurement process. The measurement strength of Bob$_1$ and Bob$_2$ are labeled as $\lambda1$ and $\lambda2$. Finally, Alice and Bob$_3$ choose their measurement directions with a combination of wave plates and a polarizing beam splitter and have two detectors $\mathrm{(D1, D2)}$ and $\mathrm{(D3, D4)}$. Four case of coincidence detection $\mathrm{(D1-D3),(D1-D4),(D2-D3),(D2-D4)}$ are observed. Each event corresponds to four possible outcomes for Alice and Bob$_3$. For a given measurement direction $\vec{\alpha}_k,\vec{\beta}_l,\vec{\gamma}_m,\vec{\delta}_n$. The measurement results are labeled as $\{a,b,c,d\}$. The joint probability is obtained as $p(a,b,c,d|\vec{\alpha}_k,\vec{\beta}_l,\vec{\gamma}_m,\vec{\delta}_n)$. The value of linear steering inequality corresponds to Alice-Bob$ _1$, Alice-Bob$ _2$, Alice-Bob$ _3$ can be given as $S_{n}={\sqrt{n}}S_n$.

To obtain the maximum value of $S_1$, the measurement direction can be setted as $\vec{\alpha}_1=\vec{\beta}_1=\vec{x}$, $\vec{\alpha}_2=\vec{\beta}_2=\vec{y}$, $\vec{\alpha}_3=\vec{\beta}_3=\vec{z}$. Similar for Bob$_2$ and Bob$_3$ as $\vec{\gamma}_1=\vec{\delta}_1=\vec{x}$, $\vec{\gamma}_2=\vec{\delta}_2=\vec{y}$, $\vec{\gamma}_3=\vec{\delta}_3=\vec{z}$.While $\vec{\gamma}_m$ and $\vec{\delta}_n$ have no limitation. Optimization is only possible when Bob$_2$ knows the value of $\lambda_1$ and Bob$_3$ knows $\lambda_1$ and $\lambda_2$. The maximum steering value is expected to be $\sqrt3$ in our three measurement cases, which can be obtained at $\lambda_1\to1$ for $S_1$, $\lambda_1\to0,\lambda_2\to1$ for $S_2$ and $\lambda_1=\lambda_2\to0,\lambda_3\to1$ for $S_3$. As the measurement strength $\lambda_1$ becomes weaker, the steering value $S_1$ decreases, while $S_2$ and $S_3$ both increase.
For a fixed value $\lambda_1$, there exists a trade-off between $S_2$ and $S_3$ depending on $\lambda_2$. 
When $\lambda_1=0.64$ and $\lambda_2=0.76$, the theoretical results for $\{S_1,S_2,S_3\}$ can be given as $\{1.113,1.107,1.124\}$ respectively, and the corresponding experimental results as $\{1.047,1.070,1.078\}$. 
These results clearly demonstrate the tripe violation of the linear steering inequality, with more than 40 standard deviations
(see Fig. \ref{choi_result_figure}).

The research demonstrated that quantum steering can be distributed among more than four observers using merely three measurement settings. Augmenting the number of measurement settings holds the potential to involve a greater number of parties concurrently sharing quantum steering. These findings offer insights into the fundamental properties of quantum correlation, paving the way for a comprehensive utilization of entangled systems across diverse applications.

\subsubsection{Witness Multi-observer Steering}


\begin{figure*}[t!]
    \centering
\includegraphics[width=0.7\textwidth]{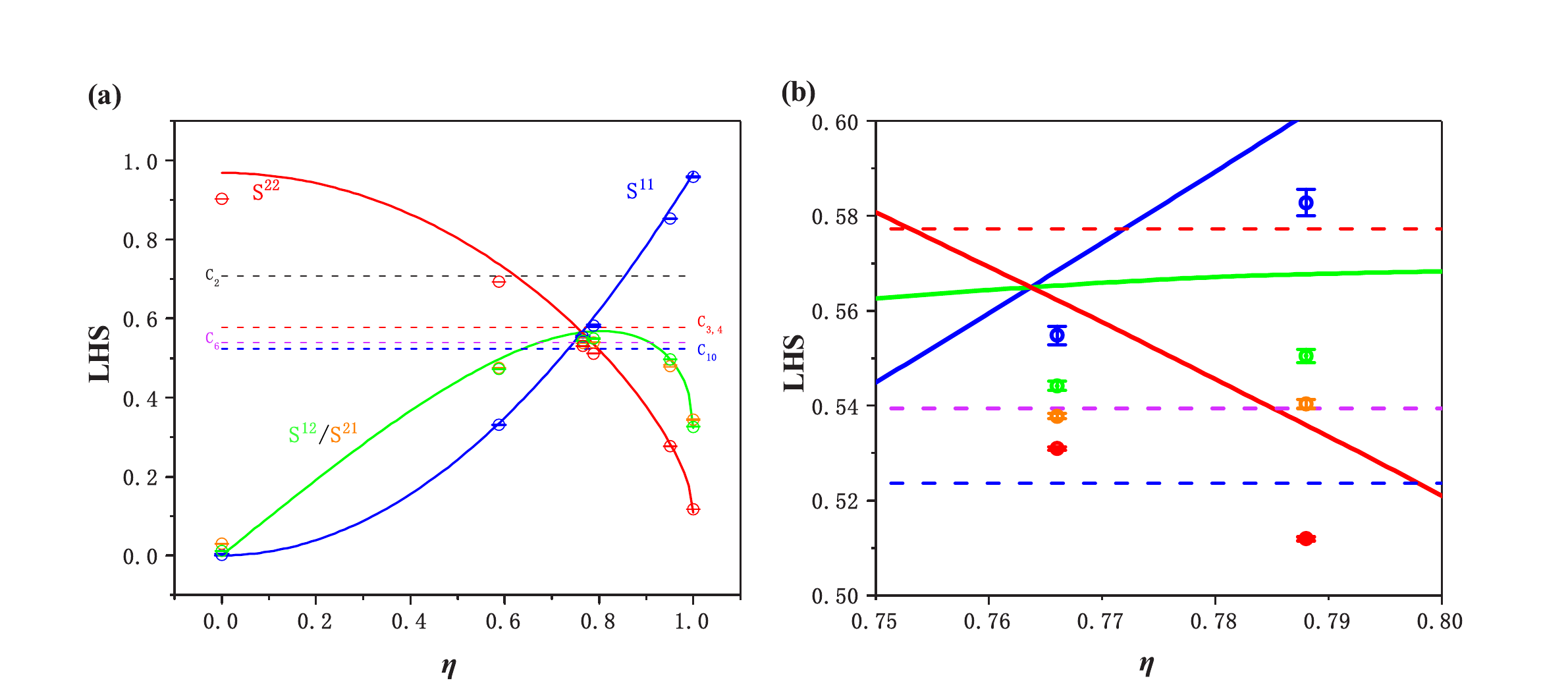}
\caption{\justifying{\small{The experimental results of quartic violation of the steering inequalities,  showcasing the steering on both sides simultaneously. This figure is adapted from \cite{Tong.Jun.Liu_OPT.OE.470229_2022}.}}}
\label{liutongjun_results_figure}
\end{figure*}

Subsequently, \citet{Tong.Jun.Liu_OPT.OE.470229_2022} conducted an experimental demonstration of the steering scenario. They observe quartic EPR-steerings of two-side sequential observers between each side of the entangled pair with a state fidelity of 97.6$\%$, compared to a maximally-entangled state. Considering a bilateral sequential scenario, the initial state is prepared in $\rho=|\psi_-\rangle \langle \psi_-|$, where $\psi=\frac{1}{\sqrt{2}}\left(\left|01\right\rangle-\left|10\right\rangle\right)$.

They experimentally verified four EPR steering, with Alice$_1$ and Bob$_1$ performing weak measurements and Alice$_2$ and Bob$_2$ performing projective measurements. If Alice$_1$ and Bob$_1$ use the same sharpness parameter $\eta=\eta_A=\eta_B$ with six independent measurements, it can be obtained that $S^{\left(12\right)}=S^{\left(21\right)}$ due to the symmetry between Alices and Bobs.

The steerabilities for the observer-pair (Alice$_1$, Bob$_1$) and (Alice$_2$, Bob$_2$) is a trade-off with $\eta$, and their steerabilities reach the peak around 0.56826 with $\eta\sim0.81$. Around $\eta\sim0.76$, it is found that four $S^{\left(ij\right)}$ can violate the linear steering inequality for the case of $n=6$. 
This means that each observer convinces others in the remote subsystem that their photons can be steered to the corresponding state without the need for any LHS model. Specifically, when $\eta=0.766,0.788$, the steering value $\{S^{\left(11\right)},S^{\left(12\right)},S^{\left(21\right)},S^{\left(22\right)}\}$ can be given as $\{0.55481,0.54416,0.5378,0.53096\}$ and $\{0.58282,0.55052,0.54034,0.51196\}$ respectively, whose error are generally below 0.00276. 
Although matched well with the theoretical curves, the LHS values when $\eta= 0.766$ succeed in violating the $C_{10}$ bound but partly fail to violate the $C_6$ bound because of the imperfections of the BD interferometers (see Fig. \ref{liutongjun_results_figure}).

The experimental results showed that four EPR steering: (Alice$ _1$,Bob$ _1$), (Alice$ _2$,Bob$ _2)$, (Alice$_1$,Bob$_2)$, and (Alice$_2$,Bob$_1$). In this scenario, the operator also used weak measurements efficiently to conserve quantum resources. The simultaneous occurrence of the four EPR steering events is crucial for testing the stability of the entire network system and eliminating the possibility of a fifth observer in the network.

\section{Sharing Of Other Kinds Of Quantum Correlations}\label{sharing_othercorrelation_section}

The series of studies on nonlocality sharing and steering sharing have also triggered similar research on other non-classical correlations, such as network nonlocality (\citealp{Hou.Wenlin_PhysRevA.105.042436_2022};
\citealp{Wang.Jian.Hui_PhysRevA.106.052412_2022};
\citealp{Pritam_2022_PhysRevA.106.052413};
\citealp{Mahato.Shyam.Sundar._PhysRevA.106.042218_2022};
\citealp{Zhang.Tinggui_FrontPhys.s11467.022.1242.6_2023};
\citealp{Kumar.Rahul_Quantum.Studies.s40509-023-00300-9_2023};
\citealp{MaoYaLi._PhysRevResearch.5.013104_2023}), 
quantum entanglement 
(\citealp{Bera.Anindita_PhysRevA.98.062304_2018};
\citealp{Maity.Ananda.G_PhysRevA.101.042340_2020};
\citealp{Foletto.Giulio_PHYS.REV.APPL.13.044008_2020};
\citealp{Srivastava.Chirag_PhysRevA.103.032408_2021};
\citealp{Pandit.Mahasweta_PhysRevA.106.032419_2022};
\citealp{Srivastava.Chirag_PhysRevA.105.062413_2022};
\citealp{Arun.Kumar.Das_Quantum.Inf.Process.s11128-022-03728-x_2022};
\citealp{Srivastava_2022sequential};
\citealp{HuMingLiang_PhysRevA.108.012423};
\citealp{limaosheng_2023arXiv_sequentially}), 
quantum contextuality (\citealp{Kumari.Asmita_PhysRevA.100.062130_2019};
\citealp{Anwer.Hammad_Quantum.q-2021-09-28-551_2021};
\citealp{Chaturvedi.Anubhav_Quantum.q-2021-06-29-484_2021};
\citealp{Kumari.A_PhysRevA.107.012615_2023}), among others. These results provide new insights into the problem of reusing relevant quantum physical resources.

\subsection{The Sharing Of Network Nonlocality}
\subsubsection{The Extended Bilocal Scenario}

Standard Bell nonlocality inspired a type of network nonlocality, which refers to a concept of nonlocality observed in the well-known entanglement-swapping scenario. A series of works have found that network nonlocality has stronger sharing properties compared to Bell nonlocality. For example, in bilateral scenarios, \citeauthor{Zhu.Jie_PhysRevA.105.032211_2022} found that Bell nonlocality sharing does not exist in the case of two-sided sequential measurements. Subsequently, \citet{Hou.Wenlin_PhysRevA.105.042436_2022} further investigated whether network nonlocality can be shared through weak measurements, and discovered the existence of network nonlocality through bilateral sequence measurements. This indicates the difference between Bell nonlocality sharing and 
network nonlocality sharing.

In the extended bilocal scenario 
Fig. \ref{network nonlocality},
there are two sources, $S_1$ and $S_2$, each of which generates a 2-qubit state. One source is shared between Alice$_1$ and Bob, while the other is shared between Charlie$_1$ and Bob. Alice$_1$ and Alice$_2$, as well as Charlie$_1$ and Charlie$_2$ on the other side, will measure their subsystems sequentially. Supposed that the two shared states emitted by the two sources are defined as $\rho_{AB}$ and $\rho_{BC}$ respectively, $\rho_{AB}=\rho_{BC}={|\psi\rangle}{\langle\psi| }$, and $|\psi \rangle=\frac{1}{\sqrt{2}}(| 00\rangle +| 11\rangle)$. The
whole initial state of the system can be described by $\rho_{ABC}=\rho_{AB}\otimes\rho_{BC}$.

Bob performs Bell state measurements on his received qubits, with four possible outputs $b=b_0b_1=\left\{ 00,01,10,11  \right\}$, which corresponding four bell state, $\left\{ \left|\phi_+\right\rangle ,\left|\phi_-\right\rangle,\left|\psi_+\right\rangle,\left|\psi_-\right\rangle,\right\}$,
where $\left|\phi^{\pm}\right\rangle=\frac{1}{\sqrt{2}}(\left|00\right\rangle \pm \left|11\right\rangle)$,
$\left|\psi^{\pm}\right\rangle=\frac{1}{\sqrt{2}}(\left|01\right\rangle \pm \left|10\right\rangle)$. Alice$_n$ chooses two different dichotomic observables independently, which are defined as $\hat{A}_{n,i}=\cos\theta_{n,i}\sigma_z-(-1)^i\sin\theta_{n,i}\sigma_x$, where $\hat{A}_{n,i}$ represents the (i)th measurement choices of Alice$ _n$.
Similarly to $\hat{C}_{t,j}$ for Charlie$_t$, and $\hat{C}_{t,j}=\cos\theta_{t,j}\sigma_z+(-1)^j\sin\theta_{t,j}\sigma_x$. After completing the measurements of all observers, the whole joint probability distribution $P(a_1,a_2,b,c_1,c_2|\hat{X}_1,\hat{X}_2,\hat{Y},\hat{Z}_1,\hat{Z}_2)$ can be fully obtained. By marginalizing the joint probability distribution, the marginal probability $P(a_n,b,c_m|\hat{X}_n,\hat{Y},\hat{Z}_t)$ of any combination of Alice$ _n$-Bob-Charlie$ _t$ can be obtained. As is known that the network nonlocality can be determined by the violation of BRGP inequality \cite{Branciard_2012_PhysRevA.85.032119}. The network nonlocality sharing can be judged when the BRGP inequalities corresponding arbitrary combination of Alice$_n$-Bob-Charlie$_t$ can be violated.

Similar to Bell nonlocality sharing, network nonlocality sharing can be also divided into two categories: passive and active network nonlocality sharing. In these two different types of network nonlocality sharing, Alice$_1$ and Charlie$_1$ will choose different unsharp measurements based on different motivations. For passive network nonlocality sharing, the maximal BRGP value can obtained under the optimal measurement settings, where $B_{11}=\sqrt{2G_1G_2},
B_{12}=\sqrt{(1+F_2)G_1},
B_{21}=\sqrt{(1+F_1)G_2},
B_{22}=\sqrt{\frac{(1+F_1)(1+F_2)}{2}}$. 
It is evident that under the optimal pointer of weak measurement, the four BRGP values can exceed 1 simultaneously within $G_1=G_2=G\in(\frac{1}{\sqrt2},\sqrt{2(\sqrt2-1)})$, indicating the passive network nonlocality sharing.
For active network nonlocality sharing, Hou, et.al give a suboptimal solution when $G_1=G_2=G$.
When $G\in [0,0.8]$, the maximal BRGP values can be given as, $B_{11}=\sqrt{2}G, B_{22}=\frac{(1+F)}{\sqrt{2}}$. When $G\in (0.8.1]$, the suboptimal values are $B_{11}=\frac{(F+\sqrt2-F(2+F))G}{\sqrt{2-2F}}, B_{22}=\sqrt{1+F^3+\frac{1}{2}F^4}\ge  1$. $B_{11}$ and $B_{22}$ can exceed 1 in the range of $G\in(\frac{1}{\sqrt2},1)$ under the optimal pointer of weak measurements.

The comparative analysis reveals that active network nonlocality sharing exhibits a wider spectrum of dual violations compared to passive network nonlocality sharing. These theoretical findings offer fresh perspectives for the exploration of network nonlocality.

\begin{figure}[t!]
	\centering
	\includegraphics[width=0.45\textwidth]{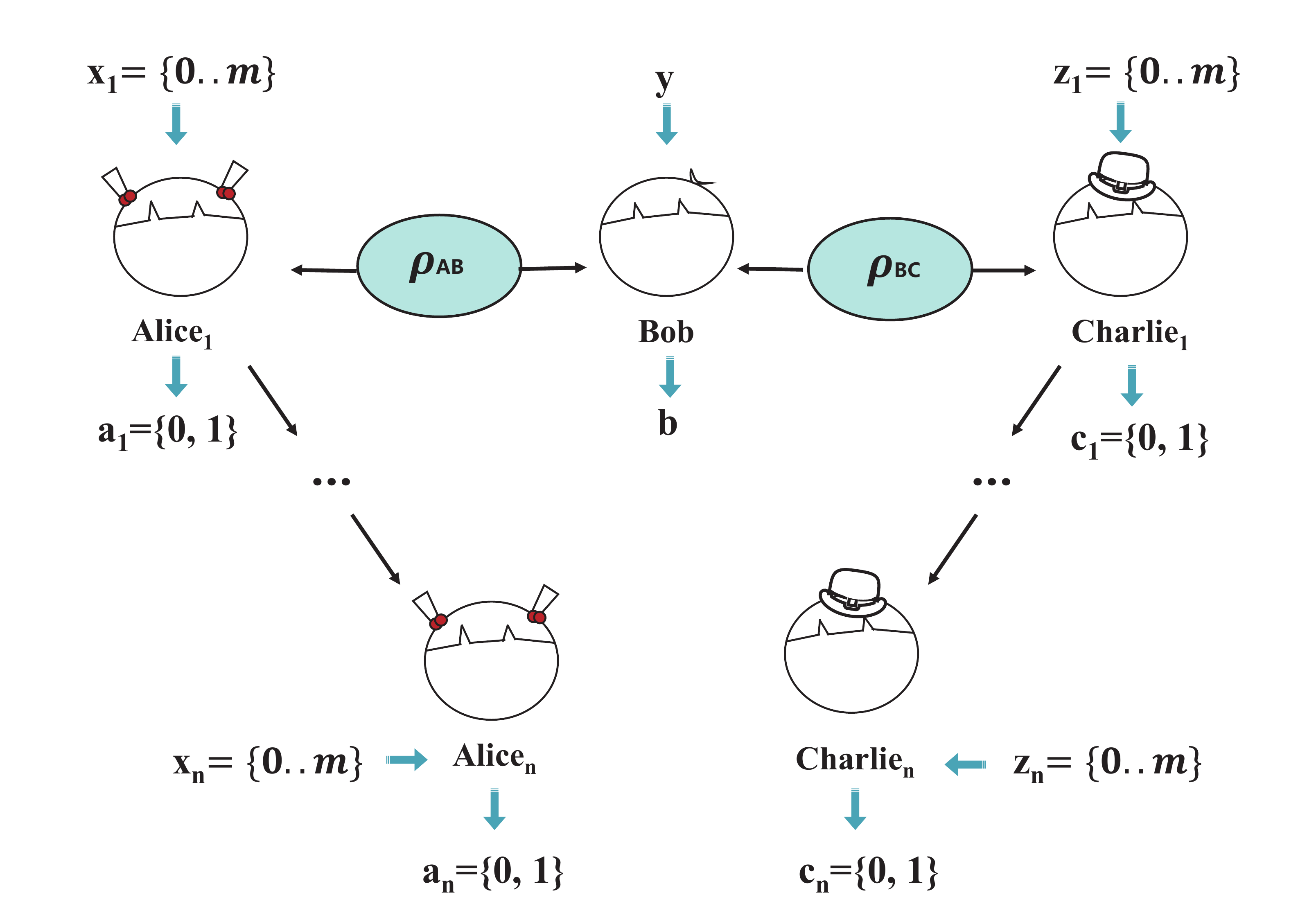}
	\caption{\justifying{\small{The extended bilocal scenario}. }}\label{network nonlocality}
\end{figure}

Subsequently, \citet{Mahato.Shyam.Sundar._PhysRevA.106.042218_2022} demonstrated that, in the symmetric case when the sharing is considered for both the edge parties, the nonlocality can be shared by, at most, two sequential observers per edge party.
\citet{Arun.Kumar.Das_Quantum.Inf.Process.s11128-022-03728-x_2022} has explored the resource-theoretic efficacy of the single copy of a 2-qubit entangled state in a sequential network.
\citet{Pritam_2022_PhysRevA.106.052413} has considered an alternative way of sharing network nonlocality when two observers initially share a maximally entangled, a nonmaximally entangled and a noisy entangled state. 
\citet{caizinuo_arXiv} has explored the recycling of full network nonlocality as quantum resources.

\subsubsection{The Extended Star Network Scenario} 

Shortly after the proposal of network nonlocality sharing \cite{Hou.Wenlin_PhysRevA.105.042436_2022}, \citet{Wang.Jian.Hui_PhysRevA.106.052412_2022} investigated the recycling of network nonlocality in the generalized star network scenario with any number of unbiased binary inputs via weak measurements.
In this scenario, $n$ independent sources $s_1,s_2...s_n$ distribute a 2-qubit state to Alice$ _{11}$... Alice$ _{n1}$ and Bob respectively. The corresponding
measurement is set as $\hat{A}_{ij^{i}}^{x^{ij^{(i)}}}$( the $j$th observer in branch $i$, $i\in\left\{1...n \right\}$, $j\in\left\{1...m \right\}$) and $\hat{B}^l$, and $\hat{A}_{ij^{i}}^{0}=\hat{A}_{ij^{i}}^{k}$. If the n-partite distribution can not be written as, 
\begin{align}
	P&(a_{1j^{(1)}},\ldots,a_{nj^{(n)}},b\mid x_{1j^{(1)}},\ldots,x_{nj^{(n)}},y)=
    \nonumber\\&\int\left[\prod_{i=1}^nd\lambda_iq_i(\lambda_i)p(a_{ij^{(i)}}\mid x_{ij^{(0)}},\lambda_i)\right]\times p(b|y,\lambda_1,\ldots,\lambda_n),
\end{align}
then the network n-nonlocality exists, where $q_i(\lambda_{i})$ is the distribution of the hidden variable $\lambda_{i}$ and the n sets of distributions of hidden variable $\lambda_{1}...\lambda_{n}$
related to the source, $\lambda=\lambda_{1}..\lambda_{n}$.
The local response function of Alice$_{ij}$ only depends on $j^{(i)}$ corresponding $\lambda_{i}$, and for Bob depends on $\lambda$.
The star network n-local scenario acknowledges the following nonlinear chain n-locality Bell inequality,
\begin{align}
	S_{j}^{(n,m,k)} =\sum_{l=1}^k|\mathcal{I}_{lj}|^{\frac{1}{n}}\leqslant k-1,
\end{align}
where 
\begin{align}
\mathcal{I}_{lj}=\frac1{2^n}\sum_{x_{1j^{(1)}},...,x_{nj^{(n)}}=l-1}^{l}\langle \hat{A}_{1j^{(1)}}^{x_{1j^{(1)}}}\ldots \hat{A}_{nj^{(n)}}^{x_{nj^{(n)}}}\hat{B}^{l-1}\rangle.
\end{align}
This inequality is a generalization of the original star network n-locality inequality, the bilocal inequality, and the chain CHSH inequality. And its upper bound only depends on the number of inputs k.

Suppose that a singlet state $\Psi=\frac{1}{\sqrt{2}}\left(\left|01\right\rangle-\left|10\right\rangle\right) $ is shared in each branch. The measurement set for Alice$_{ij}$ is $\hat{A}_{ij^{(i)}}^{x_{ij^{(i)}}}=\sin\biggl(\frac{x_{ij^{(i)}}\pi}k\biggr)\sigma_{x}+\cos\biggl(\frac{x_{ij^{(i)}}\pi}k\biggr)\sigma_{z}$. The quantum upper bound of the network inequality with any combination of observers can be given as follows, 
\begin{align}
	S_j^{(n,m,k)}=C_k\left(\prod_{i=1}^n\prod_{o=1}^{j^{(i)}}T_{io}\right)^{\frac1n},
\end{align} 
where 
\begin{align}
	T_{io}&=\begin{cases}1,&o=j^{(i)}=m,\\G_{io},&o=j^{(i)}<m,\\\frac{1+F_{io}}2,&o<j^{(i)},\end{cases}\nonumber
\end{align}
which depends only on the number of inputs k and the weak measurement parameters of the intermediate Alice involved. 
In the generalized case (n, m, k), $S_{1...1}=C_k(G_{11}...G_{n1})^{\frac{1}{n}},
S_{m...m}=C_k(\frac{1+F_{11}}{2}...\frac{1+F_{n1}}{2}...\frac{1+F_{1(m-1)}}{2}|...\frac{1+F_{n(m-1)}}{2})^{\frac{1}{n}}$.
When $G_{ij^{(i)}}$ and $F_{ij^{(i)}}$ are symmetrical, then $S_{1...1}$ and $S_{m...m}$ can simplified as $C_kG_{1}$ and $C_k(\frac{1+F_{1}}{2}...\frac{1+F_{m-1}}{2})$.

In the (n,2,k) case, choosing G$_1$=G, F$_1$=F, the quantum upper bound under the optimal pointer of weak measurement is $S_j=C_k[G^{n_1(\frac{1+F}{2})^{n_2}}]^{\frac{1}{n}}$. Here $n_1$ refers to the first observer in j, while $n_2$ refers to the second, $n_1+n_2=n$. A total of $2^n$ inequalities can be analyzed—in particular, the violation interval of G is only determined by $S_{1...1}=C_kG$ and $S_{2...2}=C_k\frac{1+F}{2}$.
When k=2, these $2^n$ inequalities can simultaneously exceed the classical bounds in the range of $G\in\left\{\frac{1}{\sqrt{2}},\sqrt{2(\sqrt{2}-1)} \right\}$.
When k=3, the quantum violation obtained can be achieved in a relatively narrow range that $G\in\left\{\frac{4}{3\sqrt{3}},\frac{4}{3}\sqrt{\frac{1}{3}(3\sqrt{3}-4)} \right\}$. 
When k $\ge3$, all inequalities in (n, 2, k) will not be violated simultaneously, as the maximum achievable value in this scenario is $\frac{4C_k}{5}$, which is lower than the classical bound (k-1) of k>3.
When m=3, it can be immediately concluded that there is no nonlocality sharing among all observers in the generalized star network.

Therefore, network nonlocality sharing exists only when m=2 and the input number is either k=2 or k=3, for any n branches where all $2^n$ star network inequalities concerning Alice$_{1j^1}$-Alice$_{nj^n}$-Bob can be violated simultaneously.

Subsequently, \citet{Zhang.Tinggui_FrontPhys.s11467.022.1242.6_2023} presented for the first time comprehensive results on the nonlocality sharing properties of quantum network states, for which the quantum nonlocality of the network quantum states has a stronger sharing ability than the Bell nonlocality.
\citet{MaoYaLi._PhysRevResearch.5.013104_2023} experimentally demonstrated nonlocality sharing in a photonic quantum three-branch star network.
\citet{Mahato.Shyam.Sundar._PhysRevA.106.042218_2022} investigated a two-input n-local scenario in the star-network scenario and demonstrated that the network nonlocality can be shared by an unbounded number of sequential observers across one edge party for a suitably large value of n in the asymmetric case.
Later, \citet{Kumar.Rahul_Quantum.Studies.s40509-023-00300-9_2023} further demonstrated the sharing of nonlocality in the star network scenario.

\subsection{The Sharing Of Entanglement}
Quantum entanglement stands as the pivotal resource that empowers both parties to conduct intriguing nonlocality protocols, conventionally recognized as an inseparable property. If the quantum state $\rho_{\mathrm{AB}}$ cannot be separated into product states $\rho_{AB}=\sum_\lambda p(\lambda)\rho^A_\lambda\otimes\rho^B_{\lambda}$, it means that the subsystem A and B are entangled. \citet{Srivastava.Chirag_PhysRevA.105.062413_2022} has proved that entanglement “nonclassicality” can be witnessed sequentially and independently by an arbitrarily large number of observers at one end of the shared state with the single observer at the other end. Furthermore,
\citet{Pandit.Mahasweta_PhysRevA.106.032419_2022} unveiled the sharing of entanglement correlation among 2-qubit quantum states through a sequential detection process conducted an arbitrary number of times. This study illustrates the distinction between the recycling of entanglement and Bell nonlocality.

Quantum entanglement can be verified by the entanglement witness. This method relies on the Hahn-Banach theorem, which guarantees that there is always a functional on a normed linear space that ``separates'' any element that falls outside a closed convex set from the set itself. The entanglement witness operator is labeled as W, and the expected value of the Hermitian operator $W$ concerning the state $\rho$ is denoted by $\left\langle \mathrm{W} \right\rangle_{\rho}$. Notably, all separable states satisfy $\left\langle \mathrm{W} \right\rangle_{\rho_{o}} \ge0$, whereas there exists at least one entangled state for which$\left\langle \mathrm{W} \right\rangle_{\rho_{e}}<0$.

The entanglement scenario is a generalized version of the sequential scenario proposed in Fig. \ref{bell nonlocality sharing}. Suppose that a bipartite entangled state is shared between a pair of observers, and the ultimate goal is to maximize the number of pairs that can independently witness entanglement. Aussmed that the k-th Alice-Bob ($AB_k$) pair shares the state $\rho_{AB_k}$, the witness operator used by this pair is as follows, $\mathrm{W}_{k}=\frac{1}{4}[\mathbb{I}_4+\sigma_z\otimes\sigma_x-\lambda_k\sigma_x\otimes\lambda_k\sigma_x-\lambda_k\sigma_y\otimes\lambda_k\sigma_y]$, where $\lambda_k$ is the sharpness parameter.
They choose that each observer has three measurement settings, $\sigma=\{ \sigma_x,\sigma_y,\sigma_z \}$, which are labelled as $\{\hat{A}^{\lambda_k^i}_0, \hat{A}^{\lambda_k^i}_1 \}$ for Alices, and $\{\hat{B}^{\lambda_k^i}_0, \hat{B}^{\lambda_k^i}_1 \}$ for Bobs. 
$\lambda_k^i$ is the i-th weak measurement parameter in the k-th observer, $i\in\{1,2,3\}$, and  $\lambda_k^{(3)}=1,\lambda_k^{(1)}=\lambda_k^{(2)}=\lambda_k$. 
If the measurement choices for all observers are unbiased, the post-measurement state of the k-th observer can be given as,
\begin{widetext}
	\begin{equation}
		\begin{aligned}
			\rho_{AB_{k}}& =\frac{1}{9}\sum_{i,j=1}^{3}\sum_{a,b=0}^{1}\sqrt{\hat{A}_{a}^{\lambda_{k-1}^{(i)}}}\otimes\sqrt{\hat{B}_{b}^{\lambda_{k-1}^{(j)}}}\rho_{AB_{k-1}}\sqrt{\hat{A}_{a}^{\lambda_{k-1}^{(i)}}}\otimes\sqrt{\hat{B}_{b}^{\lambda_{k-1}^{(j)}}}  \\&=\frac{1}{36}\sum_{i,j=1}^3\bigl[\bigl(1+\Lambda_{k-1}^i\bigr)\bigl(1+\Lambda_{k-1}^j\bigr)\rho_{AB_{k-1}}+\bigl(1+\Lambda_{k-1}^i\bigr)\bigl(1-\Lambda_{k-1}^j\bigr)\Bbb I_2\otimes\sigma_j\rho_{AB_{k-1}}\Bbb I_2\otimes\sigma_j \\&+(1-\Lambda_{k-1}^i)(1+\Lambda_{k-1}^j)\sigma_i\otimes\mathbb{I}_2\rho_{AB_{k-1}}\sigma_i\otimes\mathbb{I}_2+(1-\Lambda_{k-1}^i)(1-\Lambda_{k-1}^j)\sigma_i\otimes\sigma_j\rho_{AB_{k-1}}\sigma_i\otimes\sigma_j],
		\end{aligned}\nonumber
	\end{equation}
\end{widetext}
where $\Lambda ^i_k=\sqrt{1-\lambda^{(i)2}_k}$.
The expected values of each term present in the witness operator $W-k$ with respect to $\rho_{AB_{k}}$ can be expressed by their expected value with respect to the state $\rho_{AB_1}$,
\begin{equation}
	\begin{gathered}
		\mathrm{Tr}[\sigma_{z}\otimes\sigma_{z}\rho_{AB_{k}}]=\mathrm{Tr}[\sigma_{z}\otimes\sigma_{z}\rho_{AB_{1}}]\Pi_{l=1}^{k-1}\frac{(1+2\Lambda_{l})^{2}}{9}, \\
		\mathrm{Tr}[\sigma_{x}\otimes\sigma_{x}\rho_{AB_{k}}]=\mathrm{Tr}[\sigma_{x}\otimes\sigma_{x}\rho_{AB_{1}}]\Pi_{l=1}^{k-1}\frac{(1+\Lambda_{l})^{2}}{9}.	\nonumber 
	\end{gathered}
\end{equation}
And note that $\mathrm{Tr}[\sigma_{x}\otimes\sigma_{x}\rho_{AB_{k}}]=\mathrm{Tr}[\sigma_{y}\otimes\sigma_{y}\rho_{AB_{k}}]$, $ k\in\mathbb{N}$.
 
\citet{Pandit.Mahasweta_PhysRevA.106.032419_2022} proved that an arbitrary number of such pairs of observers can witness the entanglement, which can result from the shared entanglement state, such as all pure entangled states, a class of CHSH Bell-nonlocal mixed entangled states, and a class of CHSH-Bell local entangled states. These results show a clear distinction in the sequential witnessing of paired observers between ``entanglement'' and ``Bell nonlocality''.

Some other related work includes, \citet{Bera.Anindita_PhysRevA.98.062304_2018} explored the sharing of bipartite entanglement in a sequential scenario. \cite{Maity.Ananda.G_PhysRevA.101.042340_2020} presented the possibility of sharing genuine tripartite entanglement by considering a scenario consisting of three spin$\frac{1}{2}$ particles shared between Alice, Bob, and multiple Charlies.
\citet{Srivastava.Chirag_PhysRevA.103.032408_2021} employed a measurement-device-independent entanglement witness to detect the sharing of entanglement with two sequential observers.
\citet{Srivastava_2022sequential} has investigated the sequential detection of genuinely multipartite entanglement of quantum systems composed of an arbitrary number of qubits.
\citet{HuMingLiang_PhysRevA.108.012423} has related entanglement sharing to the entropic uncertainty relation in a (d×d)-dimensional system via weak measurements with different pointers, and consider both the scenarios of one-sided sequential measurements in which the entangled pair is distributed to multiple Alices and one Bob and two-sided sequential measurements in which the entangled pair is distributed to multiple Alices and Bobs.
\citet{limaosheng_2023arXiv_sequentially} investigated measurement strategies in a scenario where multiple pairs of Alices and Bobs independently and sequentially observe entangled states, and maximize the number of observer pairs that can witness entanglement. \citet{Arun.Kumar.Das_Quantum.Inf.Process.s11128-022-03728-x_2022} investigated the quantitative advantages, within the resource-theoretic framework, of reusing a single instance of a 2-qubit entangled state for information processing. 

\subsection{The Sharing Of Nontrivial Preparation Contextuality}

Nontrivial preparation contextuality represents a correlation of lesser strength compared to nonlocality. This concept highlights that the outcome of a quantum measurement depends not only on the physical quantity being measured but also on the manner in which the measurement is conducted. In their 2019 study, \citet{Kumari.Asmita_PhysRevA.100.062130_2019} explored the sharing of nonlocality and nontrivial preparation contextuality by investigating the quantum violation of a family of Bell's inequalities. The study establishes that nontrivial preparation contextuality can be shared among observers in any sequence within unbiased scenarios.

Utilizing a communication game and Bell's local realist inequalities as a foundation, the application of parity-oblivious conditions results in equivalent representations within the ontological model. These conditions are then regarded as assumptions of nontrivial preparation noncontextuality. Consequently, the local realist inequalities transform into nontrivial, noncontextual inequalities. As a consequence, the local realist inequalities transform into nontrivial, noncontextual inequalities. Describing this scenario, Alice possesses an n-bit input $x^\delta$, with $\delta \in {1, 2, ..., 2n}$ randomly and evenly chosen from ${0,1}^n$. Bob can select any bit $y \in {1, 2, ..., n}$ and retrieve bits $x_y^\delta$ with a certain probability. The task's condition stipulates that Bob's output must be $b=x_y^\delta$, and the y-th bit of Alice's input string x is subject to a constraint that prevents any parity information about x from being transmitted to Bob.
When the above property is not satisfied, Alice's observables are limited by a nontrivial constraint,
$\sum^{2^{n-1}}_{i=1}(-1)^{s.x^i}\hat{A}_{n,i}=0$.
The total number of nontrivial constraints on Alice's observables is $\hat{C}_n=2^{n-1}-n$. 
For $y\in\{1,2,...n\}$, Bob performs a binary measurement $\hat{M}_{n,y}$ with the outputs b. The measurements are given as,
\begin{align}
	\hat{M}_{n,y}=\begin{cases}\hat{M}_{n,y}^i,&\text{when }b=x_y^i,\\\hat{M}_{n,y}^l,&\text{when }b=x_y^l,\end{cases}\nonumber
\end{align}
\begin{align}
	\hat{M}_{n,y}^{i(l)}=\begin{cases}\hat{P}_{B_{n,y}},\quad\text{when }x_y^{i(l)}=0.\\\mathbb{I}-\hat{P}_{B_{n,y}},\quad\text{when }x_y^{i(l)}=1.\end{cases}\nonumber
\end{align}

The family of Bell expressions is formulated as follows: $\mathcal{B}_n=\sum_{y=1}^n\sum_{i=1}^{2^{n-1}}(-1)^{x_y^i}\hat{A}_{n,i}\otimes \hat{B}_{n,y}$.
Utilizing the sum of square decomposition (SOS) on the maximum quantum values, it can be deduced that $\langle\mathcal{B}_n\rangle_Q\le2^{n-1}\sqrt{n}$.
Achieving the optimal quantum value requires measurements that satisfy  $\frac{2^{n-1}}{\sqrt{n}}\sum^{2^{n-1}}_{k=1}(-1)^{x^i_y}\hat{A}_{n,i}\otimes \mathbb{I}=\hat{B}_{n,y}\otimes \mathbb{I}$. This necessitates a maximum entangled state with a dimension of  $2^{\frac{n}{2}}$, represented as $|\phi\rangle_{AB}=\frac1{\sqrt{2^{\lfloor n/2\rfloor}}}\sum_{k=1}^{2^{\lfloor n/2\rfloor}}|k\rangle_A|k\rangle_B$. The local bound of $\mathcal{B}_n$ remains $(\mathcal{B})_{\mathrm{local}}\le n\begin{pmatrix}
	n-1 \\
	[\frac{n-1}{2}]
\end{pmatrix}$.
Considering that $2^{n-1}-n$ nontrivial constraints are satisfied, a set of nontrivial preparation noncontextual inequalities can be established as $(\mathcal{B}_n)_{\mathrm{pnc}}\le 2^{n-1}$.
Notably, the nontrivial preparation noncontextual bound is lower than the local bound for $n\ge2$.

They considered a scenario with a single Alice and $k$ Bobs. Alice performs a dichotomic measurement with $2^{n-1}$ choices, and each Bob has $n$ choices. The sharpness and biasedness parameters for the $j$th Bob are denoted as $\eta _{n,j}(0<\eta_{n,j}\le1)$ and $\alpha _{n,j}(0\le|\alpha _{n,j}|\le1)$, subject to the conditions $|\alpha _{n,j}|+\eta _{n,j}\le 1$, $j=1,...,k$. Here, the sharpness and biasedness parameters for all measurements of Bobs are assumed to be the same. The Bell expression for the $k$th Bob can be expressed as,
$
\left(\mathcal{B}_{n}^{k}\right)_{\mathcal{Q}} = 2^{n-1}\sqrt{n}\eta_{n,k}\prod_{j=1}^{k-1}\gamma_{n,j},
$
where
$\gamma_{n,j}=[1+(n-1)\xi_{n,j}]/n, \quad \xi_{n,j}=[\sqrt{(1+\alpha_{n,j})^{2}-\eta_{n,j}^{2}}+\sqrt{(1-\alpha_{n,j})^{2}-\eta_{n,j}^{2}}]/2.$
If $(\mathcal{B}_n^k)_Q>(\mathcal{B}_n)_{\mathrm{local}}$ and $(\mathcal{B}_n^k)_Q>(\mathcal{B}_n)_{\mathrm{pnc}}$, then nonlocality and nontrivial prepared contextuality can be shared between Alice and the $k$th Bob. On this basis, a specific result can be derived, indicating that the sharpness parameters must satisfy $\eta_{n,k}>\frac1{\sqrt{n}\prod_{j=1}^{k-1}\gamma_{n,j}}$ for the $n$-bit case. Additionally, the sharpness parameter needs to satisfy $\eta_{n,1}\ge\frac{1}{\sqrt{n}}$ for the first Bob. For $n=2$, no nontrivial constraint exists on Alice's measurement and $(\mathcal{B}_2)_{\mathrm{local}}=(\mathcal{B}_n)_{\mathrm{pnc}}$. The discussion then commences from $n\ge3$.

In the context of one-parameter POVMs, the maximum number of Bobs sharing nontrivial preparation noncontextuality with Alice is 3 for n=3. Similarly, for n=4, this maximum number does not exceed 4. In the case of an n-bit scenario, if the k-th Bob shares nontrivial preparation noncontextuality with Alice, the unsharpness parameter must satisfy $\eta_{n,k}\geqslant\frac1{\sqrt{n}\gamma_{n,k-1}\prod_{j=1}^{k-2}\gamma_{n,j}}$. Substituting the critical value of the (k-1)-th Bob's unsharpness parameter into it yields $\eta_{n,k}\ge\frac{\eta_{n,k-1}}{\gamma_{n,k-1}}$. Through $\gamma_{n,k-1}>\sqrt{1-\eta_{n,k-1}^2}$, it follows that $\frac{\eta_{n,k-1}}{\sqrt{1-\eta_{n,k-1}^2}}>\frac{\eta_{n,k-1}}{\gamma_{n,k-1}}$. Therefore, it can be stipulated that $\eta_{n,k}\ge\frac{\eta_{n,k-1}}{\sqrt{1-\eta _{n,k-1}^2}}$. Ultimately, $\eta_{n,k}\ge\frac{1}{\sqrt{n-k+1}}$. For a given k, one can identify a condition $n\equiv n(k)\ge k-1+\frac{1}{\eta_{n,k}^2}$ under which the prepared contextuality can be shared by the k-th Bob. The final Bob (k-th) performs a sharp measurement with $\eta_{n,k}=1$, ensuring $n(k)\ge k$. Therefore, for any arbitrary number of Bobs (k), there exists an n(k) equal to or greater than k that can share nontrivial preparation contextuality.

For two-parameter positive operator-valued measures (POVMs) that satisfy $|\alpha _{n,j}|+\eta _{n,j}\le 1$, the value of $|\alpha _{n,j}|$ is determined by $\eta _{n,j}$ when $|\alpha _{n,j}|+\eta _{n,j}= 1$. Numerical evidence suggests that for $n=100$, up to 18 Bobs can share nontrivial preparation noncontextuality.

If $|\alpha _{n,j}|+\eta _{n,j}< 1$, increasing the lower value of $|\alpha _{n,j}|$ can enhance the maximum number of Bobs that share nontrivial preparation contextuality. Numeric tests have demonstrated that an arbitrary number of Bobs can share nontrivial preparation contextuality for a specific selection of two-parameter POVM.

It is intuitive to observe that one-parameter POVMs outperform two-parameter POVMs in this scenario. Additionally, it has been observed that nontrivial preparation contextuality, compared to nonlocality, allows for a larger number of Bobs to share.

\cite{Anwer.Hammad_Quantum.q-2021-09-28-551_2021} demonstrated that any number of independent observers can share the preparation of contextual outcome statistics enabled by state ensembles in quantum theory. \cite{Chaturvedi.Anubhav_Quantum.q-2021-06-29-484_2021} showcased the monogamy of preparation contextuality in a tripartite setting. \cite{Kumari.A_PhysRevA.107.012615_2023} study the sharing of nonlocality and preparation contextuality based on a bipartite Bell inequality, involving arbitrary $n$ measurements by one party and $2^{n-1}$ measurements by another party.

}

\section{Applications Based On Quantum Correlation Sharing}\label{application_section}

{\color{black}The sharing of quantum correlations is not only closely related to various important and interesting quantum information applications, but also touches upon the recycling of quantum resources \cite{Arun.Kumar.Das_Quantum.Inf.Process.s11128-022-03728-x_2022}. Here we will revisit several applications in typical tasks, such as quantum random access codes (\citealp{Li.Hong-Wei_Commun.Phys.s42005-018-0011-x_2018};
\citealp{Karthik.Mohan_NEW.J.PHYS.10.1088.1367-2630.ab3773_2019};
\citealp{Anwer.Hammad_PhysRevLett.125.080403_2020};
\citealp{Foletto.Giulio_PhysRevResearch.2.033205_2020};
\citealp{Shihui.Wei_NEW.J.PHYS.1367-2630/abf614_2021};
\citealp{Das.Debarshi_PhysRevA.104.L060602_2021};
\citealp{Xiao.Ya_PhysRevResearch.3.023081_2021};
\citealp{Xiao.Yao_Quantum.Inf.Process.s11128-023-03924-3_2023}), random number generation
(\citealp{Curchod.F.J_PhysRevA.95.020102_2017};
\citealp{BrianCoyle_EPTCS.273.2_2018};
\citealp{Xue-Bi.An_Opt.Lett.ol-43-14-3437_2018};
\citealp{Bowles_Quantum.q-2020-10-19-344_2020};
\citealp{Pan.A.K_PhysRevA.104.022212_2021};
\citealp{Foletto.Giulio_PhysRevA.103.062206_2021}), 
and self-test 
(\citealp{Karthik.Mohan_NEW.J.PHYS.10.1088.1367-2630.ab3773_2019};
\citealp{Miklin.Nikolai_PhysRevResearch.2.033014_2020};
\citealp{Armin.Tavakoli_sciadv.aaw6664_2020};
\citealp{Mukherjee.Sumit_PhysRevA.104.062214_2021};
\citealp{Pan.A.K_PhysRevA.104.022212_2021};
\citealp{Chirag_PhysRevA.103.032408_2021};
\citealp{Prabuddha.Roy_NEW.J.PHYS.10.1088.1367-2630.acb4b5_2023}).

\subsection{Quantum Random Access Codes}\label{V-A-QRAC}

Random Access Codes (RACs) are a typical and widely used communication protocol \cite{Wiesner.ACM.SIGACT.News.15.78_1983,
Ambainis2002DenseQC}. In the RAC model, the sender (Alice) prepares a set of random data, while the receiver (Bob) wishes to determine and decrypt Alice's data through communicating with her. In an $n\to m$ RAC, Alice holds $n$ number of bits, while Bob receives $m$ bits. Ideally, when Alice is permitted to send unlimited information, which is $n=m$, Bob can easily decrypt Alice's data. While, when $m<n$, Alice cannot encode all the data. Specifically,  taking $n\to 1$ RAC as an example, it can be achieved through a prepare-and-measure scenario, where Alice encodes a string of bits $\{x_0,x_1..x_n-1\}$ with $x_i\in\{0,1\}$ and transmits it to Bob, who randomly decodes one of the n bits, which defined as $x_y$, $y\in\{0,1,..,n-1\}$. Adopting appropriate strategies will enhance the applicability of RAC and be limited by the best classical strategies. Interestingly, it can be surpassed by quantum strategies \cite{Ambainis_2009arXiv_quantum,
Armin_Tavakoli_PhysRevLett.114.170502}, the probability of Bob attaining the mandatory information can be increased. The QRAC has been introduced and developed for qubit systems and even high-dimensional quantum systems. They serve as primitives for network coding \cite{Hayashi_2007_networkcoding}, random number generation \cite{lihongwei_2011_PhysRevA.84.034301}, and quantum key distribution \cite{Marcin_PhysRevA.84.010302_2011}, and are essential tools for numerous protocols in quantum information theory 
(\citealp{Wehner_PhysRevA.78.062112_2008};
\citealp{pawlowski_natureinformation_2009};
\citealp{Tavakoli_PhysRevA.93.032336_2016};
\citealp{Hameedi_PhysRevA.95.052345_2017};
\citealp{Tavakoli_PhysRevA.98.062307_2018};
\citealp{Farkas_PhysRevA.99.032316_2019};
\citealp{Armin.Tavakoli_sciadv.aaw6664_2020}
).

While the initial series of QRAC protocols are one-to-one encoding-decoding models. Due to the decoding operators for the QRAC protocol being projective measurements, the original coding state will be erased entirely, it is reasonable to only investigate one-to-one encoding-decoding models. Inspired by the work of the non-locality sharing via sequential measurements \cite{Silva.Ralph_PhysRevLett.114.250401_2015}, the constraint of projective measurements in the QRAC model can be discarded and explore more general sequential QRACs (
\citealp{Li.Hong-Wei_Commun.Phys.s42005-018-0011-x_2018};
\citealp{Karthik.Mohan_NEW.J.PHYS.10.1088.1367-2630.ab3773_2019};
\citealp{Anwer.Hammad_PhysRevLett.125.080403_2020};
\citealp{Foletto.Giulio_PhysRevResearch.2.033205_2020};
\citealp{Shihui.Wei_NEW.J.PHYS.1367-2630/abf614_2021};
\citealp{Das.Debarshi_PhysRevA.104.L060602_2021};
\citealp{Xiao.Ya_PhysRevResearch.3.023081_2021};
\citealp{Xiao.Yao_Quantum.Inf.Process.s11128-023-03924-3_2023}). \citet{Karthik.Mohan_NEW.J.PHYS.10.1088.1367-2630.ab3773_2019} explored QRAC in a three-party-prepare-transform-measure scenario, where a pair of the simplest QRAC ($2\to 1$) was achieved sequentially between Alice-Bob and Alice-Charlie. It is shown that a trade-off exists between these two QRACs.

In the three-party-prepare-transform-measure scenario \cite{Karthik.Mohan_NEW.J.PHYS.10.1088.1367-2630.ab3773_2019}, Alice prepared an arbitrary qubit state based on a random four-valued input  $x=(x_0,x_1)\in\{0,1 \}^2$. Bob will receive the states from Alice and measure it based on a random binary input $y\in\{0,1\}$.  Finally, Charlie will receive and measure the delivered state according to a random binary input $z\in\{0,1\}$. The joint probability distribution $p(b,c|x,y,z)$ can be completely obtained.

The RAC is initially considered between Alice and Bob (Charlie), where the task will succeed only when Bob (Charlie) can guess the $y$th bit $(x_0,x_1)$ of Alice's input. The average success probability is bounded $W_{AB}=\frac{1}{8}\sum_{x,y}p(b=x_y|x,y)\le\frac{3}{4}$ for any random classical strategies. While this bound will be surpassed by using quantum strategies. The optimal QRAC can achieved as $W_{AB}=\frac{1}{8}\sum_{x,y}\mathrm{Tr}[\rho_xM_{x_y|y}]=\frac{1}{2}(1+\frac{1}{\sqrt{2}})\approx 0.854$ \cite{Karthik.Mohan_NEW.J.PHYS.10.1088.1367-2630.ab3773_2019}. The four preparations are given as $\rho_{00}=\frac{1}{2}(\mathbb{I}+\frac{\sigma_x+\sigma_z}{\sqrt2})$, $\rho_{11}=\frac{1}{2}(\mathbb{I}-\frac{\sigma_x+\sigma_z}{\sqrt2})$, $\rho_{01}=\frac{1}{2}(\mathbb{I}+\frac{\sigma_x-\sigma_z}{\sqrt2})$, $\rho_{10}=\frac{1}{2}(\mathbb{I}-\frac{\sigma_x-\sigma_z}{\sqrt2})$. This QRAC shown above is independent of Charlie. Similarly, for the QRAC between Alice and Charlie, the corresponding witness of this QRAC is $W_{AC}=\frac{1}{8}\sum_{x,z}p(c=x_z|x,z)\le\frac{3}{4}$, and the QRAC is not independent of Bob.

For sequential QRACs exhibiting their optimal trade-off, the range of interest available to QRAC for $(W_{AB}, W_{AC})$ is limited in $[\frac{1}{2},\frac{(2+\sqrt{2})}{4}]$. In this range, either witness being $\frac{1}{2}-\epsilon$ is equivalent to a witness value of  $\frac{1}{2}+\epsilon$ by classically bit-flipping the outcomes. The trade-off between $W_{AB}$ and $W_{AC} $ can be formulated as: for a given $W_{AB}$ value (expressed as $\alpha $), find the maximum $W_{AC}$. The optimal QRAC can be achieved when Bob performs weak measurements with the unsharpened parameter $\eta\in[0,1]$, and Charlie performs strong measurements, $M_0=\sigma_x, M_1=\sigma_z$. The witness can be given as, $W_{AB}=\frac14(2+\eta\sqrt2), W_{AC}=\frac18(4+\sqrt2+\sqrt{s-s\eta^2})$, whose optimal trade-off corresponds to $W_{AC}^\alpha=\frac18(4+\sqrt2+\sqrt{16\alpha-16\alpha^2-2})$, $\alpha\in[\frac12,\frac{1+\frac{1}{\sqrt2}}{2}]$. This characterizes the nontrivial boundary of the quantum set in the space of witness pairs.

In a general scenario with multiple observers, the RACs between Alice and a sequence of Bobs have been investigated. It is shown that the longest sequence of QRAC is limited to 2, $W_{AB}=W_{AC}=\frac{5+2\sqrt2}{10}\approx0.7828>\frac34$.
This result is consistent with \cite{Silva.Ralph_PhysRevLett.114.250401_2015} where the sequential violations of the CHSH inequality.

Subsequently, \citet{Das.Debarshi_PhysRevA.104.L060602_2021} considered an unbiased scenario involving multiple independent pairs of observers performing sequential measurement on a singlet state, and addressed whether more than one pair of observers can demonstrate quantum advantage in some specific $n\to1 $ RAC, ($n\in\{2,3\}$).
The efficacy of the RAC protocol will quantified by minimum success probability, $P_{\mathrm{Min}}^{n\to 1}=\min_{x_{0},x_{1}\dots x_{n-1},y}P(b=x_{y})$, which is always less than or equal to $\frac{1}{2}$ for classic RAC tasks.

Moreover, these findings persist even if all observers make appropriate projective measurements and the initially shared state is separable.
Subsequently, \citet{Foletto.Giulio_PhysRevResearch.2.033205_2020} used single photons to experimentally demonstrate the feasibility of weak measurements and the achievability of nonclassical success probabilities with two decoders.
\citet{Li.Hong-Wei_Commun.Phys.s42005-018-0011-x_2018} demonstrated that double classical dimension witness violation is achievable if they choose appropriate weak
measurement parameters by applying the dimension witness inequality based on the quantum random access code and the nonlinear determinant value.
\citet{Anwer.Hammad_PhysRevLett.125.080403_2020} reported an experimental implementation of unsharp qubit measurements in a sequential communication protocol, based on a QRAC.
\citet{Shihui.Wei_NEW.J.PHYS.1367-2630/abf614_2021} conducted a three-party prepare-transform-measure experiment with unsharp measurements based on $3\to1$ sequential RAC, and derived optimal trade-off between the two correlation witnesses in $3\to1$ sequential QRACs.
\citet{Xiao.Ya_PhysRevResearch.3.023081_2021} proposed an entanglement-assisted sequential QRAC protocol which can enable device-independent tasks. \citet{Xiao.Yao_Quantum.Inf.Process.s11128-023-03924-3_2023} investigated the sequential $3\to1$ parity-oblivious QRAC using different sharpness parameters for different measurement settings.


\subsection{Random Number Generation}

Randomness refers to the uncertainty exhibited by each event in a set of events with a certain probability. The generation of randomness has a wide range of applications, such as in cryptography, simulation algorithms, and so on. The use of quantum correlation to generate and verify true randomness has received widespread attention and research (\citealp{Popescu.Found.Phys.24.3791994};
\citealp{Masanes.PhysRevA.73.012112};
\citealp{DAriano.J.Phys.A.38.5979_2005};
\citealp{Pironio.188.Nature.464.1021_2010};
\citealp{Pironio_2016_PhysRevA.93.040102}). The quantum correlation sharing model can be used to further investigate the properties of quantum random number generation (QRNG). \citet{Curchod.F.J_PhysRevA.95.020102_2017} proposed a method for generating random numbers using sequential measurements, demonstrating that an infinite number of random bits can be produced from a pair of entangled qubits.

In this work, they analyzed and quantified the randomness generation by non-local guessing games in a sequential scenario \cite{Antonio_2012_PhysRevLett.108.100402,
Torre_2015_PhysRevLett.114.160502}. 
Supposed that an additional adversary Eve (E) having a quantum system which may be related to one of A and B, E prepares a state $\rho{ABE}$ and distributes $\rho_{ABE}$ and distributes $\rho_{AB}=\mathrm{Tr}_E[\rho_{ABE}]$ to A and B respectively. The purpose of E is to attempt to guess the outcomes of Bob by performing measurements on $\rho_{ABE}$. 
The other observers treat their respective devices as black boxes, and the probability distribution $P_{\mathrm{obs}}(a,\vec{b}|x,\vec{y})$ is obtained from their inputs and outputs. This probability can be obtained from the extreme points, which is 
$P_{\mathrm{obs}}=\sum_{\mathrm{ext}}q_{\mathrm{ext}}P_{\mathrm{ext}}$, where $\sum_{\mathrm{ext}}q_{\mathrm{ext}}=1$ and $q_{\mathrm{ext}}\ge0$.
Once E knows the probability distribution at the extreme points, she can fully understand how to limit her predictive ability through decomposing $P_{\mathrm{obs}}$ into extreme points.  The predictive power is quantified through the device-independent guessing probability (DIGP), $G(\vec{y}^0,P_{\mathrm{obs}})$.
where the input string $y_1^0,y_2^0..y_n^0\equiv\vec{y}^0$, and E should guess the output $\vec{b}$. The problem changes to the optimal solution of such optimization problems,
\begin{align}
	G(\vec{\mathrm{y}}^{0},P_{\mathrm{obs}})=\operatorname*{max}_{\{q_{\mathrm{ext}},P_{\mathrm{ext}}\}}\sum_{\mathrm{ext}}q_{\mathrm{ext}}\operatorname*{max}_{\vec{b}}p_{\mathrm{ext}}(\vec{b}|\vec{\mathrm{y}}^{0}),
\end{align}
where $
p_{\mathrm{ext}}(\vec{b}|\vec{y}^{0})=\sum_{a}p_{\mathrm{ext}}(a,\vec{b}|x,\vec{y}^{0}), \forall x$; $
P_{\mathrm{obs}}=\sum_{\mathrm{ext}}q_{\mathrm{ext}}P_{\mathrm{ext}}, P_{\mathrm{ext}}\in\mathcal{Q}
$.
It is clearly shown that m random bits are returned once  $G(\vec{y}^0,P_{\mathrm{obs}})=2^{-m}$.
Subsequently, 
with the condition that $P_{\mathrm{obs}}=\sum_{\mathrm{ext}}q_{\mathrm{ext}}P_{\mathrm{ext}}$ is dissatisfied and only mandatory observed statistical $P_{\mathrm{obs}}$ give specific Bell inequality violations,
the optimal solution can be given as the upper bound of the optimal solution of the equation $P_{\mathrm{obs}}=\sum_{\mathrm{ext}}q_{\mathrm{ext}}P_{\mathrm{ext}}$.

In this case, Alice and Bobs share a pure two-qubit state and use the CHSH inequality as a criterion, $I_\theta=\beta\langle B_0\rangle+\langle A_0B_0\rangle+\langle A_1B_0\rangle+ \langle A_0B_1\rangle-\langle A_1B_1\rangle$. 
The maximal violation can be reached, $I^{\max}=2\sqrt2\sqrt{2+\beta^2/4}$, when Alice's two measurement choices are $\hat{A}_0=\cos\mu\sigma_z+\sin\mu\sigma_x,\hat{A}_1=\cos\mu\sigma_z-\sin\mu\sigma_x$, and for Bob as $\hat{B}_0=\sigma_z,\hat{B}_1=\sigma_x$.
When the maximum violation proves that Bob's second measurement choice is $y_0=1: G(\vec{y}^0=1,P_{\mathrm{obs}})=\frac12$, and then $I_\theta\to I^{\max}_{\theta}\Rightarrow G(y^0=1,P_{\mathrm{obs}})\to \frac12$. Supposed that each Bobs also has two measurement choices with $y_i=0$ and $y_i=1$, corresponding $\sigma_z$ and $\hat{\sigma}_x(\xi_i)$ respectively, where $\xi_i$ is the sharpness parameter.
If the inequality $I_{\theta_{\vec{b}_{i-1}}}$ between Bob$_i$ and Alice can approach to the maximum value infinitely, which ensures that the guessing probability $G(\vec{y}^0=1,P_{\mathrm{obs}})$ can be arbitrarily close to $\frac12$, thus verifying approximately one random bit. When $\xi$  approaching 0, DIGP(G) can be arbitrarily close to $2^{-n}$. Then all Bobs can produce m random bits by a suitable sequence $\hat{\sigma}_x(\xi_i)$, $i\in\{1,2...n\}$ measurements, where n>m.
This certification only requires that each Bob occasionally select the projective measurement $\sigma_z$ to obtain the whole statistics.
Bob can choose $\sigma_z$ with probability $\gamma_i$ and $\hat{\sigma}_x(\xi_i)$ with probability of $1-\gamma_i$. 
By setting $\gamma_i$ as close to zero as possible, Bob has complete freedom to choose. This allows the demonstration of the randomness of each Bob's measurements in the sequence, at the expense of Alice's increased measurement choices. 

Their results show that an infinite number of random bits can be derived from a pair of entangled quantum bits when one of them undergoes a series of measurements, which requires an exponential increase in the measurement choices of Alice, $\sum^n_{i=1}2^i$ measurement choices for n measurements in the sequence. It is valuable to investigate the design of device-independent protocols for generating randomness that involves sequential measurement scenarios. It is worth pointing out that, the sequential measurement scenario is a more significant challenge in practical experiments. And the efficiency of the sequential protocol compared to the standard Bell test remains unknown.  

Other related developments include, \citet{BrianCoyle_EPTCS.273.2_2018} presented a new protocol that allows an unbounded amount of randomness to be certified as being legitimately the consequence of a measurement on a quantum state.  
\citet{Xue-Bi.An_Opt.Lett.ol-43-14-3437_2018} proposed and realized a quantum random number generator among three observers based on the weak measurement technique.
\citet{Bowles_Quantum.q-2020-10-19-344_2020} considered a more general scenario in which one performs sequences of local measurements on an entangled state of arbitrary dimension with the methods of NPA hierarchy.
\citet{Foletto.Giulio_PhysRevA.103.062206_2021} showed that the weak measurements are realizable through experimental tests, but can improve the performance of randomness generation only in close-to-ideal conditions.
\citet{Pan.A.K_PhysRevA.104.022212_2021} provided an interesting two-party parity-oblivious communication game whose success probability is solely determined by the Bell expression. The parity-oblivious condition in an operational quantum theory implies the preparation of noncontextuality in an ontological model of it.

\subsection{Self-Testing}

Self-testing is a fundamental physics method for inferring quantum experiments in black box scenarios, representing the most powerful form of authentication for quantum systems (\citealp{Mayers_2003arXiv.quant.ph.0307205};
\citealp{McKague_2016_NewJ.Phys.18.045013};
\citealp{McKague_2017_selftesting};
\citealp{Supic_2020_selftesting_review};
\citealp{Supic_2021_Quantum.5.418_2021}). 
At present, the investigations of self-testing are often closely related to the Bell test
(\citealp{Yang_2013_PhysRevA.87.050102};
\citealp{Bancal_2015_PhysRevA.91.022115};
\citealp{Supic_2016_NewJ.Phys.18.075006};
\citealp{Coladangelo_2017_Naturecomms15485};
\citealp{Coopmans_2019_PhysRevA.99.052123}).

Taking a simple self-testing scheme as an example, a source generating a quantum system and distributing it between two laboratories, the observers in these two laboratories can perform different local experiments by changing the device settings. In device-independent constraint, each laboratory can be considered as a black box, which selects labels $(x, y)$ corresponding to the specific settings of the experiment as input and outputs corresponding results $(a, b)$. Repeating performing local experiments, an estimated probability P (a, b | x, y) can be obtained. The purpose of self-testing of the quantum state and measurement is to infer the source and measurement of the laboratories only from probability $P (a, b | x, y) $ in such a black box scenario (\citealp{Mayers_2003arXiv.quant.ph.0307205};
\citealp{McKague_2016_NewJ.Phys.18.045013};
\citealp{McKague_2017_selftesting}). 

\subsubsection{The Self-testing Of QRACs}
Inspired by the nonlocality sharing via sequential measurements, \citet{Karthik.Mohan_NEW.J.PHYS.10.1088.1367-2630.ab3773_2019} constructed semi-device independent self-tests of quantum instruments. They showed that self-testing is possible with the optimal trade-off between two sequential QRACs. 

In a three-party prepare-transform-measure scenario, the investigation of the ability to execute QRAC for Bob and Charlie, is equivalent to exploring the optimal trade-off of ($W_{AB},W_{AC}$). For self-testing of QRACs, it is necessary to prove that the optimal QRAC pairs can only be achieved through unique preparations, instruments, and measurements (until a collective unitary transformation).

Based on the optimal sequence QRAC $(W_{AB},W_{AC})=(\alpha, W_{AC}^\alpha)$ (as mentioned above in Sec. \ref{V-A-QRAC}), such self-testing argument can be established as follows \cite{Karthik.Mohan_NEW.J.PHYS.10.1088.1367-2630.ab3773_2019}:
(i) Alice's states are pure, pairwise antipodal on the Bloch sphere, and they form a square on the Bloch sphere.
(ii) Bob’s instruments are Kraus operators $\hat{K}_{b|y}=\hat{U}_{yb}\sqrt{\hat{M}_{b|y}}$, corresponds to unsharpen measurements along the diagonals of Alice's square of preparations, until its collective unitary. Specifically, $\forall y,b: \hat{U}_{yb}=U, \hat{M}_0=\eta\sigma_x$ and $\hat{M}_1=\eta\sigma_z$, where $\sigma=\sqrt2(2W_{AB}-1)$.
(iii) Charlie's measurement is the first order projection along the diagonal of Alice's square of preparations, until Bob's unitary, $\hat{C}_0=\hat{U}\sigma_x\hat{U}^\dagger,
\hat{C}_1=\hat{U}\sigma_z\hat{U}^\dagger$.
This self-testing remains valid until a collective reference frame is chosen.

\subsubsection{Self-testing Of Unsharp Measurements}

Subsequently, \citet{Prabuddha.Roy_NEW.J.PHYS.10.1088.1367-2630.acb4b5_2023} provided device-independent (DI) self-testing of the unsharp instrument through the quantum violation of two Bell inequalities.

In a sequential CHSH scenario, Alice performs local measurements $\hat{A}_1$ and $\hat{A}_2$ on her subsystem upon receiving inputs $x$. For Bob receives inputs $y_k\in\{1,2\}$, whose measurement operator is represented by $\hat{K}_{\pm|y_k}=\alpha_k\mathbb{I}\pm\beta_k\hat{B}_{y_k}$, where $\alpha_k=\frac12(\sqrt{\frac{1+\lambda_k}{2}}+\sqrt{\frac{1-\lambda_k}{2}})$, $\beta_k=\frac12(\sqrt{\frac{1+\lambda_k}{2}}-\sqrt{\frac{1-\lambda_k}{2}})$, and $\alpha_k^2+\beta^2_k=\frac12,\alpha_k\ge\beta_k$.
Supposed that each Bob measures the same set of measurements $\hat{B}_1$ and $\hat{B}_2$, the effective observables for Bob (un-normalized) can be written as $\tilde{B}_1=(2\alpha ^2_1+\beta^2_1)\hat{B}_1+\beta_1^2\hat{B}_2\hat{B}_1\hat{B}_2$ and $
\tilde{B}_2=(2\alpha ^2_1+\beta^2_1)\hat{B}_2+\beta_1^2\hat{B}_1\hat{B}_2\hat{B}_1$.
The CHSH value of Alice-Bob$_1$ is given as $(\mathcal{B}^1)_Q=\lambda_1{\mathcal{B}}^{\mathrm{opt}}_Q $, while for Alice-Bob$_2$ as $(\mathcal{B}^2)=\mathrm{Tr}[\rho_{AB_1}((\hat{A}_1+\hat{A}_2)\tilde{B}_1+ (\hat{A}_1-\hat{A}_2)\tilde{B}_2)] $. Here, we have $(\tilde{B}_1)^2\ne \mathbb{I}$,$(\tilde{B}_2)^2\ne \mathbb{I}$ and $\omega_1=||\tilde{B}_1)||,\omega_2=||\tilde{B}_2)||$.
By utilizing the SOS method, we can determine its maximal value with  $(\mathcal{B}^2)_Q=\max(\omega_1\tilde{\omega}_1+ \omega_2\tilde{\omega}_2)$.

For Bob’s (unnormalized) measurements require that $\{\tilde{B}_1,\tilde{B}_2\}=4\alpha^2_1(\alpha^2_1+2\beta^2_1)\{\hat{B}_1,\hat{B}_2\}+\beta^4_1\{\hat{B}_1,\hat{B}_2\}^3=0$. As $\alpha_1>0$, $\beta_1\ge0$, it imply that  $\{\hat{B}_0,\hat{B}_1 \}=0$. Hence, in order to achieve the maximum average value of the Bell expression, Bob's measurement of Bob$_2$ needs to be anticommuting. This yields  $\tilde{\omega}_1=\sqrt{(2\alpha^2_1)^2+(2\alpha^2_1+\beta^2_1)\beta^2_1\{\hat{B}_1, \hat{B}_2\}^2}=2\alpha^2_1$. The CHSH value for Alice-Bob$ _2$ reduces to $(\mathcal{B}^2 )_Q=2\alpha^2_1\max(\omega_1+\omega_2)=2\alpha^2_1(\mathcal{B})^{\mathrm{opt}}_Q$, when substituted into the value of $\alpha_1$ becomes $(\mathcal{B}^2)_Q=\frac12(1+\sqrt{1-\lambda_1^2})(\mathcal{B})^{\mathrm{opt}}_Q$.
The maximum average value of the Bell expression for two sequences of Bobs is, 
\begin{align}
	(\mathcal{B}^2)_Q=\sqrt2(1+\sqrt{1-(\frac{(\mathcal{B}^1)_Q}{(\mathcal{B})^{\mathrm{opt}}_Q})^2}).
\end{align}

Consequently, there exists a trade-off between $(\mathcal{B}^1)_Q$ and $(\mathcal{B}^2)_Q$, resulting in an optimal pair that demonstrates the certification of the unsharpness parameter.
The suboptimal value $(\mathcal{B}^1)_Q$ and $(\mathcal{B}^2)_Q$ forms an optimal pair $\{(\mathcal{B}^1)_Q,(\mathcal{B}^2)_Q\}$, which uniquely proves the shared state, the measurement set, and the unsharp parameter $\lambda_1$ between Alice and Bob$_1$.

The DI self-testing of this protocol is as follows \cite{Prabuddha.Roy_NEW.J.PHYS.10.1088.1367-2630.acb4b5_2023}:
(i) Alice makes sharp measurements of two anticommuting observable values on her local subsystem in any arbitrary local dimension.
(ii) Bob$_1$ performs unsharp measurements with two possible measurements, which are also anticommuting in any arbitrary local dimension. The measurement sets of Bob$_1$ and Bob$_2$ are the same.
(iii) Alice and Bob$_1$ share the maximum entangled state in any arbitrary dimension.
(iv) The optimal pair $\{(\mathcal{B}^1)_Q,(\mathcal{B}^2)_Q\}$ self-tests unsharp parameters, which in turn proves the shared entanglement state between Alice, Bob$_1$ and Bob$_2$.

Additionally, they broaden it to encompass the elegant Bell inequality. Suppose that it possesses two classical bounds—the local bound and the non-trivial preparation non-contextual bound, the latter being lower than the local bound. Utilizing the shared preparation contextuality among three independent sequential observers, the DI self-testing of two unsharpness parameters has been demonstrated. The robustness of their  certification has been discussed \cite{Prabuddha.Roy_NEW.J.PHYS.10.1088.1367-2630.acb4b5_2023}.

The other relative work includes: \citet{Miklin.Nikolai_PhysRevResearch.2.033014_2020} proposed a scheme for semi-device-independent self-testing of unsharp measurements and showed that all two-outcome qubit measurements can be characterized in a robust way.
\citet{Armin.Tavakoli_sciadv.aaw6664_2020} investigated self-testing of nonprojective quantum measurements theoretically and experimentally. That is, how can one certify, from observed data only, that an uncharacterized measurement device implements a desired nonprojective positive-operator-valued measure (POVM).
\citet{Mukherjee.Sumit_PhysRevA.104.062214_2021} provided semi-device-independent self-testing protocols in the prepare-measure scenario to certify multiple unsharpness parameters along with the states and the measurement settings.
\citet{Pan.A.K_PhysRevA.104.022212_2021} demonstrated device-independent self-testing of projective and non-projective measurements.
Additionally, \citet{Chirag_PhysRevA.103.032408_2021} employed a measurement-device-independent entanglement witness to detect entanglement in a scenario where half of an entangled pair is possessed by a single observer while the other half is with multiple observers performing unsharp measurements, sequentially, independently, and preserving entanglement as much as possible.

\section{Discussions And Conclusions}\label{conclusion_section}

\emph{Discussions-} In the field of quantum phenomena, nonlocal sharing is a fascinating frontier that allows us to witness the intrinsic nature of particle entanglement, challenging classical intuition. As we delve into the complexity of quantum correlations, the study of nonlocal sharing opens up another avenue for exploration. This not only provides richer perspectives on our understanding of quantum mechanics and its practical implications but also holds exciting possibilities for the development of Quantum Information Science. However, current research on quantum correlation sharing and even the reuse of quantum resources, whether in terms of models or theoretical structures, remains relatively limited. In the following discussion, we summarize some open problems that are yet to be solved and require overcoming. In-depth research on these issues may lead to new breakthroughs in our understanding of quantum correlation sharing.

Essentially, the issue of sharing quantum correlations is closely linked to the trade-off between information gain and disturbance of the system. In a broader context, the optimal measurement choices for quantum correlation sharing remain unclear \cite{Silva.Ralph_PhysRevLett.114.250401_2015}, especially for more complex high-dimensional and multi-body systems. Hence, there is an interesting question, what is the best form of pointer when making measurements in high-dimensional systems?

Currently, sporadic research findings suggest that high-dimensional quantum correlation systems seem to be more easily able to achieve correlation sharing in various scenarios. For instance, a 2-qutrit system can achieve nonlocal sharing among more observers in bilateral sequential measurements \cite{cabello_2021_bell}, which is impossible for the original 2-qubit system \cite{Mal.Shiladitya_math4030048_2016}. These conclusions align with our intuitive judgment. However, we still need more evidence to understand the nonclassical correlation sharing in high-dimensional systems.

In high-dimensional or quantum networks, the sharing involves not only the strength of correlations but may also extend to sharing from global correlations to local correlations. In multi-party correlations, exploring other important features of quantum correlations, such as entanglement depth \cite{Anders_PhysRevLett.86.4431_2001}, entanglement intactness \cite{luhe_PhysRevX.8.021072_2018}, and characteristics based on quantum measurements, are equally intriguing. These research directions have the potential to offer us fascinating perspectives.

Regarding the issue of correlation sharing, the progress of research varies for different definitions of quantum correlations. For instance, there has been considerable advancement in the study of nonlocality sharing, and some research has been conducted on quantum steering. However, compared to other nonclassical correlations, research in these areas is relatively straightforward, with many studies directly drawing from earlier research. Therefore, further exploration is needed for research on other types of correlation sharing.

From the perspective of resource theory \cite{Chitambar_RevModPhys.91.025001_2019}, it is currently unclear whether the correlation sharing based on sequential measurements indeed consumes fewer resources to handle quantum information tasks or under what scenarios it may have advantages. Classifying various information processing tasks involving sequential measurements based on the advantages of resource theory is another potentially attractive direction for future research \cite{Arun.Kumar.Das_Quantum.Inf.Process.s11128-022-03728-x_2022}. Exploring more general resource-based comparison and analysis methods between sequential and non-sequential scenarios will contribute to the advancement of this field.

The observation of quantum correlations introduced the concept of measurement, giving rise to research on quantum correlation sharing. This research began with early weak measurement models, evolved into generalized POVM models, and progressed to demonstrating quantum correlation sharing using combined strategies with strong measurements. It is crucial to explore whether other physical strategies that can achieve similar phenomena. Even with modifications to known physical models, such as investigating whether strong measurement combinations can achieve infinite sharing, remains a research question. Recent work has provided valuable insights, demonstrating correlation sharing through combined projective measurements \cite{Sasmal_2023_arXiv_unbounded}. Similarly, understanding how to achieve more generalized sharing in bilateral or even multipartite scenarios is a topic worthy of in-depth exploration.

Regarding the experimental observation of quantum correlation sharing, most current experiments are demonstrations in 2-qubit systems. Exploring more complex correlation sharing in higher-dimensional or multi-body physical systems is a topic that needs further consideration. Essentially, we need to design more explicit correlation-sharing schemes. Additionally, current experiments are limited to optical systems, and there is a need to expand experimental demonstrations to more diverse physical platforms.

In the context of random code problems under sequential measurements, especially when extended to higher dimensions and involving more sequences of observers, there are still some unclear but evidently crucial questions. It is essential to explore the optimal trade-offs among sequential Quantum Random Access Coding (QRAC) involved in the preparation scenario tests. What are the essential differences in sequential QRAC when different quantum strategies are employed? How to establish a more comprehensive set of comparison methods? These are questions that require in-depth investigation.

Exploring self-testing problems based on sequential measurements, especially in the context of noise resistance analysis, is undoubtedly a highly meaningful research avenue. Designing optimal sequential measurement schemes is a feasible path by considering different scenarios. This includes both the self-testing of quantum states and the self-testing process of generalized POVM measurement operators.

In addition to known application models, further contemplation is needed to identify scenarios suitable for non-classical correlations. For instance, exploring the possibility of extending these scenarios to encompass a broader study of numerous quantum game models. This includes investigating the application of nonlocal sharing in conflicting nonlocal games, exploring incomplete information games and perfect information games, determining the dynamics of deterministic and nondeterministic games, and analyzing symmetric and asymmetric information games, among other possibilities.

\emph{Conclusions-}This paper comprehensively reviews and summarizes the recent progress in quantum correlation sharing based on sequential measurements. Starting with the interpretation of joint probability, we introduce the study of quantum correlations considering real measurement methods. In the second part, we delve into the exploration of nonlocal sharing under different measurement strategies and scenarios. Specifically, we investigate the impact of measurement strategies on the sharing of quantum nonlocality, examining perspectives such as ``asymmetry'' and ``weak value.'' Through detailed analyses in diverse scenarios, we assess the potential of nonlocal sharing and provide a retrospective overview of experimental efforts related to nonlocal sharing. The third part introduces the research findings on steering sharing, clarifying the feasibility of steering sharing and summarizing the properties of quantum steering sharing in different scenarios. In the fourth part, we discuss the sharing of other types of quantum correlations, including network nonlocality, quantum entanglement, and quantum contextuality. In the fifth part, we review the progress of quantum correlation sharing based on quantum sequential measurement strategies in applications such as quantum random access coding, random number generation, and self-testing tasks. Finally, we deduce and analyze important open questions in this research topic and provide a summary.

}

\section{Acknowledgment}

C.R. was supported by the National Natural Science Foundation of China (Grant No. 12075245, 12247105), the Natural Science Foundation of Hunan Province (2021JJ10033), Xiaoxiang Scholars Programme
of Hunan Normal University, the Foundation Xiangjiang-Laboratory (XJ2302001) and Hunan provincial
major sci-tech program No. (2023zk1010).

\bibliographystyle{apsrmp4-2}
\bibliography{Introduction_ref,1_A,88_sharing,application}


\end{document}